\documentclass[aps,twocolumn,prd,showpacs,nofootinbib,preprintnumbers]{revtex4-2}
\usepackage{bm}
\usepackage{latexsym}
\usepackage{url}
\usepackage{color}
\usepackage{amsfonts,amssymb,amsmath}
\usepackage{mathtools}
\usepackage{graphicx,epsfig}
\usepackage{amsmath}
\usepackage{comment}
\usepackage{psfrag}
\usepackage{subfigure}
\begin{document}
\title{Probing massive gravitons in $f(R)$ with  lensed gravitational waves}
\author{Vipin Kumar Sharma}
\email[]{\textcolor[rgb]{0.00,0.00,0.00}{vipinastrophysics@gmail.com}}
\author{Sreekanth Harikumar}
\email[]{\textcolor[rgb]{0.00,0.00,0.00}{sreekanth.harikumar@ncbj.gov.pl}}

\author{Margherita Grespan}
\email[]{\textcolor[rgb]{0.00,0.00,0.00}{margherita.grespan@ncbj.gov.pl}}
\author{Marek Biesiada}
\email[]{\textcolor[rgb]{0.00,0.00,0.00}
{marek.biesiada@ncbj.gov.pl}}

\author{Murli Manohar Verma}
\email[]{\textcolor[rgb]{0.00,0.00,0.00}
{murli.manohar.verma@cern.ch}}
\affiliation{$^{*,\P}$Department of Physics, University of Lucknow, Lucknow 226 007, India \\
$^{\dag, \ddag, \S}$National Centre for Nuclear Research, Andrzeja Sołtana 7, Otwock, 100190, Poland\\
$^{\P}$Theoretical Physics Department, CERN, CH-1211 Geneva 23, Switzerland.}

\begin{abstract}
We investigate the novel features of gravitational wave solutions in $f(R)$ gravity under proper gauge considerations in the shifted Ricci scalar background curvature ($R^{1+\epsilon}$). The solution is further explored to study the modified dispersion relations for massive modes at local scales and to derive constraints on $\epsilon$. Our analysis yields new insights as we scrutinize these dispersion effects on the polarization (modified Newman-Penrose content) and lensing properties of gravitational waves. It is discovered that the existing longitudinal scalar mode, and transverse breathing scalar mode are both independent of the mass parameter for $\epsilon<<1$. Further, by analysing the lensing amplification factor for the point mass lens model, we show that lensing of gravitational wave is highly sensitive to these dispersion effects in the milli-Hertz frequency (wave optics regime). It is expected that ultra-light modes, having mass about $\mathcal{O} (10^{-15})$ eV  for $\epsilon<<1 (\approx 10^{-7})$ lensed by ($10^3\leq M_{Lens}\leq 10^6$)$M_\odot$ compact objects are likely to be detected by the advanced gravitational wave space-borne detectors, particularly within LISA's (The Laser Interferometer Space Antenna) sensitivity band.
\end{abstract}
\maketitle{}

\section{\label{1}Introduction}

General Relativity (GR) is considered as one of the most accurate descriptions of gravity and has undergone rigorous testing over the past century. The most recent detection of gravitational waves (GWs)  by LIGO Science Collaboration represents a significant test of  GR in the strong-field regime \textcolor[rgb]{0.00,0.00,1.00}{\cite{LIGOScientific:2016aoc, LIGOScientific:2016lio,LIGOScientific:2017vwq,LIGOScientific:2017zic}}. However, despite all these tests, the domain of validity of GR has been questioned by the existing open issues such as the nature of dark energy, dark matter, cosmological tensions like  Hubble Constant, $H_0$ and matter fluctuation amplitude, $\sigma_8$ along with the non-renormalizability of Einstein-Hilbert (E-H) action \textcolor[rgb]{0.00,0.00,1.00}{\cite{Green:2021jrr,Riess:2019cxk,Silk:2016srn,DelPopolo:2016emo,Planck:2018vyg,Macaulay:2013swa,Charnock:2017vcd,Heisenberg:2018vsk}}. To address these existing problems,  several modifications to GR have been proposed in the literature \textcolor[rgb]{0.00,0.00,1.00}{\cite{Heisenberg:2018vsk,Wang:2020dsc,Saridakis:2023pzo,DeFelice:2010aj,Nojiri:2010wj,Nojiri:2017ncd}}. Some of them stand out as potential alternatives to GR. One such  modified theory is  $f(R)$  gravity \textcolor[rgb]{0.00,0.00,1.00}{\cite{DeFelice:2010aj,Faraoni:2010pgm,Nojiri:2017ncd,Nojiri:2010wj}}. These classes of theories in which the gravitational action is generalized to be a function of the Ricci Scalar are widely used to address the above-mentioned issues, such as in explaining the early and late time cosmic acceleration, large-scale structure formation of the universe, and galactic dynamics. In particular,  power-law modification has been phenomenologically investigated in the literature \textcolor[rgb]{0.00,0.00,1.00}{\cite{DeFelice:2010aj,Faraoni:2010pgm,Clifton:2005aj,Boehmer:2007kx,Nojiri:2017ncd,Nojiri:2010wj}}, and more recently, the authors of this paper have conducted investigations through precise modeling of deviations in the standard Ricci scalar at various scales \textcolor[rgb]{0.00,0.00,1.00}{\cite{KumarSharma:2022qdf,Sharma:2020vex,Sharma:2019yix,Yadav:2018llv,Sharma:2022fiw,Sharma:2022tce}}. These results show that the predictions of such $f(R)$ models differ significantly from GR as well as address the existing cosmological and astrophysical conundrums.

While significant advancements have been achieved in experimental endeavors, the development of phenomenological models to explain certain characteristic aspects like dispersion, polarization, and lensing of GWs in the framework of modified gravity theories is still at an early stage \textcolor[rgb]{0.00,0.00,1.00}{\cite{Casado-Turrion:2023rni,Capozziello:2017vdi,Katsuragawa:2019uto,Will:2014kxa,Berry:2011pb,Ezquiaga:2020dao,Goyal:2020bkm,Katsuragawa:2019uto,Will:2018gku, Liang:2017ahj,Nishizawa:2016kba,Alves:2009eg,Will:1997bb,Odintsov:2022cbm}}. In the era of multi-messenger astronomy \textcolor[rgb]{0.00,0.00,1.00}{\cite{LIGOScientific:2017zic}}, such phenomenological investigations of theories beyond GR are crucial in observational astronomy \textcolor[rgb]{0.00,0.00,1.00}{\cite{Mukherjee:2019wcg,Mukherjee:2020mha,Liao:2017ioi,meena2023gravitational,Takahashi_2017,Hou_2020}}. Furthermore, the utilization of GWs as a tool to test theories of gravity is well established in the literature \textcolor[rgb]{0.00,0.00,1.00}{\cite{Eardley73,Newman:1961qr,Nishizawa:2017nef,Will:2014kxa,Arun:2013bp,Alves:2009eg}}. The recent polarization tests conducted by the LIGO/Virgo scientific collaboration have paved the way to carry out more stringent polarization tests \textcolor[rgb]{0.00,0.00,1.00}{\cite{LIGOScientific:2017ous,Fesik2017LIGOVirgoEL,Takeda:2020tjj}}. This will become especially pertinent as additional detectors are integrated into the network, which is expected to enhance the precision of the constraints on modified gravity models.
These potential avenues have rekindled interest in modified theories of gravity, introducing novel aspects of polarizations \textcolor[rgb]{0.00,0.00,1.00}{\cite{Hyun:2018pgn,Goyal:2020bkm}}. Tests focusing on propagation effects are designed to investigate theories that predict GWs to be nearly identical to that of GR but differ in the way the waves propagate in (non)dispersive background. This phenomenon is particularly relevant in theories such as massive graviton theories (like scalar-tensor, scalar-tensor-vector, etc.). For instance the first multi-messenger event GW170817 helped us obtain precise constraints of graviton mass and rule out certain theories \textcolor[rgb]{0.00,0.00,1.00}{\cite{Shoom:2022cmo,Piorkowska-Kurpas:2022xmb,Jana:2018djs,Svidzinsky:2018hnx}}.

The next most awaited propagating phenomenon in GWs is the detection of lensed signals. Like electromagnetic waves, the presence of massive astrophysical sources along the line of sight between the source and the detector is likely to lens the GW signal \textcolor[rgb]{0.00,0.00,1.00}{\cite{Nakamura:1999uwi,Takahashi:2003ix}}. This leads to the formation of multiple signals with specific time-delay which has several applications ranging from precision cosmology \textcolor[rgb]{0.00,0.00,1.00}{\cite{liao2017precision,Sereno:2011ty,Li:2019rns}} to detection of Primodial Black Holes \textcolor[rgb]{0.00,0.00,1.00}{\cite{Diego:2019rzc,Oguri:2020ldf}} and Intermediate Mass Black Holes \textcolor[rgb]{0.00,0.00,1.00}{\cite{Lai:2018rto,meena2023gravitational}} and, also gives us an opportunity to test GR \textcolor[rgb]{0.00,0.00,1.00}{\cite{LISACosmologyWorkingGroup:2022jok,Ezquiaga:2020dao,Goyal:2020bkm, Fan:2016swi, Collett:2016dey}}. Furthermore, the next-generation GW detectors, such as LISA \textcolor[rgb]{0.00,0.00,1.00}{\cite{amaro2017laser}}, Einstein Telescope (ET)
 \textcolor[rgb]{0.00,0.00,1.00}{\cite{Maggiore:2019uih}},  Cosmic Explorer(CE) \textcolor[rgb]{0.00,0.00,1.00}{\cite{Evans:2021gyd}}, and DECIGO \textcolor[rgb]{0.00,0.00,1.00}{\cite{Kawamura:2020pcg}}, are highly sensitive and capable of observing a larger volume, reaching up to redshift z $\approx$ 20 and beyond. As a result, one can anticipate signals from cosmologically distant sources and thereby increasing the probability of lensed GW events. Lensing estimates  for different type of detectors has been studied in the literature \textcolor[rgb]{0.00,0.00,1.00}{\cite{Biesiada:2014kwa,Piorkowska2021}}, for instance, ET is likely to observe about 50 lensed events per year. Unlike EM lensing, the wavelength of GW signals spans from a few km to parsec scales which are comparable to the size of astrophysical sources serving as lenses. As a result, there are two regimes in which lensing is studied: Geometric Optics (GO) and Wave Optics (WO) \textcolor[rgb]{0.00,0.00,1.00}{\cite{Takahashi:2003ix,Nakamura:1999uwi}}. Theories in which the graviton is massive show a dispersion relation, which would impact the amplification factor used in the lensing studies \textcolor[rgb]{0.00,0.00,1.00}{\cite{Chung:2021rcu}}. It has been reported that observable deviations could be obtained in the low-frequency regime accessible to LISA and  DECIGO. For recent reviews on the lensing of GWs in the literature and by the authors of this paper we refer the readers to the following works \textcolor[rgb]{0.00,0.00,1.00}{\cite{Grespan:2023cpa,Biesiada:2021pzo}}

These theoretical insights are reflected in our analysis. The subject of this paper delves into the examination of power-law deviations from GR, which is done through the assessment of various propagating facets of GWs including dispersion, polarization, and lensing in modified gravity background.
Therefore, in this study, we have examined the aforementioned facets through  $f(R) \propto R^{1+\epsilon}$ model. It is a general feature of metric $f(R)$  theories to have propagating scalar degree of freedom (also known as scalaron) along with the tensor contribution present in GR.  The scalarons present in $f(R)$ can have different modes, viz., massive and massless modes, depending on the functional form of $f(R)$. To distinguish independent propagating solutions in modified gravity, Newman-Penrose (NP) formalism is employed as a tool which is well known in the literature \textcolor[rgb]{0.00,0.00,1.00}{\cite{Eardley:1973br}}. However, the modes here are massive and therefore a modified version of NP formalism is required which has been explored recently by \textit{Hyun et al.} \textcolor[rgb]{0.00,0.00,1.00}{\cite{Hyun:2018pgn}} which has been used in our study.

Hence, the non-negligible mass of the scalaron exerts a notable influence on dispersion relations and further, it have a significant effect on the modified NP polarization contents, and on the amplification factor involved in lensing.
Earlier research conducted by the LVK Collaboration also examined various strong and microlensing indications for events during the first part of the third observing run (O3a) \textcolor[rgb]{0.00,0.00,1.00}{\cite{LIGOScientific:2023bwz,LIGOScientific:2021izm}}. Although, these investigations did not produce definitive evidence of GW lensing.
Maintaining a positive outlook, we anticipate the identification of lensing signatures  when the next generation of GW  observatories, such as LISA, ET and CE becomes operational. This will also enable us to explore the implications of GW lensing (especially as a diagnostics) in modified gravity more comprehensively.

The contents of the paper are organized as follows: In Section II, we formulate our model and discuss the field equations. Followed by in Section III, we discuss the GW solutions. In Section IV, we investigate the scalarons as massive gravitons through the discussion of modified dispersion relations and place bounds on the propagating massive scalar mode in different backgrounds.
Further, for the study of propagating massive scalar modes, we explore the polarization contents in the modified N-P formalism with the discussion of gauge artifacts for distinguishing the massless scalar modes from the massless tensor modes and obtain the modified NP quantities for the $f(R)\propto R^{1+\epsilon}$ model under section V. In Section VI, we have analytically discussed under wave optics, the gravitational lensing of perturbed signals as a diagnostic tool to characterize massive and massless signals. In Section VII, we conclude our work with a summary and discussion of results with a future outlook.
Throughout the text, natural units of $c=\hbar=1$ are assumed.

\section{\label{2}$f(R)$ gravity and field equations}
$f(R)$ gravity fulfills the necessary conditions stipulated by Lovelock's theorem, thereby providing a suitable framework to extend Einstein's General Relativity theory \textcolor[rgb]{0.00,0.00,1.00}{\cite{Lovelock:1971yv}}.
We consider the 4-dimensional action integral,
 \begin{equation}
\mathcal{A}= \frac{1}{2}\int \sqrt{-g} \left[\frac{1}{8\pi G}f(R)\right]d^{4}x+ \mathcal{A}_m(g_{\mu\nu}, \Psi_m) \label{a1},\end{equation}
where $f(R)$ is an arbitrary function of the Ricci scalar $R$, $g$ is the determinant of metric $g_{\mu\nu}$, 
$G$ is the Newtonian gravitational constant, and $\mathcal{A}_m$ is the action of the 
matter fields $\Psi_m$.

By varying the action \eqref{a1} with respect to $g_{\mu\nu}$, we derive the modified Einstein field equations, 
\begin{equation}
f_{R} R_{\mu\nu}-\frac{1}{2}g_{\mu\nu}f(R)-\nabla_{\mu} \nabla_{\nu} f_{R}+  g_{\mu\nu}\Box f_{R}=\kappa^2 T_{\mu\nu} \label{a020},
\end{equation}
where $f_{R}=\frac{\partial f}{\partial R}$, $T_{\mu\nu}$ is the energy-momentum tensor for the standard matter
and $\kappa^2=8\pi G=M_{pl}^{-2}$ where $M_{pl}$ is the Planck mass.
The trace of the field equation \textcolor[rgb]{0.00,0.00,1.00}{(\ref{a020})} gives
\begin{equation}
Rf_R(R)-2f(R)+3\Box f_R(R)= \kappa^2 T \label{a2},
\end{equation}
where
$T=g^{\mu\nu}T_{\mu\nu} = -\rho_m+3P_m$ is the trace of the (perfect fluid) matter energy-momentum tensor in Friedmann-Lemaitre-Robertson-Walker (FLRW) metric background ($P_m=0$ for dust matter).
Recently, some of us \textcolor[rgb]{0.00,0.00,1.00}{\cite{KumarSharma:2022qdf,Yadav:2018llv}} explored the contribution of dynamical $f(R)$ cosmological background geometry for dark matter and dark energy interpretation.  The de-Sitter stage in $f(R)$ gravity is just a vacuum solution with constant background curvature ($R_d$)  which is assumed to be homogeneous and static. Thus, we have from equation \textcolor[rgb]{0.00,0.00,1.00}{(\ref{a2})}
\begin{equation}
Rf_R(R)=2f(R)_{\mid_{R=R_d}} \label{a3},\end{equation}

From equation \textcolor[rgb]{0.00,0.00,1.00}{(\ref{a020})} for de-Sitter stage, we get
\begin{equation}
f_R(R)\big[R_{\mu\nu}-\frac{1}{4} R g_{\mu\nu}\big]_{\mid_{R=R_d}}=0 \label{a4},\end{equation}

As $f_R(R)_{\mid_{R=R_d}}\neq 0$, so equation  \textcolor[rgb]{0.00,0.00,1.00}{(\ref{a4})} gives on using \textcolor[rgb]{0.00,0.00,1.00}{(\ref{a3})}
\begin{equation}
 {\left[ R_{\mu\nu}=\frac{g_{\mu\nu} R}{4}=\frac{g_{\mu\nu} f(R)}{2f_R(R)}\right] }_{\mid_{R=R_d}}.  \label{b3}\end{equation}

It is useful to rewrite equation  \textcolor[rgb]{0.00,0.00,1.00}{(\ref{a2})} in the form of canonical Klein-Gordon scalar wave equation as
\begin{equation}
\Box \phi = \frac{dV_{eff.}}{d\phi}\label{a6},\end{equation}
where we have identified $\phi = f_R(R)$ and $\frac{dV_{eff.}}{d\phi} = \frac{2f(R)-Rf_R(R)-\kappa^2\rho_m}{3}$ in the weak field approximation.
The calculation of scalaron mass profile requires the stability analysis through its effective potential. The effective potential has an extremum at
\begin{equation}
2f(R)-Rf_R(R)=\kappa^2\rho_m \label{b5}.
\end{equation}
For the power-law $f(R)$ model, i.e., for   $f(R)=\frac{R^{1+\epsilon}}{R_c^\epsilon}$ with $\epsilon$ as a dimensionless parameter, the extremum lies at the general-relativistic expectation of $R$, i.e., $R\approx\kappa^2\rho_m$ for very small deviations in the model parameter.
Thus the curvature at the extremum is analysed through,
\begin{equation}
\frac{dV_{eff.}}{d\phi}\bigm|_{\phi_0} =0 \; \text{and} \; m^2_\phi\equiv\frac{d^2V_{eff.}}{d\phi^2}\bigm|_{\phi_0}>0\label{a7},\end{equation}
where $m_\phi$ is the effective mass of scalaron (or scalar field particle) at $\phi_0$ associated with the scalar degree of freedom in $f(R)$ gravity.

To study the novel features of GWs in $f(R)$, one must linearize \textcolor[rgb]{0.00,0.00,1.00}{(\ref{a020})}, and \textcolor[rgb]{0.00,0.00,1.00}{(\ref{a2})}, which will be discussed in the next section.

\section{\label{3} Linearized  $f(R)$ field equations and solutions}
Under the linearized perturbations (up to the first order) of the modified field equations, it is useful to investigate the combined solutions of perturbed scalar and tensor modes for the study of the polarisation contents of GWs.
The perturbations in the scalar field $\phi$ in $f(R)$ theory, and in the metric tensor $g_{\mu\nu}$ can be written as
\begin{equation}
\phi=\phi_0+\delta\phi\ \; ; \;g_{\mu\nu}={g}^{(B)}_{\mu\nu}+ h_{\mu\nu}\   \label{a8},
\end{equation}
where the perturbation $|h_{\mu\nu}|<<|{g}^{(B)}_{\mu\nu}|$  in modified background with ${g}^{(B)}_{\mu\nu}$ as the background metric vacuum solutions of the modified field equations, and  the background $\phi_0$
satisfies equation \textcolor[rgb]{0.00,0.00,1.00}{(\ref{a7})} with ${\delta}\phi = {\delta} f_R(R)=(\partial f_R/\partial R)\mid_{R^{(B)}}\delta R$ as the perturbation of the scalar field. Similarly,  the curvature tensor, Ricci tensor, Ricci scalar and its functional i.e., $f(R)$ can be also expanded up to the first order in perturbations about the background curvature (derived from the metric ${g}^{(B)}_{\mu\nu}$) \textcolor[rgb]{0.00,0.00,1.00}{\cite{Weinberg:2008zzc}} as,
\begin{equation}
R^\rho\ _{\mu\sigma\nu} = {R^{(B)}}^\rho\ _{\mu\sigma\nu}+\delta R^\rho\ _{\mu\sigma\nu}   \label{b9},
\end{equation}
where $\delta R^\rho\ _{\mu\sigma\nu}= \frac{1}{2}[\nabla_{\sigma}\nabla_{\mu}h^\rho\ _\nu+\nabla_{\sigma}\nabla_\nu h^\rho_{\mu}-\nabla_{\sigma}\nabla^\rho h_{\mu \nu}-\nabla_{\nu}\nabla_\mu h^\rho_\sigma - \nabla_{\nu}\nabla_\sigma h^\rho_{\mu}+\nabla_{\nu}\nabla^\rho h_{\mu \sigma}]+  \text{higher order terms}.$
Contracting the curvature tensor, we obtain the Ricci tensor,
\begin{equation}
R_{\mu\nu} = {R^{(B)}}_{\mu\nu}+\delta R_{\mu\nu}\label{b10},\end{equation}
where $ \delta R_{\mu\nu} = - \frac{1}{2} [\nabla_{\mu}\nabla_{\nu}h-\nabla_{\mu}\nabla^\lambda h_{\lambda \nu}-\nabla_{\nu}\nabla^\lambda h_{\lambda \mu}+ \Box h_{\mu\nu}] +  \mathcal{O}(h^2)$
and
\begin{equation}
R = {R^{(B)}}+\delta R  \label{c9},
\end{equation}
where $\delta R =  \nabla^{\mu}\nabla^{\nu}h_{\mu\nu}-\Box h-{R^{(B)}}_{\mu\nu}h^{\mu\nu} + \mathcal{O}(h^2).$
All differential operators (both covariant and contravariant) above have background curvature coupling through the connection.

Similarly, it is possible to expand the functional $f(R)$ in the background curvature as,
\begin{equation}
f(R) = f({R^{(B)}})+f_{R}({R^{(B)}})\delta R+ \mathcal{O}(h^{2})  \label{d9},
\end{equation}
and
\begin{equation}
f_R(R) = f_R({R^{(B)}})+f_{RR}({R^{(B)}})\delta R+ \mathcal{O}(h^{2})  \label{e9},
\end{equation}

We now explore the linearized scalar perturbation \textcolor[rgb]{0.00,0.00,1.00}{(\ref{a8})}, through the perturbed trace field equation \textcolor[rgb]{0.00,0.00,1.00}{(\ref{a2})} around a non-zero constant background curvature $R^{(B)}$ (or $R_{d}$ i.e. de Sitter background) as,
\begin{equation}
R^{(B)} \ \delta f_R-2\ \delta f(R)+3\Box \ \delta f_R+f_R(R^{(B)})\ \delta R=0 \label{f9},
\end{equation}
which yields the linearized scalar field equation given by

\begin{equation}
\left( \Box-m^2_\phi\right) \frac{{\delta}\phi}{\phi_0}=0\label{a9},
\end{equation}
with
\begin{equation}
m^2_\phi\equiv\frac{1}{3}\left(\frac{f_R-Rf_{RR}}{f_{RR}}\right)_{\bigm|_{R=R^{(B)}}},\label{aa9a}\end{equation}
where $m^2_\phi$ satisfies the equation \textcolor[rgb]{0.00,0.00,1.00}{(\ref{a7})}, and $\delta f(R)=f_R(R^{(B)})\delta R$, $\delta f_R$=$f_{RR}\delta R$. Here $m^2_\phi$ corresponds to the perturbed scalar mode mass of the oscillating scalar field around the minimum of the background potential. The linearized wave-like equation \textcolor[rgb]{0.00,0.00,1.00}{(\ref{a9})} suggests the existence of a propagating perturbed massive scalar mode due to propagating GWs and its mass profile depends on $f(R)$ and background curvature which is known as the chameleon mechanism in $f(R)$ gravity \textcolor[rgb]{0.00,0.00,1.00}{\cite{Brax:2008hh}}.

For the model $f(R) \propto R^{1+\epsilon}$, which was previously investigated at different spatial scales \textcolor[rgb]{0.00,0.00,1.00}{\cite{KumarSharma:2022qdf,Yadav:2018llv,Sharma:2022tce,Boehmer:2007kx,Clifton:2005aj}}, from equation \textcolor[rgb]{0.00,0.00,1.00}{(\ref{aa9a})} we obtain the perturbed  mass of the scalaron as
\begin{equation}
m^2_\phi=\frac{(1-\epsilon)}{3\epsilon}R^{(B)}\label{aa9}.\end{equation}

FIG. 1 shows the dependence of massive scalar mode on the model parameter $\epsilon$ for the unit value of background curvature.
\begin{figure}[h]
\centering  \begin{center}  \end{center}
\includegraphics[width=0.44 \textwidth,origin=c,angle=0]{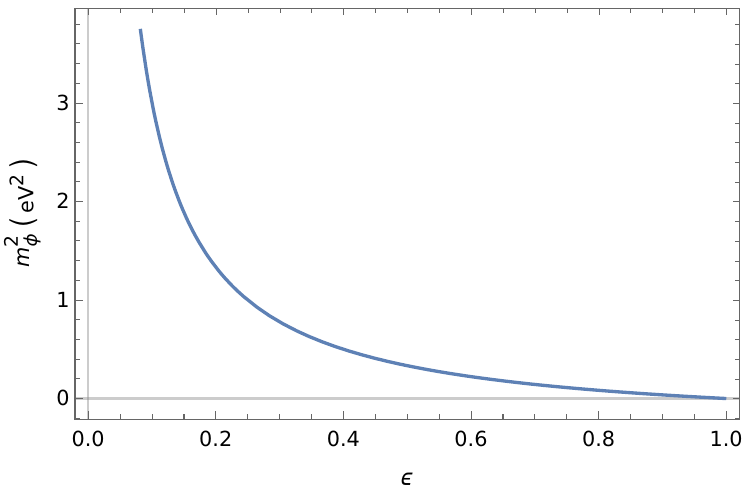}

\caption{\label{fig:p1} The plot shows $m^2_\phi$ as a function of $\epsilon$ in $f(R)\propto R^{1+ \epsilon}$ model with a unit value of the background curvature $R_0$. For $\epsilon$=1, the massive scalar mode vanishes i.e., we have a massless scalar field. For massive scalar modes to exist we must have $0<\epsilon<1$. Beyond this range the tachyonic instability occurs.
}
\end{figure}
Also, we see that $\epsilon<1$ is required for avoiding the tachyonic instability, whereas $\epsilon<<1$ is required for constraining the viable range of the mass profile of the scalar mode in different backgrounds at a local scale. As shown in \textcolor[rgb]{0.00,0.00,1.00}{\cite{Muller:1987hp}}, the condition $m^2_\phi>0$ is needed for the stability of cosmological perturbation for any viable $f(R)$ model.
Now, to obtain the constraint on $m_\phi$, we need to investigate the dispersion relation through the discussion of solutions for the perturbed wave equation.
We now explore the linearized tensor perturbation \textcolor[rgb]{0.00,0.00,1.00}{(\ref{a8})} for the $f(R)\propto R^{1+\epsilon}$ field equation \textcolor[rgb]{0.00,0.00,1.00}{(\ref{a020})}. Under equations \textcolor[rgb]{0.00,0.00,1.00}{(\ref{b9})}, \textcolor[rgb]{0.00,0.00,1.00}{(\ref{d9})}, \textcolor[rgb]{0.00,0.00,1.00}{(\ref{e9})}, and \textcolor[rgb]{0.00,0.00,1.00}{(\ref{b3})} and also by neglecting the higher order terms  we have 
\begin{equation}
\delta R_{\mu\nu}-{g}^{(B)}_{\mu\nu}\frac{\delta R}{2(1+\epsilon)}+\frac{\epsilon}{{R^{(B)}}}\left[ {g}^{(B)}_{\mu\nu}\Box-\nabla_\mu \nabla_\nu\right] \delta R=0 \label{a_10}.
\end{equation}
To completely resolve the  above linearized field equation 
and to  obtain the wave-like equation, we consider a variable (perturbed trace-reverse) motivated by the gauge conditions as in GR \textcolor[rgb]{0.00,0.00,1.00}{\cite{Berry:2011pb}},
\begin{equation}
\bar{h}_{\mu\nu}=h_{\mu\nu}-{g}^{(B)}_{\mu\nu}\left( \frac{h}{2}+h_f\right)  \label{a19},
\end{equation}
where $h_f=[f_{RR}(R)\delta R/f_R(R)]_{\mid_{R=R^{(B)}}} $.
The trace of 
\textcolor[rgb]{0.00,0.00,1.00}{(\ref{a19})} is given as
\begin{equation}
\bar{h}=-h-4h_f\label{a20}.
\end{equation}
We can eliminate $h$ from $\bar{h}_{\mu\nu}$, by making use of equation \textcolor[rgb]{0.00,0.00,1.00}{(\ref{a20})} in the equation \textcolor[rgb]{0.00,0.00,1.00}{(\ref{a19})}. Also, an interesting feature ($\bar{\bar{h}}_{\mu\nu}=h_{\mu\nu}$) can be seen by comparing the equation \textcolor[rgb]{0.00,0.00,1.00}{(\ref{a19})} with
\begin{equation}
h_{\mu\nu}=\bar{h}_{\mu\nu}-{g}^{(B)}_{\mu\nu}\left( \frac{\bar{h}}{2}+h_f\right) \label{a21}.
\end{equation}
Thus, the normal metric perturbation $h_{\mu\nu}$ and the trace-reversed perturbation $\bar{h}_{\mu\nu}$ contain exactly the same information.
But in contrast to GR, the vanishing of the trace $\bar{h}$ is not obvious from the traceless nature of $h$ because of the presence of the second term in equation \textcolor[rgb]{0.00,0.00,1.00}{(\ref{a20})}.
Also, for such perturbed metric $h_{\mu\nu}$ or $\bar{h}_{\mu\nu}$  which in a realistic situation contains (i) gauge degrees of freedom; (ii) radiative degrees of freedom; and (iii)  non-radiative degrees of freedom tied to the modified spacetime background, one can always choose gauges like the Lorenz gauge in which the non-radiative part of the metric perturbation also obeys the wave equations \textcolor[rgb]{0.00,0.00,1.00}{\cite{Berry:2011pb}}.

Since the Lorenz condition does not fix the gauge freedom completely, it leaves some localized coordinate transformation, therefore,  one must bother about the gauge conditions that can also be satisfied by some appropriate choice of vector field $\boldsymbol{\xi}$ in perturbed coordinate system, $x^{\mu\prime}=x^\mu+\xi^\mu$. It is possible to obtain the traceless condition that is satisfied by some appropriate choice of parametric vector field $\boldsymbol{\xi}$. Thus, an infinitesimal change of coordinates affects the metric perturbation according to
\begin{equation}
 h_{\mu\nu}^\prime={h}_{\mu\nu}-2\nabla_{( \mu}\xi_{ \nu)}\label{ab21}.
\end{equation}
The divergence of the trace-reversed metric perturbation thus transforms as
\begin{equation}
\nabla^\mu \bar{h}_{\mu\nu}^\prime=\nabla^\mu \bar{h}_{\mu\nu} -\Box \xi_\nu\label{abb21}.
\end{equation}
We can enforce in the new gauge the transverse condition
\begin{equation}
\nabla^\mu \bar{h}_{\mu\nu}^\prime=0, \label{abbb21}
\end{equation}
by requiring that $\xi_\nu$ satisfies the wave equation $\Box \xi_\nu=\nabla^\mu \bar{h}_{\mu\nu}$.
We can further specialize the gauge to satisfy\footnote{For any function $f$, there always exists a function $F$ such that $\Box F=f$. Consequently, there is a variety of gauge choices available that satisfy the Lorenz condition, as stated in equation \textcolor[rgb]{0.00,0.00,1.00}{(\ref{abbb21})}. In fact, there exist multiple functions that fulfill this condition, indicating that the Lorenz gauge is not uniquely determined. We have the flexibility to apply additional transformations using $\xi$, where $\Box \xi=0$, while still remaining within the Lorenz gauge.} $h^{\prime}=0$.  Dropping the primes, the metric perturbation is thus transverse and traceless,
\begin{equation}
\nabla^\mu \bar{h}_{\mu\nu}=\bar{h}=0. \label{aabbb21}
\end{equation}
Thus, the vanishing of the trace in modified gravity is just a gauge artifact. In \textcolor[rgb]{0.00,0.00,1.00}{\cite{Berry:2011pb,Liang:2017ahj}}, it was explored that for the null signals in the modified gravity background, the trace $\bar{h}$ is not a physical degree of freedom. Also, in a few modified gravity theories (like Brans-Dicke, $R^2$ \textcolor[rgb]{0.00,0.00,1.00}{\cite{Kehagias:2015ata}}, and many other), there exists a massless scalar field apart from the massless tensor field. Therefore, to distinguish such characteristic features of gravity theory, we discuss the effect of gauge artifacts on the polarization contents of the perturbed signals through the generalized Newman-Penrose (NP) formalism in section V.

From the transverse-traceless (TT) gauge conditions (equation \textcolor[rgb]{0.00,0.00,1.00}{\ref{aabbb21}}), we get from equation \textcolor[rgb]{0.00,0.00,1.00}{(\ref{b9})} on substituting the equation \textcolor[rgb]{0.00,0.00,1.00}{(\ref{a21})}, the perturbed linearized Ricci tensor as
\begin{equation}
\delta R_{\mu\nu} = \frac{\epsilon}{{R^{(B)}}}\nabla_{\mu} \nabla_{\nu}\delta R -\frac{1}{2}\Box \bar{h}_{\mu\nu}+ \frac{\epsilon}{2{R^{(B)}}}{g}^{(B)}_{\mu\nu}\Box \delta R \label{a22}.
\end{equation}
Inserting equation \textcolor[rgb]{0.00,0.00,1.00}{(\ref{a22})} into equation \textcolor[rgb]{0.00,0.00,1.00}{(\ref{a_10})}, we obtain
\begin{equation}
\frac{3\epsilon}{2{R^{(B)}}}{g}^{(B)}_{\mu\nu} (\Box - m^2_{\phi})\delta R-\frac{1}{2}\Box \bar{h}_{\mu\nu} = 0 \label{a23},
\end{equation}
where $m^2_\phi$ is exactly given by equation \textcolor[rgb]{0.00,0.00,1.00}{(\ref{aa9})}. The first term on the left-hand side of equation \textcolor[rgb]{0.00,0.00,1.00}{(\ref{a23})} denotes the propagator of a massive scalar field, and the second term denotes the propagator of a massless tensor field.
Taking the trace of equation \textcolor[rgb]{0.00,0.00,1.00}{(\ref{a23})}, we get
\begin{equation}
(\Box-m^2_\phi)\delta R=0\label{a17}.
\end{equation}
The standard massless tensor wave equation can be derived by plugging the equation \textcolor[rgb]{0.00,0.00,1.00}{(\ref{a17})} into equation \textcolor[rgb]{0.00,0.00,1.00}{(\ref{a23})}, otherwise by substituting $\epsilon=0$. Therefore, we have
\begin{equation}
\Box\bar{h}_{\mu\nu}\simeq 0 \label{a24}.
\end{equation}
In physical applications, we are considering the monochromatic wave solutions and its deviation from GR counterparts. Also, only the real part ($\mathfrak{R}$) of the  wave solutions are useful for the study.
The standard massless tensor solution of the above wave equation is given as
\begin{equation}
\bar{h}_{\mu\nu} = \mathfrak{R}( A_{\mu\nu} e^{ik_\alpha x^\alpha}) = \mathfrak{R}( A_{\mu\nu} e^{ik_i x^i}e^{-i\omega t}) \label{a25},
\end{equation}
where $A_{\mu\nu}$  is a constant symmetric tensor, the polarization tensor, in which information about the amplitude and the polarization of the waves is encoded, while $k^\alpha$ is a constant vector, the wave vector that determines the propagation direction of the wave and its frequency, i.e., $\omega= k^0=- k_0$. The d’Alembertian operator acting on a complex exponential in Fourier space is $\Box=(i k_\alpha)(i k^\alpha)= - k_\alpha k^\alpha$. So we have a solution according to equation \textcolor[rgb]{0.00,0.00,1.00}{(\ref{a24})} if $k^{\alpha}$ is a null vector, i.e. $k_\alpha k^\alpha=0$.
In other words,
\begin{equation}
\omega=\pm\sqrt{k_1^2+k_2^2+k_3^2}  \label{ab25}.
\end{equation}
From this dispersion relation, one can see that all perturbations have phase and group velocities both equal to the speed of light.
The solution of scalar wave equation \textcolor[rgb]{0.00,0.00,1.00}{(\ref{a17})} or \textcolor[rgb]{0.00,0.00,1.00}{(\ref{a9})}
leads to a simple plane wave equation given as
\begin{equation}
\delta R\simeq\mathfrak{R}\left( A(p^{\mu})e^{i p_\alpha x^\alpha}\right) \label{a26},
\end{equation}
where $ A(p^{\mu})$ represents the amplitude, and $g_{\mu\nu}^{(B)}p^\mu p^\nu=-m^2_\phi$. The solution of the scalar wave equation suggests that the scalar field is oscillating rapidly in the background and can be expressed in terms of frequency $\omega$ and wave-vector $k^i$ by virtue of the dispersion relation.
This solution is further discussed in the next section.

As the field equations are linearized to the first order one can superpose the two wave solutions. By assuming that perturbed wave propagates along $z$ direction, the generalized minimal coupling solution of the equation \textcolor[rgb]{0.00,0.00,1.00}{(\ref{a23})} can be constructed from equation \textcolor[rgb]{0.00,0.00,1.00}{(\ref{a21})} under the Lorenz gauge condition as
\begin{equation}
h_{\mu\nu}=\bar{h}_{\mu\nu}(ct-z)-\frac{\epsilon}{R^{(B)}}g_{\mu\nu}^{(B)}\delta R\label{a210}.
\end{equation}
In the above expression, the first term represents the propagation of tensor modes of GWs with the speed of light, and the second term represents the propagation of background scalar modes in $f(R)$ theory. The mass profile of a scalar field depends on its dispersive or non-dispersive nature in different backgrounds of gravity theories. It is even possible that in some gravity theories the second term is massless (like in Brans-Dicke gravity, and $f(R)=R^2$ gravity theory).
It should be noted that the scalar mode is coupled with the background metric which affects the polarization contents of the propagating perturbed signals, and clearly for the vanishing value of deviation parameter ($\epsilon$), the solution approaches to GR. Such couplings have the effect of causing gradual evolution in the properties of waves \textcolor[rgb]{0.00,0.00,1.00}{\cite{Ezquiaga:2020dao,Dalang_2021}}. The solution \textcolor[rgb]{0.00,0.00,1.00}{(\ref{a210})}  is useful for exploring the polarization properties of perturbed waves for massless ($\epsilon=1$) and massive ($\epsilon<<1$ with $\epsilon/{R^{(B)}} \approx 1/{3m_{\phi}^2}$) scalars. Before exploring it, we first discuss the propagating properties of the scalar modes through dispersion relations.

\section{\label{4}Scalaron as massive graviton and its dispersion relation}

To investigate the scalarons as spin-zero mode of gravitons (or massive gravitons), we need to explore the modified dispersion relation. In GR, gravitational waves are locally Lorenz invariant, obeying the dispersion relation $\omega = k$, where $\omega$ is the angular frequency,  and $k$ is the wavenumber. This relation implies that gravitons are massless \textcolor[rgb]{0.00,0.00,1.00}{\cite{Misner:1973prb,Will:2018bme,Will:2014kxa}} travelling with the speed of light. However, modified theories could have a massive degree of freedom like  scalarons in $f(R)$ gravity which obey a modified dispersion relation leading to different propagation speeds \textcolor[rgb]{0.00,0.00,1.00}{\cite{Ezquiaga:2020dao,Dalang_2021,Mirshekari:2011yq,Nishizawa:2014zna,Nishizawa:2016kba,Yunes:2016jcc}}.
In physical applications, we are considering monochromatic wave solutions and their deviation from GR counterparts. Also, only the real part of the  wave solution is useful for the study. Therefore, according to the equation \textcolor[rgb]{0.00,0.00,1.00}{(\ref{a26})}, the  dispersion relation for the propagating massive scalar mode of gravitational waves takes the following form
\begin{equation}
p^\mu p_\mu=-m^2_\phi \label{a27},
\end{equation}
where $p^\mu=(p^0, \vec{p})=\left( \omega,0,0, \sqrt{\left|\omega^2-m^2_\phi\right|} \right)$ is the four-wave vector of the scalar GWs and $m_\phi$ is given by equation \textcolor[rgb]{0.00,0.00,1.00}{(\ref{aa9})}.

The propagation speed $v_{group}$ of dispersive scalar modes of GWs is given by the following equation,
\begin{equation}
{v_{group}} = 1- \frac{m^2_\phi}{2\omega^2} = 1-\frac{(\omega^2-k^2)}{2\omega^2}\label{a30}.
\end{equation}
This equation suggests that group velocity of the scalar mode of GWs deviates from the speed of light $c=1$, and clearly for $\omega=k$, the propagating mode follows null geodesics with $v_{group}=1$. Considering equation \textcolor[rgb]{0.00,0.00,1.00}{(\ref{aa9})}, equation \textcolor[rgb]{0.00,0.00,1.00}{(\ref{a30})} can be rewritten as,
\begin{equation}
{v_{group}} \approx 1- \frac{1}{6\epsilon}\frac{R^{(B)}}{\omega^2}\label{abab30}.
\end{equation}
The above equation conveys the message that the propagation speed of scalarons depend on the background curvature. Let us start with the Solar System background since already operating GW detectors are located on Earth and future space-borne detectors like LISA will also be parked in the Earth-like heliocentric orbit. The LISA space-based GW detector will explore the constraints on the massive nature (if any), within its sensitive operational frequency range of $\mathcal{O} (10^{-4})Hz \leq f \leq 1 Hz$. In terms of energy units (eV) corresponding to this frequency range, the massive nature (like scalar mode mass profile) should fall within the range of $\mathcal{O} (10^{-15}) eV\geq m_\phi \geq \mathcal{O} (10^{-19}) eV$.

Relevant information is contained in equation \textcolor[rgb]{0.00,0.00,1.00}{(\ref{aa9})} and in FIG. 2 we show the variation of $m_\phi$ with $\epsilon$ for the Solar System background.
\begin{figure}[h]
	\centering \begin{center}  \end{center}
{\includegraphics[width=0.44 \textwidth,origin=c,angle=0]{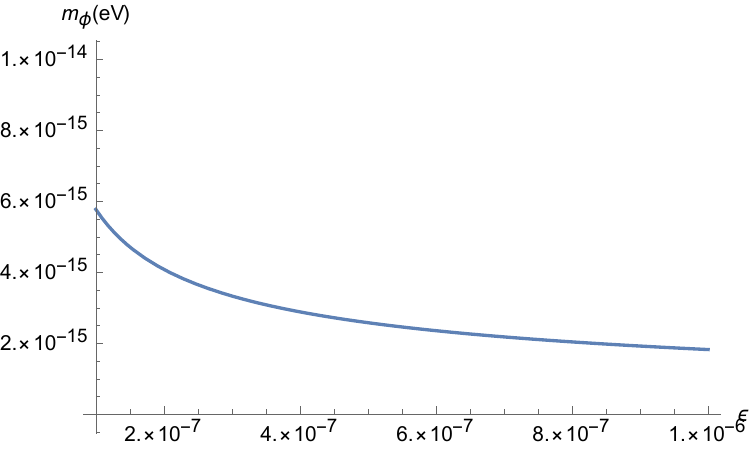} }
\caption{\label{fig:p2} The dependency of $m_{\phi}$ on $\epsilon$ in the solar system background ($\rho_{\odot} \sim 10^{19}eV^4$) with $R^{(B)}\approx 10^{-35}eV^2$ is shown.
For $\epsilon \approx \mathcal{O}(10^{-7})$, $m_\phi$ attains approximately constant order $\mathcal{O}(10^{-15})$ in $eV$, falling within the operational range of LISA.}

\end{figure}
One can see that, for $f(R)\propto R^{1+\epsilon}$ in the Solar System background with $\epsilon \approx \mathcal{O}(10^{-7})$ \textcolor[rgb]{0.00,0.00,1.00}{\cite{KumarSharma:2022qdf}}, the scalar mode mass profile $m_\phi$ attains approximately constant value of $\mathcal{O}(10^{-15}) eV$.
These considerations do not violate the bounds on graviton mass obtained from the detection of various GW events \textcolor[rgb]{0.00,0.00,1.00}{\cite{Morisaki:2018htj,Brito:2017zvb}}.
\begin{figure}[h]
	\centering \begin{center}  \end{center}
	{{\includegraphics[width=0.44 \textwidth,origin=c,angle=0]{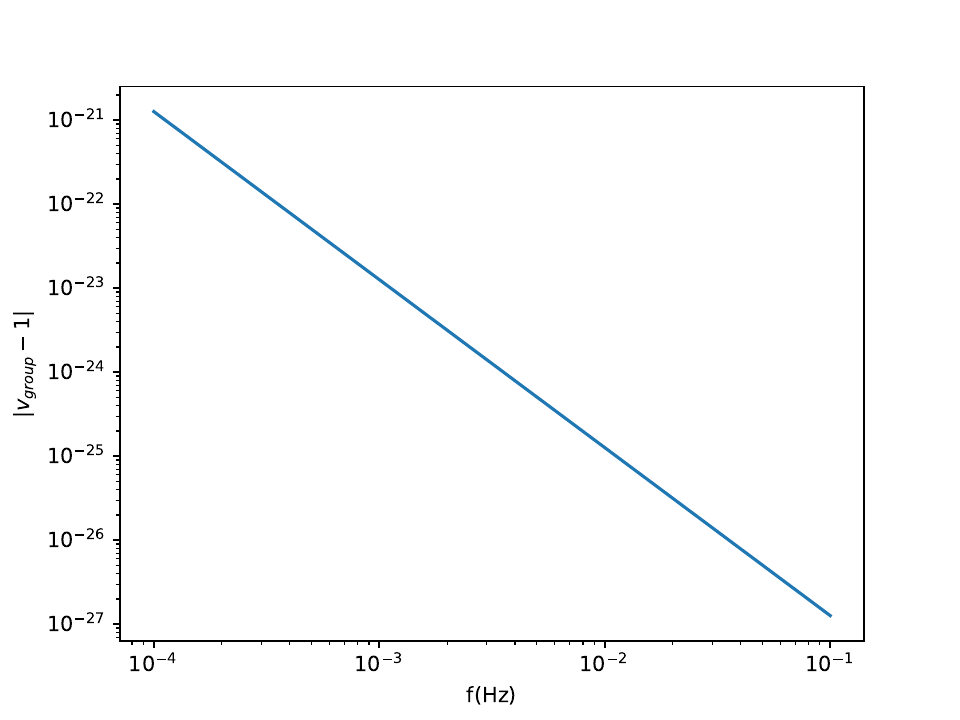} }}%
	\caption{In the Solar System background with $\epsilon\approx\mathcal{O}(10^{-7})$, the fractional deviation in the dispersive signal speed is plotted against the sensitivity band of LISA. }
	\label{fig:3}
\end{figure}
Quite stringent and prominent direct constraint $-3\times 10^{-15}<\frac{v_{GW}-c}{c}<+7\times 10^{-16}$ on the fractional deviation of GW propagation speed  was obtained from the GW170817  and its EM counterpart GRB170817A 
\textcolor[rgb]{0.00,0.00,1.00}{\cite{Harry:2022zey}}.
LISA is expected to provide a significant improvement of constraints over  GW170817-like bound.
With the same assumptions of FIG. 2, FIG. 3 shows the variation of the fractional deviation $\mid v_{group}-1\mid$ of the speed of perturbed signal from $c$ (assumed here as a unit), as a function of GW frequency $f$. Let us recall that the sensitivity window of LISA would be $\mathcal{O} (10^{-4})Hz \leq f \leq 1 Hz$.
One can see that possible  constraints on such signal speed would be $\mathcal{O} (10^{-27}) \leq  |v_{group}-1| \leq \mathcal{O} (10^{-21})$ and at the extreme sensitivity $\mathcal{O} (10^{-4})$ or even one order less than this, the deviation in speed will be around $\mathcal{O} (10^{-20})$. Several authors have also explored the constraints on the GW speed  with LISA and ground based detectors together \textcolor[rgb]{0.00,0.00,1.00}{\cite{Harry:2022zey,Yagi:2009zm,LISACosmologyWorkingGroup:2022wjo,Nishizawa:2017nef}}. We will delve deeper into LISA's lower sensitivity threshold concerning intermediate and massive lens mass objects in Section VI.

There have been proposals to constrain the graviton mass from the future observations of a compact binary with space-based GW detectors such as LISA \textcolor[rgb]{0.00,0.00,1.00}{\cite{amaro2017laser}} in the milli-Hertz band and DECIGO \textcolor[rgb]{0.00,0.00,1.00}{\cite{Sato:2017dkf}} in the deci-Hertz band.
Hence, in the dispersive background, the perturbed signals will reflect such small deviation by following the non-null geodesics. It is therefore necessary to investigate the effect of such small deviations on the  polarization contents of perturbed signals and on the amplification of the wave amplitude by lensing.
In the next section, we use the Newman Penrose (NP) formalism \textcolor[rgb]{0.00,0.00,1.00}{\cite{Eardley:1973zuo,Eardley:1973br,RizwanaKausar:2016zgi,Alves:2009eg}} in order to investigate the behaviour of different polarisation contents. Further, in our case of $f(R)$ gravity, we use its modified version \textcolor[rgb]{0.00,0.00,1.00}{\cite{Hyun:2018pgn}} and investigate its few distinguishing features to characterise the massless scalar modes from the massless tensor modes, as well as the massive modes with $\epsilon<<1$.

\section{\label{5}Polarization properties  of $f(R)$   signals as diagnostics  via modified NP approach and gauge artifacts}

In the context of gravity theories incorporating a scalar degree of freedom, there is a notable interest in re-examining the polarization modes, specifically concerning the modified dispersion relation  \textcolor[rgb]{0.00,0.00,1.00}{\cite{Hyun:2018pgn,Isi:2017equ,Ezquiaga:2020dao,Dalang_2021}}. The dispersion relation as explored in Section \ref{4}, presents an opportunity to characterise the polarization modes within these theories.
At first glance, from equation \textcolor[rgb]{0.00,0.00,1.00}{(\ref{a24})} one could naively say that there are ten polarisations of gravitational waves since there are ten wave equations.
However, this is not the case. The harmonic Lorenz gauge condition $\nabla^\mu\bar{h}_{\mu\nu}=0$ tells us that $k^\mu A_{\mu\nu}=0$. This restriction eliminates four of the degrees of freedom, so there are only six legal degrees of freedom in $A_{\mu\nu}$. That implies, we have six physical and four gauge degrees of freedom. Similarly, by considering the generalized wave solutions in the context of metric-compatible theories, using the NP formalism \textcolor[rgb]{0.00,0.00,1.00}{\cite{Hyun:2018pgn}}
generalized the six  polarization modes: the breathing(b), longitudinal ($l$), vector-x (x), vector-y (y), plus (+), and cross($\times$).
Accordingly, the six polarization contents ($p_n$) are given as \textcolor[rgb]{0.00,0.00,1.00}{\cite{Hyun:2018pgn}},
\begin{equation}
p_1^{(l)}=\frac{1}{2}\left(\frac{\omega^2-k^2}{\omega^2+k^2} \right)\omega^2(h_{tt}+h_{zz})-\frac{1}{2}(\omega^2-k^2)h_{tt}\label{ab31}.
\end{equation}
\begin{equation}
p_2^{(x)}=\frac{1}{2}\left({\omega^2-k^2} \right)h_{xz}\label{abc31},
\end{equation}
\begin{equation}
p_3^{(y)}=\frac{1}{2}\left({\omega^2-k^2} \right)h_{yz}\label{abcd31},
\end{equation}
\begin{equation}
p_4^{(+)}=-\frac{1}{2}\left(\frac{\omega^2-k^2}{\omega^2+k^2} \right)\omega^2(h_{tt}+h_{zz})-\omega^2h_{yy}\label{abcde31},
\end{equation}
\begin{equation}
p_5^{(\times)}=\frac{1}{2}\omega^2h_{xy}\label{abcdef31},
\end{equation}
\begin{equation}
p_6^{(b)}=-\frac{1}{2}\left(\frac{\omega^2-k^2}{\omega^2+k^2} \right)\omega^2(h_{tt}+h_{zz})\label{abcdefg31},
\end{equation}
The set of equations \textcolor[rgb]{0.00,0.00,1.00}{(\ref{ab31})}-\textcolor[rgb]{0.00,0.00,1.00}{(\ref{abcdefg31})} distinguishes the gravity theories having non dispersive signals ($\omega=k$) from those having dispersive signals ($\omega\neq k$).
For theories that have massive modes of polarization, the magnitude of such modes is quantitatively determined by the specific form of the dispersion relation, $\omega=\omega(k)$ as in the case of $f(R)$ theories. However, there are certain modified version of gravity theories, like the Brans-Dicke gravity theory and  $f(R)\propto R^2, \epsilon=1$ \textcolor[rgb]{0.00,0.00,1.00}{\cite{Brans:1961sx,Vilhena:2021bsx}} that have non-dispersive (massless) characteristics.
Therefore, the above set of polarisation contents cannot distinguish gravity theories which have non dispersive signal characteristics (i.e.,  massless scalar modes and massless tensor modes). The reason behind this lies in the gauge artifacts. The authors of \textcolor[rgb]{0.00,0.00,1.00}{\cite{Hyun:2018pgn}}
used the transverse-traceless (TT) gauge conditions for obtaining the above set of polarization contents. As we have discussed in Section \ref{3}, in modified gravity theory, the traceless condition is just a gauge artifact for trace-reversed perturbed metric, i.e.,  the vanishing of the trace  of $\bar{h}_{\mu\nu}$  does not imply the traceless nature of $h_{\mu\nu}$ (equation \textcolor[rgb]{0.00,0.00,1.00}{\ref{a20}}). Hence, if we investigate this simple consequence for the massless waves, or if we abandon the traceless condition in \textcolor[rgb]{0.00,0.00,1.00}{\cite{Hyun:2018pgn}}, then under the transverse gauge condition, the set (equations \textcolor[rgb]{0.00,0.00,1.00}{(\ref{ab31})}-\textcolor[rgb]{0.00,0.00,1.00}{(\ref{abcdefg31})}) of polarisation contents (or amplitudes) can be written as \textcolor[rgb]{0.00,0.00,1.00}{\cite{Hyun:2018pgn}}
\begin{equation}
p_1^{(l)}=\frac{1}{2}\left[k^2(h_{xx}+h_{yy})+(\omega^2-k^2)h_{zz} \right]\label{ab311}.
\end{equation}
\begin{equation}
p_2^{(x)}=\frac{1}{2}\left({\omega^2-k^2} \right)h_{xz}\label{abc312},
\end{equation}
\begin{equation}
p_3^{(y)}=\frac{1}{2}\left({\omega^2-k^2} \right)h_{yz}\label{abcd313},
\end{equation}
\begin{equation}
p_4^{(+)}=\frac{1}{2}\omega^2(h_{xx}-h_{yy})\label{abcde3144},
\end{equation}
\begin{equation}
p_5^{(\times)}=\frac{1}{2}\omega^2h_{xy}\label{abcdef3155},
\end{equation}
\begin{equation}
p_6^{(b)}=\frac{1}{2}\omega^2(h_{xx}+h_{yy})\label{abcdefg3166},
\end{equation}
Now, for the non-dispersive ($\omega=k$) or massless signals, the non-vanishing polarization contents from the set of equations \textcolor[rgb]{0.00,0.00,1.00}{(\ref{ab311})}-\textcolor[rgb]{0.00,0.00,1.00}{(\ref{abcdefg3166})} are $p_1^{(l)}$, $p_4^{(+)}$, $p_5^{(\times)}$, and $p_6^{(b)}$. In contrast to TT gauge conditions, here we have two more non-vanishing contents, $p_1^{(l)}$ and $p_6^{(b)}$. Under the sole imposition of the transverse gauge condition and for non-dispersive signal,  $p_1^{(l)}$ which  is the
amplitude of the longitudinal scalar wave coincides  with $p_6^{(b)}$, the breathing mode of scalar wave. Hence they can be considered as a single polarised state. Thus, the total of three polarization states exist for the massless modes  under transverse gauge conditions.
This distinguishes such modified gravity theories from the GR.

With reference to the set of equations \textcolor[rgb]{0.00,0.00,1.00}{(\ref{ab311})}-\textcolor[rgb]{0.00,0.00,1.00}{(\ref{abcdefg3166})}, the generalized NP quantities under the transverse gauge read as \textcolor[rgb]{0.00,0.00,1.00}{\cite{Hyun:2018pgn}}
\begin{equation}
\Psi_2=\frac{1}{24}\left({3k^2-\omega^2}  \right)(h_{xx}+h_{yy})+\frac{1}{12}\left({\omega^2-k^2} \right)h_{zz}\label{ab316},
\end{equation}
\begin{equation}
\Psi_3=\frac{1}{8}\frac{(\omega-k)(\omega+k)^2}{\omega}(h_{xz}-ih_{yz})\label{abcde3145},
\end{equation}
\begin{equation}
\Psi_4=\frac{1}{8}(\omega+k)^2(h_{xx}+h_{yy})-\frac{1}{4}(\omega+k)^2(h_{yy}+ih_{xy})\label{abcdef3156},
\end{equation}
\begin{equation}
\Phi_{22}=\frac{1}{8}(\omega+k)^2(h_{xx}+h_{yy})\label{abcdefg3167}.
\end{equation}
Now, for the $f(R)$ model, the minimal wave solution is given by the equation  \textcolor[rgb]{0.00,0.00,1.00}{(\ref{a210})}.
Because of the traceless property of the trace-reversed perturbed variable ($\bar{h}_{\mu\nu}$), the traceless nature of the perturbed variable ($h_{\mu\nu}$) cannot be demanded by equation \textcolor[rgb]{0.00,0.00,1.00}{(\ref{a20})}.
As a result, it is reflected in the polarisation contents.
Hence, the polarization contents are calculated from the set of equations  \textcolor[rgb]{0.00,0.00,1.00}{(\ref{ab311})}-\textcolor[rgb]{0.00,0.00,1.00}{(\ref{abcdefg3166})} for our $f(R)(\propto R^{1+\epsilon})$ model and then substituted with $\epsilon=1$ (or massless scalarons in $f(R)$ (see equation \textcolor[rgb]{0.00,0.00,1.00} {(\ref{aa9})})) for exploring the effects on the polarization contents as

\begin{equation}\label{ab3119}
\begin{split}
p_1^{(l)}&=\frac{1}{2} \frac{\epsilon}{R^{(B)}}\left[-\left( g_{xx}^{(B)}+g_{yy}^{(B)}\right) k^2 - g_{zz}^{(B)} m_\phi^2 \right]\delta R \\&= -\frac{1}{2R^{(B)}}\left( g_{xx}^{(B)}+g_{yy}^{(B)}\right) \delta \ddot{R}
\end{split}
\end{equation}
\begin{equation}
p_2^{(x)}=0\label{abc3129},
\end{equation}
\begin{equation}
p_3^{(y)}=0\label{abcd3139},
\end{equation}
\begin{equation}
\begin{split}
p_4^{(+)}&=-\frac{1}{2}\left( \ddot{\bar{h}}_{xx}-\ddot{\bar{h}}_{yy}\right)-\frac{1}{2}\frac{\epsilon}{R^{(B)}}\left(g_{xx}^{(B)}-g_{yy}^{(B)} \right) \omega^2 \delta R\\&= -\frac{1}{2}\left( \ddot{\bar{h}}_{xx}-\ddot{\bar{h}}_{yy}\right)-\frac{1}{2R^{(B)}}\left(g_{xx}^{(B)}-g_{yy}^{(B)} \right) \delta \ddot{R}\label{abcde31449},
\end{split}
\end{equation}
\begin{equation}
p_5^{(\times)}=-\frac{1}{2}\ddot{\bar{h}}_{xy}\label{abcdef31559},
\end{equation}
\begin{equation}\label{abcdefg31669}
\begin{split}
    p_6^{(b)}&=-\frac{1}{2}\frac{\epsilon}{R^{(B)}} \left( g_{xx}^{(B)}+g_{yy}^{(B)}\right) \omega^2 \delta R\\&= -\frac{1}{2R^{(B)}} \left( g_{xx}^{(B)}+g_{yy}^{(B)}\right)  \delta \ddot{R} .
\end{split}
\end{equation}
Consequently, the generalized NP quantities (or amplitudes of wave) from the set of equations \textcolor[rgb]{0.00,0.00,1.00}{(\ref{ab316})}-\textcolor[rgb]{0.00,0.00,1.00}{(\ref{abcdefg3167})} reads as
\begin{equation}
\Psi_2= -\frac{1}{12 R^{(B)}} \left(g_{xx}^{(B)}+g_{yy}^{(B)} \right) \delta \ddot{R}\label{ab3160},
\end{equation}
\begin{equation}
\Psi_3=0\label{abcde31450},
\end{equation}
\begin{equation}
\Psi_4= \left( \ddot{\bar{h}}_{yy}+i\ddot{\bar{h}}_{xy}\right) -\frac{4}{R^{(B)}} \left( g_{xx}^{(B)}-g_{yy}^{(B)} \right) \delta \ddot{R} \label{abcdefg31670},
\end{equation}
\begin{equation}
\Phi_{22}=  -\frac{1}{2 R^{(B)}} \left( g_{xx}^{(B)}+g_{yy}^{(B)} \right)\delta \ddot{R}\label{abcdef31560}.
\end{equation}
Clearly, the GW modes are modified from their GR counterparts to include a contribution from the shifted ($\epsilon=1$) Ricci scalar background curvature. Consequently, the set of equations \textcolor[rgb]{0.00,0.00,1.00}{(\ref{ab3160})}-\textcolor[rgb]{0.00,0.00,1.00}{(\ref{abcdef31560})} can be further reduced according to the nature of spacetime background curvature (for example at local scales $g_{\mu\nu}^{(B)}\approx \eta_{\mu\nu}$). As a result, we can observe a distinguishable impact of the massless scalar modes in contrast to the massless tensor modes.

However, when dealing with the massive scalar modes, we utilize modified NP quantities within the TT gauge. This analysis focuses on the regime where $\epsilon<<1$, with the condition $\epsilon/{R^{(B)}} \approx 1/{3m_{\phi}^2(=\omega^2-k^2)}$ as

\begin{equation}
\begin{split}
\Psi_2=\frac{\epsilon}{24R^{(B)}}\frac{(\omega^2-k^2)}{(\omega^2+k^2)}  & \left\{ \left(3g_{tt}^{(B)}+g_{zz}^{(B)}\right)k^2 \right.\\
   - & \left. \left(3g_{zz}^{(B)}+g_{tt}^{(B)}\right)\omega^2 \right\}\delta R\\ \approx \frac{1}{72}\frac{1}{(\omega^2+k^2)} & \left\{ \left(3g_{tt}^{(B)}+g_{zz}^{(B)}\right)k^2\right.\\- & \left. \left(3g_{zz}^{(B)}+g_{tt}^{(B)}\right)\omega^2 \right\}\delta R\label{ab31601}
\end{split}
 \end{equation}
\begin{equation}
\Psi_3=0\label{abcde314502},
\end{equation}
\begin{equation}\label{abcdefg316703}
\begin{split}
\Psi_4&= \left( \ddot{\bar{h}}_{yy}+i\ddot{\bar{h}}_{xy}\right) +\frac{\epsilon}{8R^{(B)}} \\& \left\{\left( g_{tt}^{(B)}+g_{zz}^{(B)} \right)\frac{(\omega^2-k^2)}{(\omega^2+k^2)} +2g_{yy}^{(B)}\right\} (\omega+k)^2\delta R
\end{split}
\end{equation}

\begin{equation}\label{abcdef315604}
\begin{split}
\Phi_{22}= \frac{\epsilon}{8R^{(B)}}\frac{(\omega^2-k^2)(\omega+k)^2}{\omega^2+k^2} \left( g_{tt}^{(B)}+g_{zz}^{(B)} \right)\delta R \\ \approx  \frac{1}{24}\frac{(\omega+k)^2}{\omega^2+k^2} \left( g_{tt}^{(B)}+g_{zz}^{(B)} \right)\delta R
\end{split}
\end{equation}
Again, for the vanishing value of $\epsilon$, the GR prediction is exactly recovered. The above sets of generalized NP quantities (equations \textcolor[rgb]{0.00,0.00,1.00}{(\ref{ab3160})}-\textcolor[rgb]{0.00,0.00,1.00}{(\ref{abcdef31560})} and \textcolor[rgb]{0.00,0.00,1.00}{(\ref{ab31601})}-\textcolor[rgb]{0.00,0.00,1.00}{(\ref{abcdef315604})}) would clearly serve as a diagnostic tool to distinguish $f(R)$ (massless or massive scalar modes) from its GR counterparts. Now, since the mass of the scalar mode  is given by equation \textcolor[rgb]{0.00,0.00,1.00}{(\ref{aa9})}, therefore, for $\epsilon<<1$, the longitudinal as well as breathing mode (equations \textcolor[rgb]{0.00,0.00,1.00}{(\ref{ab31601})} and \textcolor[rgb]{0.00,0.00,1.00}{(\ref{abcdef315604})}) both are independent of $f(R)$ model parameter and  the scalaron mass. However, $\Psi_4$ depends on the $f(R)$  parameter, i.e., there is an extra contribution due to the massive scalar mode attached to the massless tensor modes. Also, we can recover the GR-based NP quantity  by switching off the  $f(R)$ model parameter ($\epsilon$).
Space-based GW detectors make it possible to explore frequency ranges where the influence on polarization amplitudes is of a similar magnitude for both massive (scalarons) and massless (tensor) modes. These detectors are also well-positioned to detect deviations in the amplitude of the tensor field caused by the presence of a massive scalar mode. This effect may contribute to slight deviations in the speed of tensor signals from their typical behavior. It has been reported recently in the literature that in the event of lensing one might observe additional polarizations \textcolor[rgb]{0.00,0.00,1.00}{\cite{Dalang:2021qhu}}, therefore one must investigate further to understand if the non-vanishing modes are physical or not. Recently, \textcolor[rgb]{0.00,0.00,1.00}{\cite{Casado-Turrion:2023rni}} explored $f(R)$ solutions that exhibit a variety of unphysical properties. This is beyond the scope of our paper.
We anticipate that our work will contribute insights to deeper understanding of the polarization characteristics within the context of alternative theories of gravity and their implications for lensed GW observations \textcolor[rgb]{0.00,0.00,1.00}{\cite{Ezquiaga:2020dao,LISACosmologyWorkingGroup:2022jok}}. In next section we study the effects of massive gravitons in the lensing amplification factor.

\section{\label{6} Lensing of massive scalarons}

With the space-based GW detector LISA, one can explore the dispersive characteristics (including massive 
modes like gravitons and scalarons) within a mass range of approximately  $10^{-19}eV\leq m\leq 10^{-15}eV$.
Such mass range is typical for the strongly motivated dark matter candidates -- ultralight particles (like scalarons, or axion-like particles) manifesting at local scales as fuzzy dark matter  \textcolor[rgb]{0.00,0.00,1.00}{\cite{Miller:2023kkd,Brito:2022lmd,Yu:2023iog,Garoffolo:2019mna,Morisaki:2018htj,Brito:2017zvb}}.
However, in this mass range, the scalaron acts more like a wave than a particle (since $\lambda_\phi \propto 1/m_\phi$ ) \textcolor[rgb]{0.00,0.00,1.00}{\cite{Hui:2021tkt}}. Therefore, we will discuss gravitational lensing of (non)dispersive perturbed signals in the wave optics formalism presented e.g. in
\textcolor[rgb]{0.00,0.00,1.00}{\cite{Takahashi_2017,Takahashi:2003ix,Nakamura:1999uwi,Gravitational_lenses1992}}. \\
We consider the signal from a distant source propagating in the background spacetime $g_{\mu\nu}^{(L)}$ of the lens object characterized by the gravitational potential $U(\textbf{r}) << 1$. The total metric, including the perturbation (due to the GW signal)  is given by $g_{\mu\nu} =  g_{\mu\nu}^{(L)}+ h_{\mu\nu}$, where $h_{\mu \nu} = \tilde{h}e^{-2 \pi i f t} A_{\mu \nu}$, where $A_{\mu \nu}$ is the polarization tensor and $\tilde{h}$ is the amplitude.
Under the weak field approximation for the lens, the propagation equation can be cast to the form of the Helmholtz equation, whose solution can be given in terms of the Kirchhoff integral \textcolor[rgb]{0.00,0.00,1.00}{\cite{Gravitational_lenses1992}}. It is convenient to introduce the dimensionless amplification factor:
\begin{equation}
    F(f) =  \frac{\tilde{h}_{L}(f)}{\tilde{h}(f)}
\end{equation} \label{amplification factor}
which is the ratio of wave amplitudes with and without lensing. Then, the Kirchhoff integral allows us to calculate the amplification factor at the observer \textcolor[rgb]{0.00,0.00,1.00}{\cite{Gravitational_lenses1992}} as:
\begin{equation}
\label{Kirchoff}
    F(f,\beta) =  \frac{1+z_l}{c} \frac{D_{s}}{D_l D_{ls}} \frac{f}{i} \int d^{2}\theta \exp\left[ 2\pi i f \Delta t(\theta, \beta) \right]
\end{equation}
where $D_s, D_l, D_{ls}$ are angular diameter distances to the source, to the lens ($z_l$ is the redshift of the lens) and between the lens and the source respectively. Cosmological setting is invoked here because all GW signals registered so far came from cosmological distances. In a local scenario, involving e.g. rotating pulsars or binary stellar systems (observable by LISA) redshifts should be set to zero and distances would become ordinary Euclidean distances. Measured from the axis connecting the observer and the center of the lens, angle $\beta$ refers to the direction to the source (unobservable, but representing the mismatch in the optical system) while $\theta$ is actual direction from which lensed signal comes.
$\Delta t$ is the time delay introduced by gravitational lensing at the angular position $\theta$ from the lens. It is given by
\begin{equation}
\Delta t(\theta, \beta) = \frac{1+z_l}{c} \frac{D_lD_s}{D_{ls}} \left[ \frac{(\theta - \beta)^2}{2} - \phi(\theta) + \phi_m(\beta) \right]
\end{equation}
where $\phi(\theta)$ is the lens potential (essentially a 2D projection of the full 3D potential of the lens) determining the deflection angle $\alpha(\theta) = \nabla_{\theta} \phi(\theta)$. Term  $\phi_m(\beta)$ corresponds to the arrival time in a non-lensed case, and in practice, it is a constant adjusted to ensure the extreme value of the time delay functional. Note,  in this section, instead of geometric units, we reintroduce SI units hence $c$ and $h$ are explicitly present in the equations.

The calculation of the integral \textcolor[rgb]{0.00,0.00,1.00}{(\ref{Kirchoff})} is much easier after switching from angles to dimensionless variables $x = \theta / \theta_E$ and $y = \beta / \theta_E$, where $\theta_E$ is the Einstein radius of the lens determined by its mass $M_l$ and relative distances in the system and $y$ the impact factor. It is also useful to introduce the dimensionless frequency
\begin{equation}
\label{ref:dim_freq}
     w = \frac{8 \pi G}{c^3} M_l  (1+z_l) f
\end{equation}
rewriting equation \textcolor[rgb]{0.00,0.00,1.00}{(\ref{Kirchoff})} using the newly introduced terms we get:
\begin{equation} \label{Fw}
    F(w,y) = \frac{w}{2 \pi i} \int d^2y \left[ \exp[ i w T(x,y)] \right]
\end{equation}
where $T(x,y) = \frac{(x-y)^2}{2} - \frac{\phi(x)}{\theta_E^2}$.

\begin{figure}[h]
	\centering
	\includegraphics[width=0.44 \textwidth,origin=c,angle=0]{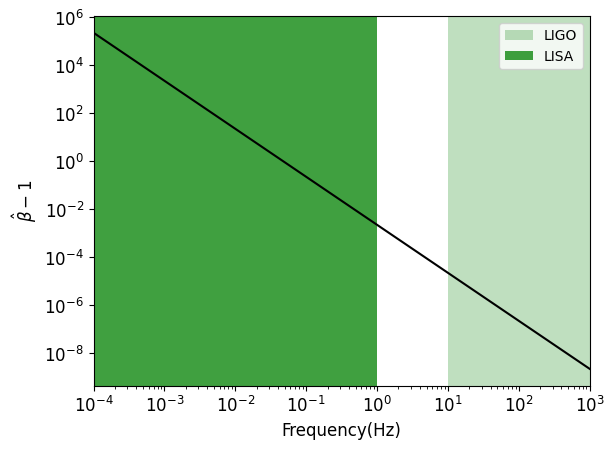}
	\caption{Dependency of $\hat{\beta}-1$ with respect to the frequency. The vertical bars are drawn to show the operating windows of LIGO and LISA. The scalaron mass used is $m_{\phi} =10^{-15} eV$.}
	\label{f4}
\end{figure}

For the purpose of illustration, we consider gravitational lensing by a point mass, e.g. a black hole of a mass of a few hundred $M_\odot$. We also consider the dispersion relations in shifted ($R^{1+\epsilon}$) background as discussed in Section \ref{4}.
Following the treatment of \textcolor[rgb]{0.00,0.00,1.00}{\cite{Chung:2021rcu}} and 
 \textcolor[rgb]{0.00,0.00,1.00}{\cite{Takahashi:2003ix}}, the resulting waveform of perturbed signals in $f(R)$ dispersive background can be expressed as
 \begin{equation}
\tilde{h}_L(f)=F(f, y; M_{l},m_\phi)\tilde{h}_{disp}(f),\label{abcdef3156042100}
\end{equation}
where the amplification factor, for a point mass lens, can be analytically evaluated as \textcolor[rgb]{0.00,0.00,1.00}{\cite{Chung:2021rcu}}

\begin{equation}
\begin{split}
F(w, y; M_l, m_\phi) = \exp\left[\frac{\pi}{4} w \hat{\beta}\right] \left(\frac{w}{2} \hat{\beta}\right)^{i \frac{w}{2} \hat{\beta}} \\ \Gamma\left(1 - i \frac{w}{2} \hat{\beta}\right) {}_1F_1\left(i \frac{w}{2} \hat{\beta}, 1, i \frac{w}{2} \hat{\beta} y^2\right)
\end{split}
\label{amp_fact_an}
\end{equation}

where $_1F_1$ is the confluent hypergeometric function and $\Gamma$ is the Euler gamma function. It reduces to GR when the dimensionless factor  $\hat{\beta}=1$. Therefore, $\hat{\beta}$ distinguishes the amplification factor with or without dispersion. For the massive scalarons ($m_\phi \neq 0$) it is
\begin{equation}
\hat{\beta}(f) \equiv\frac{c}{v_{group}(f)} =  1+\frac{m_\phi^2 c^4}{8\pi h^2 f^2}.
\label{eq:beta}
\end{equation}
Equation \textcolor[rgb]{0.00,0.00,1.00}{(\ref{eq:beta})} is plotted in Fig. \ref{f4}, in which we evidence the LISA and LIGO interferometers sensitivity. For the purpose of illustration, we consider the scalaron mass to be 10$^{-15}$ eV.
The effect of massive scalarons ($\hat{\beta}(f)$) gets more and more important at lower frequencies at which LISA will be operating, at LIGO's frequencies, however, it is negligible and $\hat{\beta}(f) \rightarrow 1$.

\begin{figure}[h]
\centering
\includegraphics[width=0.5 \textwidth,origin=c,angle=0]{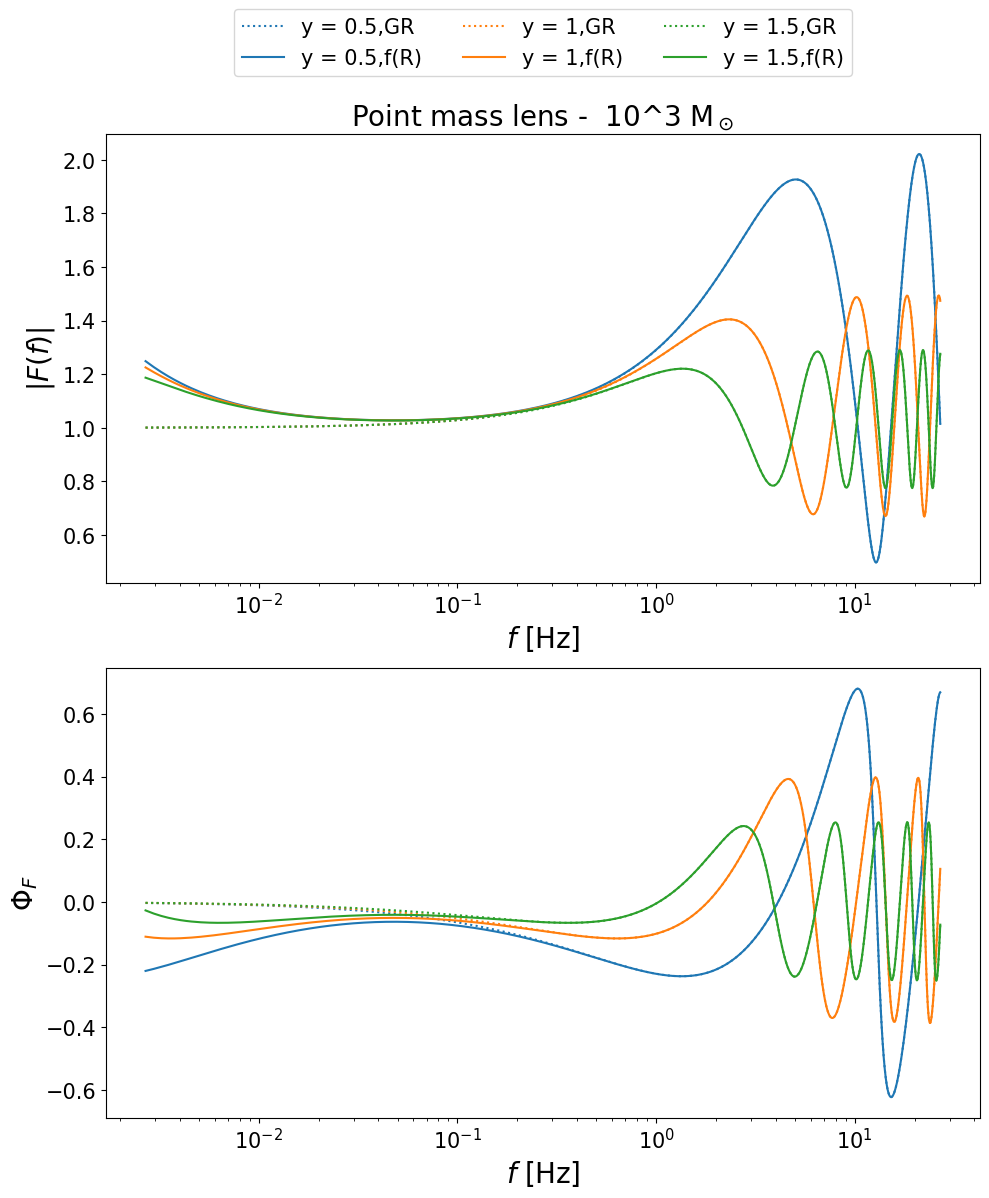}
\caption{\label{fig:amp_fac_10^3} Amplification factor $F(f)$ phase (bottom plot) and amplitude (top plot) for a point mass lens ($M_{l}=10^3 M_\odot$). The solid lines are used for the massive scalaron ($\hat{\beta} \neq1$) and the dashed ones in the GR regime ($\hat{\beta}=1$). At low frequencies, the two theories slightly differ.}
\end{figure}

\begin{figure}[h]
\centering
\includegraphics[width=0.5 \textwidth,origin=c,angle=0]{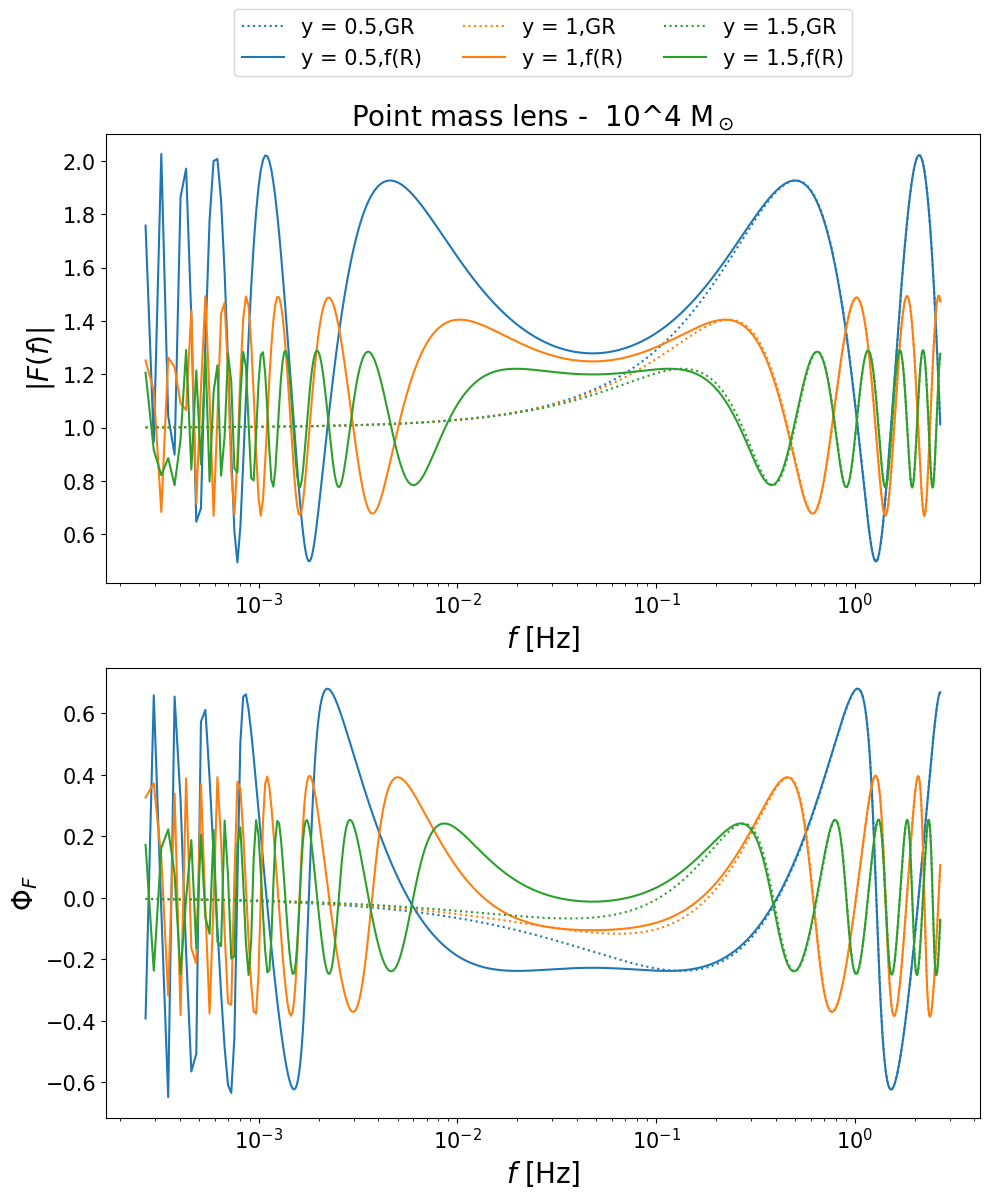}
\caption{\label{fig:amp_fac_10^4}
Amplification factor $F(f)$ phase (bottom plot) and amplitude (top plot) for a point mass lens ($M_{l}=10^4 M_\odot$).
 The solid lines are used for the massive scalaron ($\hat{\beta} \neq1$) and the dashed ones in the GR regime ($\hat{\beta}=1$). At low frequencies, there are significant differences between GR and massive scalarons. This frequency range can be explored with space-based interferometers.}
\end{figure}

\begin{figure}[h]
\includegraphics[width=8cm]{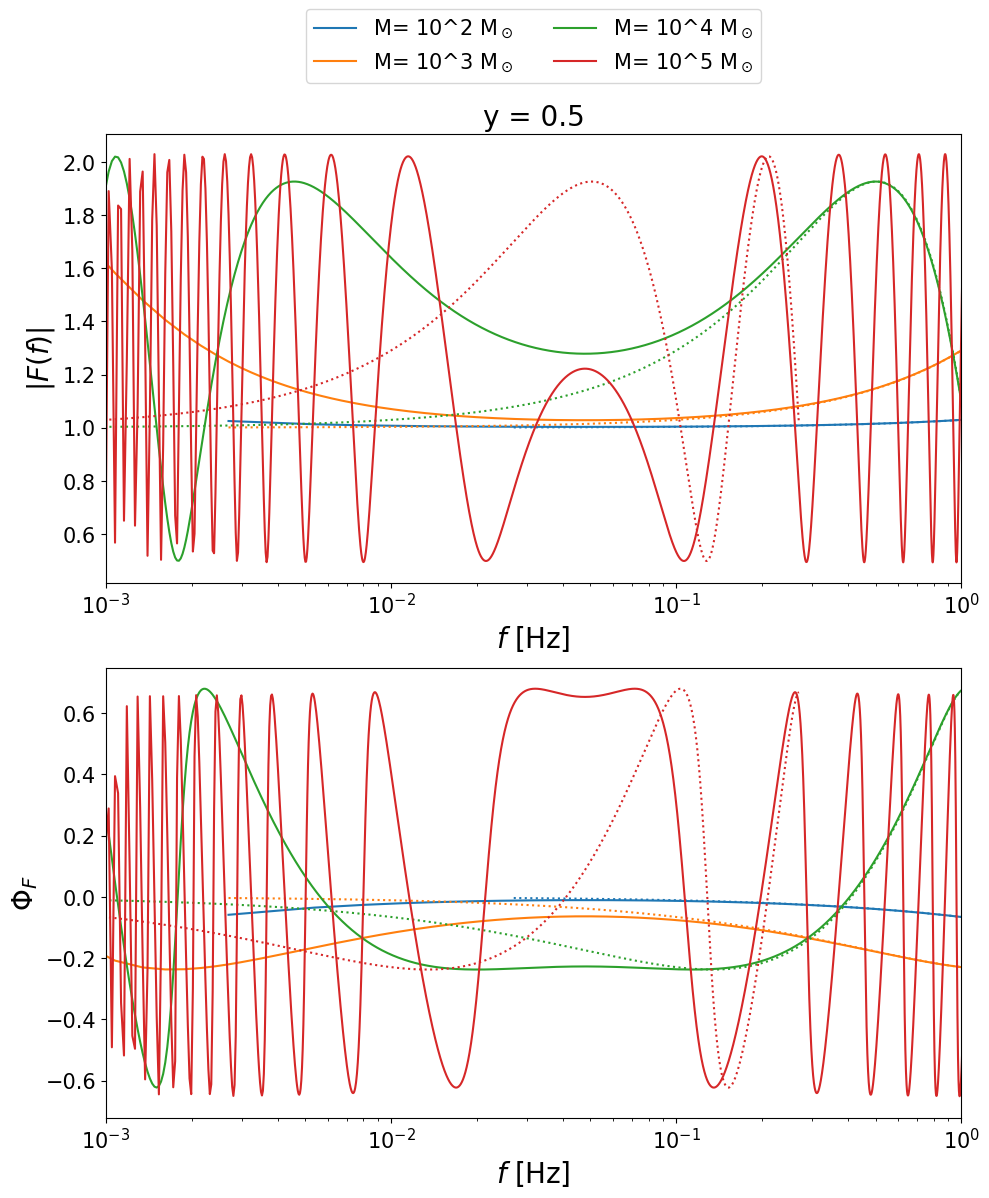}
\centering
\caption{Amplification factor $F(f)$ phase (bottom plot) and amplitude (top plot) for different point mass lenses with fixed impact factor y=0.5. The solid lines are used for the massive scalaron ($\hat{\beta} \neq1$) and the dashed ones in the GR regime ($\hat{\beta}=1$).}
\label{fig:amp_fac_diff_mass}
\end{figure}

\begin{figure}[h]
\includegraphics[width=8cm]{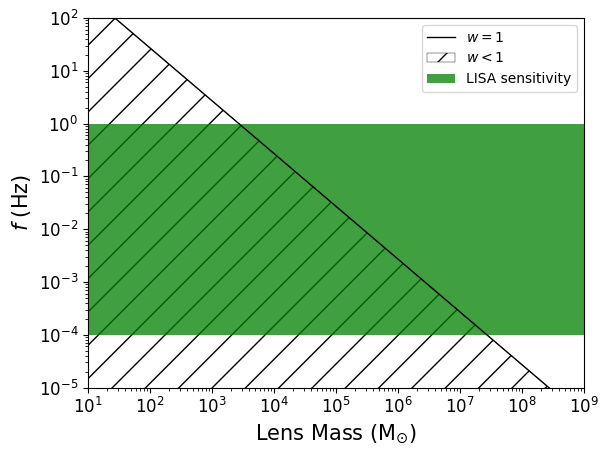}
\centering
\caption{The green block represents LISA working frequencies. The diagonal black line follows $f_T$ (i.e. $w=1$) for a given lens mass. The dashed part of the plot indicates $w<1$, where the $f(R)$ effects in the amplification factor become visible.}
\label{fig:transition_freq}
\end{figure}

In Fig. \ref{fig:amp_fac_10^3} and Fig. \ref{fig:amp_fac_10^4}, we plot the amplification factor amplitude and phase for 10$^3$ $M_{\odot}$ and 10$^4$ $M_{\odot}$ for different impact factors at $z_L=2$. In these figures, the solid line follows the amplification factor for the case with massive scalarons($\hat{\beta} \neq1$) and the dashed one for the GR ($\hat{\beta} =1$) case.  The amplification factor, in the two theories, matches at high frequencies, making the massive scalaron effects negligible and beyond the scope of ground-based detectors like LIGO. At lower frequencies and higher lens masses, particularly in the range of intermediate-mass BHs and higher, we have a significant deviation from GR. For our choice of scalaron mass ($m_{\phi} =  \mathcal{O}(10^{-15})eV$), it has been found that the lens mass of $10^{3} M_{\odot}$ shows very small deviations (Fig. \ref{fig:amp_fac_10^3}) from GR at low frequencies. However, for lens mass $10^{4} M_{\odot}$ the deviations from GR are significant at low frequencies and accessible to space-based detectors like LISA.
In Fig. \ref{fig:amp_fac_diff_mass} we zoom in on the frequencies at which these differences are significant. Keeping the impact factor constant at y=0.5, we plot for different $M_l$, the amplification factor phase and amplitude. The higher the mass of the lens the bigger the differences. With a $M_l$=10$^2M_{\odot}$ the differences are undetectable by any interferometer, with $M_l$=10$^3M_{\odot}$ and higher masses the discrepancies between f(R) and GR would be detectable by future generations space interferometers.
As the lens mass decreases, the frequency at which both curves (solid and dashed) get distinguished tends to much lower limits. The lens mass is the only reason for lensing effects in GR backgrounds.
All the above-mentioned figures displayed the amplification factor as a function of frequency $f$ in [Hz]. In the Appendix, some more plots for intermediate-mass black holes in the role of lenses are shown with respect to dimensionless frequency $w$ instead of $f$.
\\
For the sake of discussion, we call the frequency at which GR and massive sclarons lensing effects deviate noticeably, the transition frequency $f_T$. This happens for the dimensionless frequency $w < 1 $, hence by virtue of the equation \textcolor[rgb]{0.00,0.00,1.00}{  \eqref{ref:dim_freq}}, the transition frequency for a given lens mass $M_l$ can be expressed as:
\begin{equation}
\label{eq:ft}
f_{T} = \frac{c^{3}}{8\pi G M_{l}(1 + z_{l})}.
\end{equation}
This relation is plotted in Fig. \ref{fig:transition_freq}, in which we highlighted the LISA operating frequencies. The solid black line marks the transition frequency $f_T$ as a function of the lens mass, while the shaded region under the curve indicates the region in which GR and f(R) lensing effects differ. Let us stress that $f_T$ is not a sharp transition but for masses from 10$^3 M_{\odot}$ to $\sim$ 10$^6 M_{\odot}$ we should be able to observe deviations from GR of gravitationally lensed GWs.

\section{\label{7}Summary and Conclusions}

Building on previous investigations conducted by  authors of this paper, regarding the dynamics of $f(R)$ gravity and lensing \textcolor[rgb]{0.00,0.00,1.00} {\cite{KumarSharma:2022qdf,Sharma:2022fiw,Yadav:2018llv,Grespan:2023cpa,Biesiada:2021pzo,Biesiada:2014kwa}}, we undertook a comprehensive examination of novel aspects of   $f(R)\propto R^{(1+\epsilon)}$ gravity in the context of the propagation and lensing of GWs.  The GW  solutions are derived for the model parameter $\epsilon<<1$. Further, the scalaron mass and the dispersion relations are constrained  for the Solar System background characterized by $R^{(B)}\approx 10^{-35}eV^2$. It is to be noted that for the choice of  $\epsilon\approx\mathcal{O}(10^{-7})$ the minimum bound on the scalar mode mass turned out to be $m_\phi \approx \mathcal{O} (10^{-15}) eV$. These considerations do not violate the bounds on graviton mass obtained from the detection of various GW events \textcolor[rgb]{0.00,0.00,1.00}{\cite{Morisaki:2018htj,Brito:2017zvb}}. 
The  fractional variation in the speed of this massive modes  is expected to fall within the operational range of the LISA detector.

An important feature of $f(R)\propto R^{(1+\epsilon)}$ model is the presence of different polarization states depending on the values of the parameter $\epsilon$. We have investigated the polarization properties and its dependence on scalaron mass using the modified NP formalism.
For example, the absence of massive scalar degrees of freedom in $f(R)$ gravity for $\epsilon=1$ does not necessitate the disappearance of NP scalars $\Psi_2$ and $\Phi_{22}$ which are responsible for longitudinal and breathing modes of polarization. Also, it is worth noting that for  $\epsilon<<1$, both the modes are independent of the scalaron mass. However, in $\Psi_4$ scalar, we observe an additional contribution from the massive scalar mode depending on  the background curvature scalar $R^{(B)}$.
These sets of modified NP quantities (equations \textcolor[rgb]{0.00,0.00,1.00}{(\ref{ab3160})}-\textcolor[rgb]{0.00,0.00,1.00}{(\ref{abcdef31560})} and \textcolor[rgb]{0.00,0.00,1.00}{(\ref{ab31601})}-\textcolor[rgb]{0.00,0.00,1.00}{(\ref{abcdef315604})}) would serve as a diagnostic tool for distinguishing the characteristics of massless and massive modes of propagation from their GR counterparts in shifted Ricci scalar model of $f(R)$.

Among other probes, gravitational lensing could emerge as an important tool to scrutinize theories of gravity, bound the mass of graviton (and possibly scalaron), and constrain models related to dark matter.
We showed that at low frequencies (see Fig. \ref{fig:amp_fac_10^3}, \ref{fig:amp_fac_10^4}), GR and $f(R)$ lensing effects vary. While at low frequencies, in GR, the amplification factor tends to 1. Instead, for massive scalarons in $f(R)$ with $\epsilon<<1$ and $m_\phi \approx \mathcal{O} (10^{-15})eV$, in the same frequency regime, there are deviations from unity which are not negligible for compact lenses.
Such deviations from GR can be explored with the space-based detectors with lens mass lying in the range, ($10^3\leq M_{l}\leq 10^6$)$M_\odot$, see Fig. \ref{fig:transition_freq}.
Our work establishes the groundwork for a novel approach to study the massive  graviton effects at local scales when the lensed GWs are detected by space-based low-frequency detectors.
Typically, investigations of GWs focus on the non-dispersive vacuum modes that propagate at the speed of light, as these modes are also relevant ones in the context of current observations. However, the coupling between GWs and matter could be significant in the early Universe and possibly near compact objects. Understanding this coupling holds potential importance \textcolor[rgb]{0.00,0.00,1.00}{\cite{Garg:2022wdm,Garg:2021jss,Will:2018gku}}. Moreover, a crucial aspect is the dispersion relation of GWs due to the existence of a massive mode that alters the time delay of waves with different frequencies in various directions. This alteration leads to the emergence of additional features within the lensing pattern. These supplementary features create an entirely new possibility of measuring the scalaron mass through the detection of lensed GWs.

\newpage

\begin{appendix}
\section{Additional plots}
\begin{figure}[h]
\includegraphics[width=8cm]{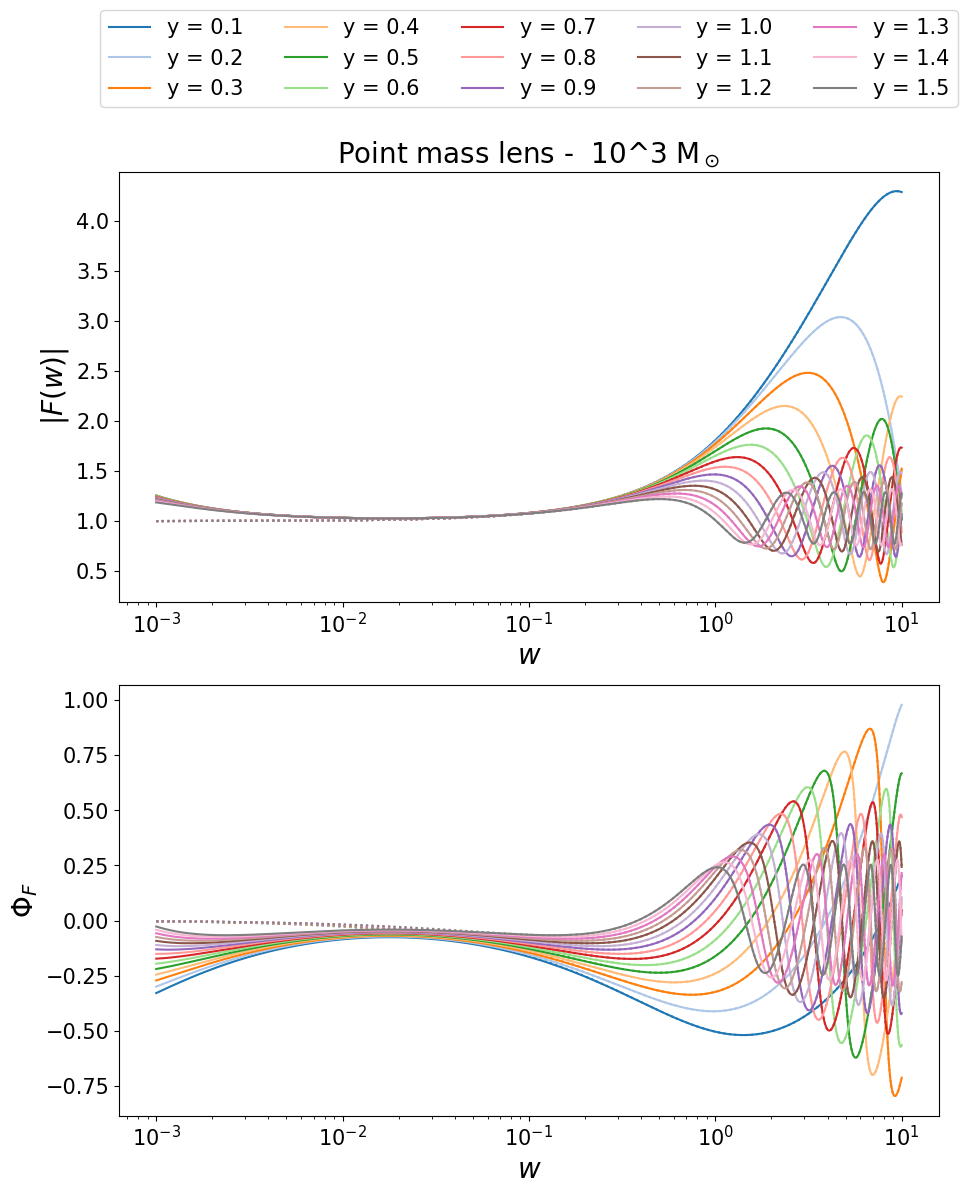}
\centering
\caption{Amplification factor amplitude (top) and phase (bottom) dependency on dimensionless frequency for a point mass lens of $M_l$= 10$^3$. Solid lines for $f(R)$ scalarons and dashed lines for GR.}
\label{fig:diff_y_103}
\end{figure}

\begin{figure}[h]
\includegraphics[width=8cm]{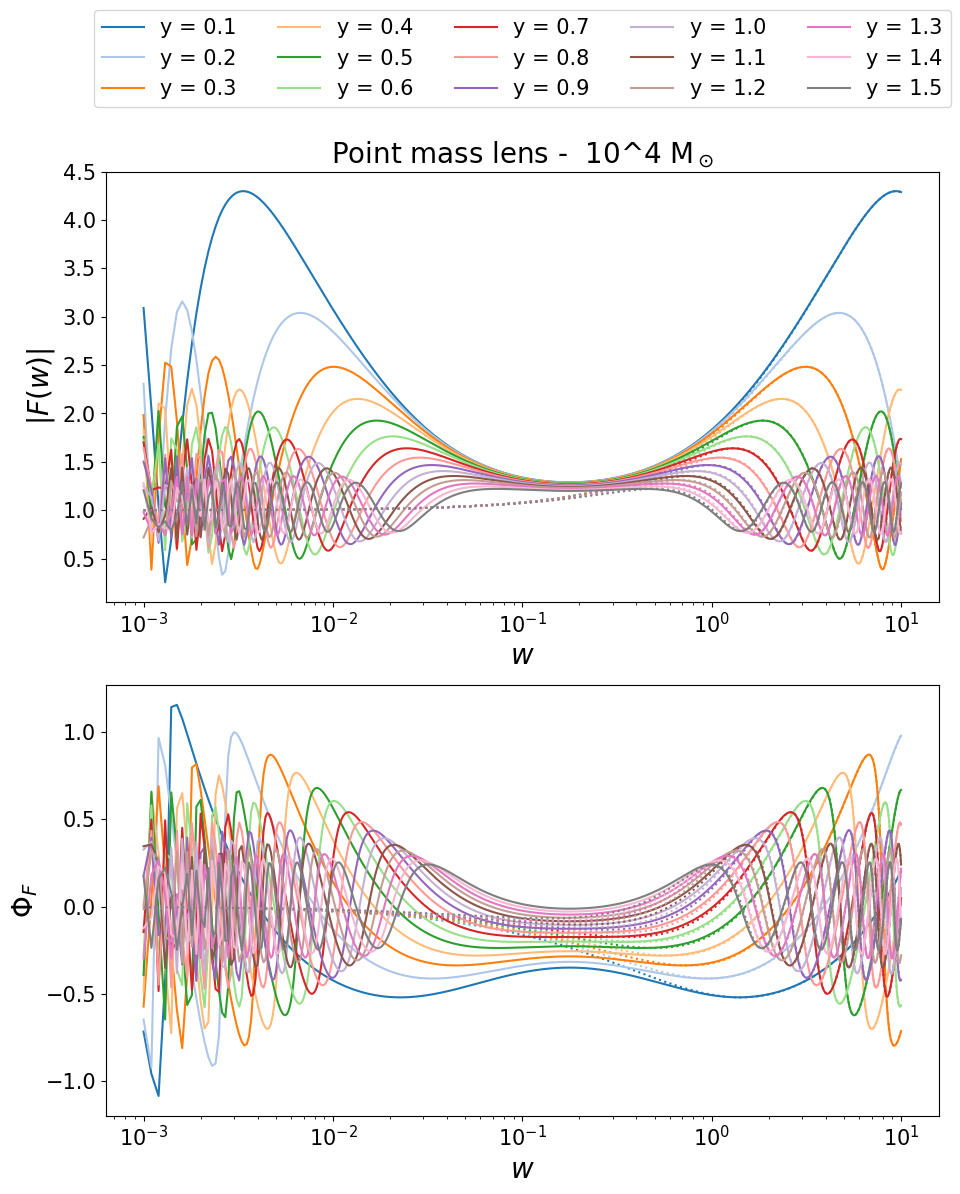}
\centering
\caption{Amplification factor amplitude (top) and phase (bottom) dependency on dimensionless frequency for a point mass lens of $M_l$= 10$^4$. Solid lines for $f(R)$ scalarons and dashed lines for GR.}
\label{fig:diff_y_104}
\end{figure}

In the main body of the paper, we show the amplification factor evolution in terms of frequency, mass and impact factor. Here we put additional, and maybe a bit more intricate, figures of the amplification factor (in respect to the dimensionless frequency) for two point mass lenses, $M_l$= 10$^3$, 10$^4$ $M_{\odot}$ in Fig. \ref{fig:diff_y_103} and \ref{fig:diff_y_104} respectively. These plots show how much $F(w)$ depends on $y$ as well as on $M_l$. In the GR regime, the amplification factors, for a point mass lens, in dimensionless frequency depend only on the impact factor and not on the mass of the lens. However, in f(R), the mass of the lens comes into play in Eq. \ref{eq:beta} through the frequency $f$ (Eq. \ref{ref:dim_freq}).

\end{appendix}

\newpage
\section*{Acknowledgments}
VKS thanks  Varun Sahni and Swagat S. Mishra for their motivation for the present work. VKS and MMV both thank Inter-University Centre for Astronomy and Astrophysics (IUCAA), Pune, India for the facilities provided under the associateship programme.

\bibliography{p_4} 

\begin{thebibliography}{111}%
\makeatletter
\providecommand \@ifxundefined [1]{%
 \@ifx{#1\undefined}
}%
\providecommand \@ifnum [1]{%
 \ifnum #1\expandafter \@firstoftwo
 \else \expandafter \@secondoftwo
 \fi
}%
\providecommand \@ifx [1]{%
 \ifx #1\expandafter \@firstoftwo
 \else \expandafter \@secondoftwo
 \fi
}%
\providecommand \natexlab [1]{#1}%
\providecommand \enquote  [1]{``#1''}%
\providecommand \bibnamefont  [1]{#1}%
\providecommand \bibfnamefont [1]{#1}%
\providecommand \citenamefont [1]{#1}%
\providecommand \href@noop [0]{\@secondoftwo}%
\providecommand \href [0]{\begingroup \@sanitize@url \@href}%
\providecommand \@href[1]{\@@startlink{#1}\@@href}%
\providecommand \@@href[1]{\endgroup#1\@@endlink}%
\providecommand \@sanitize@url [0]{\catcode `\\12\catcode `\$12\catcode
  `\&12\catcode `\#12\catcode `\^12\catcode `\_12\catcode `\%12\relax}%
\providecommand \@@startlink[1]{}%
\providecommand \@@endlink[0]{}%
\providecommand \url  [0]{\begingroup\@sanitize@url \@url }%
\providecommand \@url [1]{\endgroup\@href {#1}{\urlprefix }}%
\providecommand \urlprefix  [0]{URL }%
\providecommand \Eprint [0]{\href }%
\providecommand \doibase [0]{https://doi.org/}%
\providecommand \selectlanguage [0]{\@gobble}%
\providecommand \bibinfo  [0]{\@secondoftwo}%
\providecommand \bibfield  [0]{\@secondoftwo}%
\providecommand \translation [1]{[#1]}%
\providecommand \BibitemOpen [0]{}%
\providecommand \bibitemStop [0]{}%
\providecommand \bibitemNoStop [0]{.\EOS\space}%
\providecommand \EOS [0]{\spacefactor3000\relax}%
\providecommand \BibitemShut  [1]{\csname bibitem#1\endcsname}%
\let\auto@bib@innerbib\@empty
\bibitem [{\citenamefont {Abbott}\ \emph
  {et~al.}(2016{\natexlab{a}})\citenamefont {Abbott} \emph
  {et~al.}}]{LIGOScientific:2016aoc}%
  \BibitemOpen
  \bibfield  {author} {\bibinfo {author} {\bibfnamefont {B.~P.}\ \bibnamefont
  {Abbott}} \emph {et~al.} (\bibinfo {collaboration} {LIGO Scientific,
  Virgo}),\ }\bibfield  {title} {\bibinfo {title} {{Observation of
  Gravitational Waves from a Binary Black Hole Merger}},\ }\href
  {https://doi.org/10.1103/PhysRevLett.116.061102} {\bibfield  {journal}
  {\bibinfo  {journal} {Phys. Rev. Lett.}\ }\textbf {\bibinfo {volume} {116}},\
  \bibinfo {pages} {061102} (\bibinfo {year} {2016}{\natexlab{a}})},\ \Eprint
  {https://arxiv.org/abs/1602.03837} {arXiv:1602.03837 [gr-qc]} \BibitemShut
  {NoStop}%
\bibitem [{\citenamefont {Abbott}\ \emph
  {et~al.}(2016{\natexlab{b}})\citenamefont {Abbott} \emph
  {et~al.}}]{LIGOScientific:2016lio}%
  \BibitemOpen
  \bibfield  {author} {\bibinfo {author} {\bibfnamefont {B.~P.}\ \bibnamefont
  {Abbott}} \emph {et~al.} (\bibinfo {collaboration} {LIGO Scientific,
  Virgo}),\ }\bibfield  {title} {\bibinfo {title} {{Tests of general relativity
  with GW150914}},\ }\href {https://doi.org/10.1103/PhysRevLett.116.221101}
  {\bibfield  {journal} {\bibinfo  {journal} {Phys. Rev. Lett.}\ }\textbf
  {\bibinfo {volume} {116}},\ \bibinfo {pages} {221101} (\bibinfo {year}
  {2016}{\natexlab{b}})},\ \bibinfo {note} {[Erratum: Phys.Rev.Lett. 121,
  129902 (2018)]},\ \Eprint {https://arxiv.org/abs/1602.03841}
  {arXiv:1602.03841 [gr-qc]} \BibitemShut {NoStop}%
\bibitem [{\citenamefont {Abbott}\ \emph
  {et~al.}(2017{\natexlab{a}})\citenamefont {Abbott} \emph
  {et~al.}}]{LIGOScientific:2017vwq}%
  \BibitemOpen
  \bibfield  {author} {\bibinfo {author} {\bibfnamefont {B.~P.}\ \bibnamefont
  {Abbott}} \emph {et~al.} (\bibinfo {collaboration} {LIGO Scientific,
  Virgo}),\ }\bibfield  {title} {\bibinfo {title} {{GW170817: Observation of
  Gravitational Waves from a Binary Neutron Star Inspiral}},\ }\href
  {https://doi.org/10.1103/PhysRevLett.119.161101} {\bibfield  {journal}
  {\bibinfo  {journal} {Phys. Rev. Lett.}\ }\textbf {\bibinfo {volume} {119}},\
  \bibinfo {pages} {161101} (\bibinfo {year} {2017}{\natexlab{a}})},\ \Eprint
  {https://arxiv.org/abs/1710.05832} {arXiv:1710.05832 [gr-qc]} \BibitemShut
  {NoStop}%
\bibitem [{\citenamefont {Abbott}\ \emph
  {et~al.}(2017{\natexlab{b}})\citenamefont {Abbott} \emph
  {et~al.}}]{LIGOScientific:2017zic}%
  \BibitemOpen
  \bibfield  {author} {\bibinfo {author} {\bibfnamefont {B.~P.}\ \bibnamefont
  {Abbott}} \emph {et~al.} (\bibinfo {collaboration} {LIGO Scientific, Virgo,
  Fermi-GBM, INTEGRAL}),\ }\bibfield  {title} {\bibinfo {title} {{Gravitational
  Waves and Gamma-rays from a Binary Neutron Star Merger: GW170817 and GRB
  170817A}},\ }\href {https://doi.org/10.3847/2041-8213/aa920c} {\bibfield
  {journal} {\bibinfo  {journal} {Astrophys. J. Lett.}\ }\textbf {\bibinfo
  {volume} {848}},\ \bibinfo {pages} {L13} (\bibinfo {year}
  {2017}{\natexlab{b}})},\ \Eprint {https://arxiv.org/abs/1710.05834}
  {arXiv:1710.05834 [astro-ph.HE]} \BibitemShut {NoStop}%
\bibitem [{\citenamefont {Green}(2022)}]{Green:2021jrr}%
  \BibitemOpen
  \bibfield  {author} {\bibinfo {author} {\bibfnamefont {A.~M.}\ \bibnamefont
  {Green}},\ }\bibfield  {title} {\bibinfo {title} {{Dark matter in
  astrophysics/cosmology}},\ }\href
  {https://doi.org/10.21468/SciPostPhysLectNotes.37} {\bibfield  {journal}
  {\bibinfo  {journal} {SciPost Phys. Lect. Notes}\ }\textbf {\bibinfo {volume}
  {37}},\ \bibinfo {pages} {1} (\bibinfo {year} {2022})},\ \Eprint
  {https://arxiv.org/abs/2109.05854} {arXiv:2109.05854 [hep-ph]} \BibitemShut
  {NoStop}%
\bibitem [{\citenamefont {Riess}\ \emph {et~al.}(2019)\citenamefont {Riess},
  \citenamefont {Casertano}, \citenamefont {Yuan}, \citenamefont {Macri},\ and\
  \citenamefont {Scolnic}}]{Riess:2019cxk}%
  \BibitemOpen
  \bibfield  {author} {\bibinfo {author} {\bibfnamefont {A.~G.}\ \bibnamefont
  {Riess}}, \bibinfo {author} {\bibfnamefont {S.}~\bibnamefont {Casertano}},
  \bibinfo {author} {\bibfnamefont {W.}~\bibnamefont {Yuan}}, \bibinfo {author}
  {\bibfnamefont {L.~M.}\ \bibnamefont {Macri}},\ and\ \bibinfo {author}
  {\bibfnamefont {D.}~\bibnamefont {Scolnic}},\ }\bibfield  {title} {\bibinfo
  {title} {{Large Magellanic Cloud Cepheid Standards Provide a 1\% Foundation
  for the Determination of the Hubble Constant and Stronger Evidence for
  Physics beyond $\Lambda$CDM}},\ }\href
  {https://doi.org/10.3847/1538-4357/ab1422} {\bibfield  {journal} {\bibinfo
  {journal} {Astrophys. J.}\ }\textbf {\bibinfo {volume} {876}},\ \bibinfo
  {pages} {85} (\bibinfo {year} {2019})},\ \Eprint
  {https://arxiv.org/abs/1903.07603} {arXiv:1903.07603 [astro-ph.CO]}
  \BibitemShut {NoStop}%
\bibitem [{\citenamefont {Silk}(2017)}]{Silk:2016srn}%
  \BibitemOpen
  \bibfield  {author} {\bibinfo {author} {\bibfnamefont {J.}~\bibnamefont
  {Silk}},\ }\bibfield  {title} {\bibinfo {title} {{Challenges in Cosmology
  from the Big Bang to Dark Energy, Dark Matter and Galaxy Formation}},\ }\href
  {https://doi.org/10.7566/JPSCP.14.010101} {\bibfield  {journal} {\bibinfo
  {journal} {JPS Conf. Proc.}\ }\textbf {\bibinfo {volume} {14}},\ \bibinfo
  {pages} {010101} (\bibinfo {year} {2017})},\ \Eprint
  {https://arxiv.org/abs/1611.09846} {arXiv:1611.09846 [astro-ph.CO]}
  \BibitemShut {NoStop}%
\bibitem [{\citenamefont {Del~Popolo}\ and\ \citenamefont
  {Le~Delliou}(2017)}]{DelPopolo:2016emo}%
  \BibitemOpen
  \bibfield  {author} {\bibinfo {author} {\bibfnamefont {A.}~\bibnamefont
  {Del~Popolo}}\ and\ \bibinfo {author} {\bibfnamefont {M.}~\bibnamefont
  {Le~Delliou}},\ }\bibfield  {title} {\bibinfo {title} {{Small scale problems
  of the $\Lambda$CDM model: a short review}},\ }\href
  {https://doi.org/10.3390/galaxies5010017} {\bibfield  {journal} {\bibinfo
  {journal} {Galaxies}\ }\textbf {\bibinfo {volume} {5}},\ \bibinfo {pages}
  {17} (\bibinfo {year} {2017})},\ \Eprint {https://arxiv.org/abs/1606.07790}
  {arXiv:1606.07790 [astro-ph.CO]} \BibitemShut {NoStop}%
\bibitem [{\citenamefont {Aghanim}\ \emph {et~al.}(2020)\citenamefont {Aghanim}
  \emph {et~al.}}]{Planck:2018vyg}%
  \BibitemOpen
  \bibfield  {author} {\bibinfo {author} {\bibfnamefont {N.}~\bibnamefont
  {Aghanim}} \emph {et~al.} (\bibinfo {collaboration} {Planck}),\ }\bibfield
  {title} {\bibinfo {title} {{Planck 2018 results. VI. Cosmological
  parameters}},\ }\href {https://doi.org/10.1051/0004-6361/201833910}
  {\bibfield  {journal} {\bibinfo  {journal} {Astron. Astrophys.}\ }\textbf
  {\bibinfo {volume} {641}},\ \bibinfo {pages} {A6} (\bibinfo {year} {2020})},\
  \bibinfo {note} {[Erratum: Astron.Astrophys. 652, C4 (2021)]},\ \Eprint
  {https://arxiv.org/abs/1807.06209} {arXiv:1807.06209 [astro-ph.CO]}
  \BibitemShut {NoStop}%
\bibitem [{\citenamefont {Macaulay}\ \emph {et~al.}(2013)\citenamefont
  {Macaulay}, \citenamefont {Wehus},\ and\ \citenamefont
  {Eriksen}}]{Macaulay:2013swa}%
  \BibitemOpen
  \bibfield  {author} {\bibinfo {author} {\bibfnamefont {E.}~\bibnamefont
  {Macaulay}}, \bibinfo {author} {\bibfnamefont {I.~K.}\ \bibnamefont
  {Wehus}},\ and\ \bibinfo {author} {\bibfnamefont {H.~K.}\ \bibnamefont
  {Eriksen}},\ }\bibfield  {title} {\bibinfo {title} {{Lower Growth Rate from
  Recent Redshift Space Distortion Measurements than Expected from Planck}},\
  }\href {https://doi.org/10.1103/PhysRevLett.111.161301} {\bibfield  {journal}
  {\bibinfo  {journal} {Phys. Rev. Lett.}\ }\textbf {\bibinfo {volume} {111}},\
  \bibinfo {pages} {161301} (\bibinfo {year} {2013})},\ \Eprint
  {https://arxiv.org/abs/1303.6583} {arXiv:1303.6583 [astro-ph.CO]}
  \BibitemShut {NoStop}%
\bibitem [{\citenamefont {Charnock}\ \emph {et~al.}(2017)\citenamefont
  {Charnock}, \citenamefont {Battye},\ and\ \citenamefont
  {Moss}}]{Charnock:2017vcd}%
  \BibitemOpen
  \bibfield  {author} {\bibinfo {author} {\bibfnamefont {T.}~\bibnamefont
  {Charnock}}, \bibinfo {author} {\bibfnamefont {R.~A.}\ \bibnamefont
  {Battye}},\ and\ \bibinfo {author} {\bibfnamefont {A.}~\bibnamefont {Moss}},\
  }\bibfield  {title} {\bibinfo {title} {{Planck data versus large scale
  structure}: {Methods to quantify discordance}},\ }\href
  {https://doi.org/10.1103/PhysRevD.95.123535} {\bibfield  {journal} {\bibinfo
  {journal} {Phys. Rev. D}\ }\textbf {\bibinfo {volume} {95}},\ \bibinfo
  {pages} {123535} (\bibinfo {year} {2017})},\ \Eprint
  {https://arxiv.org/abs/1703.05959} {arXiv:1703.05959 [astro-ph.CO]}
  \BibitemShut {NoStop}%
\bibitem [{\citenamefont {Heisenberg}(2019)}]{Heisenberg:2018vsk}%
  \BibitemOpen
  \bibfield  {author} {\bibinfo {author} {\bibfnamefont {L.}~\bibnamefont
  {Heisenberg}},\ }\bibfield  {title} {\bibinfo {title} {{A systematic approach
  to generalisations of General Relativity and their cosmological
  implications}},\ }\href {https://doi.org/10.1016/j.physrep.2018.11.006}
  {\bibfield  {journal} {\bibinfo  {journal} {Phys. Rept.}\ }\textbf {\bibinfo
  {volume} {796}},\ \bibinfo {pages} {1} (\bibinfo {year} {2019})},\ \Eprint
  {https://arxiv.org/abs/1807.01725} {arXiv:1807.01725 [gr-qc]} \BibitemShut
  {NoStop}%
\bibitem [{\citenamefont {Wang}(2021)}]{Wang:2020dsc}%
  \BibitemOpen
  \bibfield  {author} {\bibinfo {author} {\bibfnamefont {D.}~\bibnamefont
  {Wang}},\ }\bibfield  {title} {\bibinfo {title} {{Can $f(R)$ gravity relieve
  $H_0$ and $\sigma _8$ tensions?}},\ }\href
  {https://doi.org/10.1140/epjc/s10052-021-09264-9} {\bibfield  {journal}
  {\bibinfo  {journal} {Eur. Phys. J. C}\ }\textbf {\bibinfo {volume} {81}},\
  \bibinfo {pages} {482} (\bibinfo {year} {2021})},\ \Eprint
  {https://arxiv.org/abs/2008.03966} {arXiv:2008.03966 [astro-ph.CO]}
  \BibitemShut {NoStop}%
\bibitem [{\citenamefont {Saridakis.}(2023)}]{Saridakis:2023pzo}%
  \BibitemOpen
  \bibfield  {author} {\bibinfo {author} {\bibfnamefont {E.~N.}\ \bibnamefont
  {Saridakis.}},\ }\href {https://doi.org/10.1142/9789811269776_0139} {\bibinfo
  {title} {Solving both $h_0$ and $\sigma_8$ tensions in $f(t)$ gravity}}
  (\bibinfo {year} {2023}),\ \Eprint {https://arxiv.org/abs/2301.06881}
  {arXiv:2301.06881 [gr-qc]} \BibitemShut {NoStop}%
\bibitem [{\citenamefont {De~Felice}\ and\ \citenamefont
  {Tsujikawa}(2010)}]{DeFelice:2010aj}%
  \BibitemOpen
  \bibfield  {author} {\bibinfo {author} {\bibfnamefont {A.}~\bibnamefont
  {De~Felice}}\ and\ \bibinfo {author} {\bibfnamefont {S.}~\bibnamefont
  {Tsujikawa}},\ }\bibfield  {title} {\bibinfo {title} {{f(R) theories}},\
  }\href {https://doi.org/10.12942/lrr-2010-3} {\bibfield  {journal} {\bibinfo
  {journal} {Living Rev. Rel.}\ }\textbf {\bibinfo {volume} {13}},\ \bibinfo
  {pages} {3} (\bibinfo {year} {2010})},\ \Eprint
  {https://arxiv.org/abs/1002.4928} {arXiv:1002.4928 [gr-qc]} \BibitemShut
  {NoStop}%
\bibitem [{\citenamefont {Nojiri}\ and\ \citenamefont
  {Odintsov}(2011)}]{Nojiri:2010wj}%
  \BibitemOpen
  \bibfield  {author} {\bibinfo {author} {\bibfnamefont {S.}~\bibnamefont
  {Nojiri}}\ and\ \bibinfo {author} {\bibfnamefont {S.~D.}\ \bibnamefont
  {Odintsov}},\ }\bibfield  {title} {\bibinfo {title} {{Unified cosmic history
  in modified gravity: from F(R) theory to Lorentz non-invariant models}},\
  }\href {https://doi.org/10.1016/j.physrep.2011.04.001} {\bibfield  {journal}
  {\bibinfo  {journal} {Phys. Rept.}\ }\textbf {\bibinfo {volume} {505}},\
  \bibinfo {pages} {59} (\bibinfo {year} {2011})},\ \Eprint
  {https://arxiv.org/abs/1011.0544} {arXiv:1011.0544 [gr-qc]} \BibitemShut
  {NoStop}%
\bibitem [{\citenamefont {Nojiri}\ \emph {et~al.}(2017)\citenamefont {Nojiri},
  \citenamefont {Odintsov},\ and\ \citenamefont {Oikonomou}}]{Nojiri:2017ncd}%
  \BibitemOpen
  \bibfield  {author} {\bibinfo {author} {\bibfnamefont {S.}~\bibnamefont
  {Nojiri}}, \bibinfo {author} {\bibfnamefont {S.~D.}\ \bibnamefont
  {Odintsov}},\ and\ \bibinfo {author} {\bibfnamefont {V.~K.}\ \bibnamefont
  {Oikonomou}},\ }\bibfield  {title} {\bibinfo {title} {{Modified Gravity
  Theories on a Nutshell: Inflation, Bounce and Late-time Evolution}},\ }\href
  {https://doi.org/10.1016/j.physrep.2017.06.001} {\bibfield  {journal}
  {\bibinfo  {journal} {Phys. Rept.}\ }\textbf {\bibinfo {volume} {692}},\
  \bibinfo {pages} {1} (\bibinfo {year} {2017})},\ \Eprint
  {https://arxiv.org/abs/1705.11098} {arXiv:1705.11098 [gr-qc]} \BibitemShut
  {NoStop}%
\bibitem [{\citenamefont {Faraoni}\ and\ \citenamefont
  {Capozziello}(2011)}]{Faraoni:2010pgm}%
  \BibitemOpen
  \bibfield  {author} {\bibinfo {author} {\bibfnamefont {V.}~\bibnamefont
  {Faraoni}}\ and\ \bibinfo {author} {\bibfnamefont {S.}~\bibnamefont
  {Capozziello}},\ }\href {https://doi.org/10.1007/978-94-007-0165-6} {\emph
  {\bibinfo {title} {{Beyond Einstein Gravity}: {A Survey of Gravitational
  Theories for Cosmology and Astrophysics}}}}\ (\bibinfo  {publisher}
  {Springer},\ \bibinfo {address} {Dordrecht},\ \bibinfo {year}
  {2011})\BibitemShut {NoStop}%
\bibitem [{\citenamefont {Clifton}\ and\ \citenamefont
  {Barrow}(2005)}]{Clifton:2005aj}%
  \BibitemOpen
  \bibfield  {author} {\bibinfo {author} {\bibfnamefont {T.}~\bibnamefont
  {Clifton}}\ and\ \bibinfo {author} {\bibfnamefont {J.~D.}\ \bibnamefont
  {Barrow}},\ }\bibfield  {title} {\bibinfo {title} {{The Power of General
  Relativity}},\ }\href {https://doi.org/10.1103/PhysRevD.72.103005} {\bibfield
   {journal} {\bibinfo  {journal} {Phys. Rev. D}\ }\textbf {\bibinfo {volume}
  {72}},\ \bibinfo {pages} {103005} (\bibinfo {year} {2005})},\ \bibinfo {note}
  {[Erratum: Phys.Rev.D 90, 029902 (2014)]},\ \Eprint
  {https://arxiv.org/abs/gr-qc/0509059} {arXiv:gr-qc/0509059} \BibitemShut
  {NoStop}%
\bibitem [{\citenamefont {Boehmer}\ \emph {et~al.}(2008)\citenamefont
  {Boehmer}, \citenamefont {Harko},\ and\ \citenamefont
  {Lobo}}]{Boehmer:2007kx}%
  \BibitemOpen
  \bibfield  {author} {\bibinfo {author} {\bibfnamefont {C.~G.}\ \bibnamefont
  {Boehmer}}, \bibinfo {author} {\bibfnamefont {T.}~\bibnamefont {Harko}},\
  and\ \bibinfo {author} {\bibfnamefont {F.~S.~N.}\ \bibnamefont {Lobo}},\
  }\bibfield  {title} {\bibinfo {title} {{Dark matter as a geometric effect in
  f(R) gravity}},\ }\href {https://doi.org/10.1016/j.astropartphys.2008.04.003}
  {\bibfield  {journal} {\bibinfo  {journal} {Astropart. Phys.}\ }\textbf
  {\bibinfo {volume} {29}},\ \bibinfo {pages} {386} (\bibinfo {year} {2008})},\
  \Eprint {https://arxiv.org/abs/0709.0046} {arXiv:0709.0046 [gr-qc]}
  \BibitemShut {NoStop}%
\bibitem [{\citenamefont {Kumar~Sharma}\ and\ \citenamefont
  {Verma}(2022)}]{KumarSharma:2022qdf}%
  \BibitemOpen
  \bibfield  {author} {\bibinfo {author} {\bibfnamefont {V.}~\bibnamefont
  {Kumar~Sharma}}\ and\ \bibinfo {author} {\bibfnamefont {M.~M.}\ \bibnamefont
  {Verma}},\ }\bibfield  {title} {\bibinfo {title} {{Unified f(R) gravity at
  local scales}},\ }\href {https://doi.org/10.1140/epjc/s10052-022-10329-6}
  {\bibfield  {journal} {\bibinfo  {journal} {Eur. Phys. J. C}\ }\textbf
  {\bibinfo {volume} {82}},\ \bibinfo {pages} {400} (\bibinfo {year} {2022})},\
  \Eprint {https://arxiv.org/abs/2201.01058} {arXiv:2201.01058 [astro-ph.CO]}
  \BibitemShut {NoStop}%
\bibitem [{\citenamefont {Sharma}\ \emph {et~al.}(2021)\citenamefont {Sharma},
  \citenamefont {Yadav},\ and\ \citenamefont {Verma}}]{Sharma:2020vex}%
  \BibitemOpen
  \bibfield  {author} {\bibinfo {author} {\bibfnamefont {V.~K.}\ \bibnamefont
  {Sharma}}, \bibinfo {author} {\bibfnamefont {B.~K.}\ \bibnamefont {Yadav}},\
  and\ \bibinfo {author} {\bibfnamefont {M.~M.}\ \bibnamefont {Verma}},\
  }\bibfield  {title} {\bibinfo {title} {{Light deflection angle through
  velocity profile of galaxies in $f(R)$ model}},\ }\href
  {https://doi.org/10.1140/epjc/s10052-021-08908-0} {\bibfield  {journal}
  {\bibinfo  {journal} {Eur. Phys. J. C}\ }\textbf {\bibinfo {volume} {81}},\
  \bibinfo {pages} {109} (\bibinfo {year} {2021})},\ \Eprint
  {https://arxiv.org/abs/2011.02878} {arXiv:2011.02878 [astro-ph.CO]}
  \BibitemShut {NoStop}%
\bibitem [{\citenamefont {Sharma}\ \emph {et~al.}(2020)\citenamefont {Sharma},
  \citenamefont {Yadav},\ and\ \citenamefont {Verma}}]{Sharma:2019yix}%
  \BibitemOpen
  \bibfield  {author} {\bibinfo {author} {\bibfnamefont {V.~K.}\ \bibnamefont
  {Sharma}}, \bibinfo {author} {\bibfnamefont {B.~K.}\ \bibnamefont {Yadav}},\
  and\ \bibinfo {author} {\bibfnamefont {M.~M.}\ \bibnamefont {Verma}},\
  }\bibfield  {title} {\bibinfo {title} {{Extended galactic rotational velocity
  profiles in $f(R)$ gravity background}},\ }\href
  {https://doi.org/10.1140/epjc/s10052-020-8186-1} {\bibfield  {journal}
  {\bibinfo  {journal} {Eur. Phys. J. C}\ }\textbf {\bibinfo {volume} {80}},\
  \bibinfo {pages} {619} (\bibinfo {year} {2020})},\ \Eprint
  {https://arxiv.org/abs/1912.12206} {arXiv:1912.12206 [gr-qc]} \BibitemShut
  {NoStop}%
\bibitem [{\citenamefont {Yadav}\ and\ \citenamefont
  {Verma}(2019)}]{Yadav:2018llv}%
  \BibitemOpen
  \bibfield  {author} {\bibinfo {author} {\bibfnamefont {B.~K.}\ \bibnamefont
  {Yadav}}\ and\ \bibinfo {author} {\bibfnamefont {M.~M.}\ \bibnamefont
  {Verma}},\ }\bibfield  {title} {\bibinfo {title} {{Dark matter as scalaron in
  $f(R)$ gravity models}},\ }\href
  {https://doi.org/10.1088/1475-7516/2019/10/052} {\bibfield  {journal}
  {\bibinfo  {journal} {JCAP}\ }\textbf {\bibinfo {volume} {10}},\ \bibinfo
  {pages} {052}},\ \Eprint {https://arxiv.org/abs/1811.03964} {arXiv:1811.03964
  [gr-qc]} \BibitemShut {NoStop}%
\bibitem [{\citenamefont {Sharma}\ and\ \citenamefont
  {Verma}(2022{\natexlab{a}})}]{Sharma:2022fiw}%
  \BibitemOpen
  \bibfield  {author} {\bibinfo {author} {\bibfnamefont {A.~K.}\ \bibnamefont
  {Sharma}}\ and\ \bibinfo {author} {\bibfnamefont {M.~M.}\ \bibnamefont
  {Verma}},\ }\bibfield  {title} {\bibinfo {title} {{Effect of the Modified
  Gravity on the Large-scale Structure Formation}},\ }\href
  {https://doi.org/10.3847/1538-4357/ac7b8e} {\bibfield  {journal} {\bibinfo
  {journal} {Astrophys. J.}\ }\textbf {\bibinfo {volume} {934}},\ \bibinfo
  {pages} {13} (\bibinfo {year} {2022}{\natexlab{a}})},\ \Eprint
  {https://arxiv.org/abs/2203.06741} {arXiv:2203.06741 [astro-ph.CO]}
  \BibitemShut {NoStop}%
\bibitem [{\citenamefont {Sharma}\ and\ \citenamefont
  {Verma}(2022{\natexlab{b}})}]{Sharma:2022tce}%
  \BibitemOpen
  \bibfield  {author} {\bibinfo {author} {\bibfnamefont {A.~K.}\ \bibnamefont
  {Sharma}}\ and\ \bibinfo {author} {\bibfnamefont {M.~M.}\ \bibnamefont
  {Verma}},\ }\bibfield  {title} {\bibinfo {title} {{Power-law Inflation in the
  $f$(R) Gravity}},\ }\href {https://doi.org/10.3847/1538-4357/ac3ed7}
  {\bibfield  {journal} {\bibinfo  {journal} {Astrophys. J.}\ }\textbf
  {\bibinfo {volume} {926}},\ \bibinfo {pages} {29} (\bibinfo {year}
  {2022}{\natexlab{b}})}\BibitemShut {NoStop}%
\bibitem [{\citenamefont {Casado-Turri\'on}\ \emph {et~al.}(2023)\citenamefont
  {Casado-Turri\'on}, \citenamefont {de~la Cruz-Dombriz},\ and\ \citenamefont
  {Dobado}}]{Casado-Turrion:2023rni}%
  \BibitemOpen
  \bibfield  {author} {\bibinfo {author} {\bibfnamefont {A.}~\bibnamefont
  {Casado-Turri\'on}}, \bibinfo {author} {\bibfnamefont {A.}~\bibnamefont
  {de~la Cruz-Dombriz}},\ and\ \bibinfo {author} {\bibfnamefont
  {A.}~\bibnamefont {Dobado}},\ }\bibfield  {title} {\bibinfo {title}
  {{Physical nonviability of a wide class of f(R) models and their
  constant-curvature solutions}},\ }\href
  {https://doi.org/10.1103/PhysRevD.108.064006} {\bibfield  {journal} {\bibinfo
   {journal} {Phys. Rev. D}\ }\textbf {\bibinfo {volume} {108}},\ \bibinfo
  {pages} {064006} (\bibinfo {year} {2023})},\ \Eprint
  {https://arxiv.org/abs/2303.02103} {arXiv:2303.02103 [gr-qc]} \BibitemShut
  {NoStop}%
\bibitem [{\citenamefont {Capozziello}\ \emph {et~al.}(2017)\citenamefont
  {Capozziello}, \citenamefont {De~Laurentis}, \citenamefont {Nojiri},\ and\
  \citenamefont {Odintsov}}]{Capozziello:2017vdi}%
  \BibitemOpen
  \bibfield  {author} {\bibinfo {author} {\bibfnamefont {S.}~\bibnamefont
  {Capozziello}}, \bibinfo {author} {\bibfnamefont {M.}~\bibnamefont
  {De~Laurentis}}, \bibinfo {author} {\bibfnamefont {S.}~\bibnamefont
  {Nojiri}},\ and\ \bibinfo {author} {\bibfnamefont {S.~D.}\ \bibnamefont
  {Odintsov}},\ }\bibfield  {title} {\bibinfo {title} {{Evolution of gravitons
  in accelerating cosmologies: The case of extended gravity}},\ }\href
  {https://doi.org/10.1103/PhysRevD.95.083524} {\bibfield  {journal} {\bibinfo
  {journal} {Phys. Rev. D}\ }\textbf {\bibinfo {volume} {95}},\ \bibinfo
  {pages} {083524} (\bibinfo {year} {2017})},\ \Eprint
  {https://arxiv.org/abs/1702.05517} {arXiv:1702.05517 [gr-qc]} \BibitemShut
  {NoStop}%
\bibitem [{\citenamefont {Katsuragawa}\ \emph {et~al.}(2019)\citenamefont
  {Katsuragawa}, \citenamefont {Nakamura}, \citenamefont {Ikeda},\ and\
  \citenamefont {Capozziello}}]{Katsuragawa:2019uto}%
  \BibitemOpen
  \bibfield  {author} {\bibinfo {author} {\bibfnamefont {T.}~\bibnamefont
  {Katsuragawa}}, \bibinfo {author} {\bibfnamefont {T.}~\bibnamefont
  {Nakamura}}, \bibinfo {author} {\bibfnamefont {T.}~\bibnamefont {Ikeda}},\
  and\ \bibinfo {author} {\bibfnamefont {S.}~\bibnamefont {Capozziello}},\
  }\bibfield  {title} {\bibinfo {title} {{Gravitational Waves in $F(R)$
  Gravity: Scalar Waves and the Chameleon Mechanism}},\ }\href
  {https://doi.org/10.1103/PhysRevD.99.124050} {\bibfield  {journal} {\bibinfo
  {journal} {Phys. Rev. D}\ }\textbf {\bibinfo {volume} {99}},\ \bibinfo
  {pages} {124050} (\bibinfo {year} {2019})},\ \Eprint
  {https://arxiv.org/abs/1902.02494} {arXiv:1902.02494 [gr-qc]} \BibitemShut
  {NoStop}%
\bibitem [{\citenamefont {Will}(2014)}]{Will:2014kxa}%
  \BibitemOpen
  \bibfield  {author} {\bibinfo {author} {\bibfnamefont {C.~M.}\ \bibnamefont
  {Will}},\ }\bibfield  {title} {\bibinfo {title} {{The Confrontation between
  General Relativity and Experiment}},\ }\href
  {https://doi.org/10.12942/lrr-2014-4} {\bibfield  {journal} {\bibinfo
  {journal} {Living Rev. Rel.}\ }\textbf {\bibinfo {volume} {17}},\ \bibinfo
  {pages} {4} (\bibinfo {year} {2014})},\ \Eprint
  {https://arxiv.org/abs/1403.7377} {arXiv:1403.7377 [gr-qc]} \BibitemShut
  {NoStop}%
\bibitem [{\citenamefont {Berry}\ and\ \citenamefont
  {Gair}(2011)}]{Berry:2011pb}%
  \BibitemOpen
  \bibfield  {author} {\bibinfo {author} {\bibfnamefont {C.~P.~L.}\
  \bibnamefont {Berry}}\ and\ \bibinfo {author} {\bibfnamefont {J.~R.}\
  \bibnamefont {Gair}},\ }\bibfield  {title} {\bibinfo {title} {{Linearized
  f(R) Gravity: Gravitational Radiation and Solar System Tests}},\ }\href
  {https://doi.org/10.1103/PhysRevD.83.104022} {\bibfield  {journal} {\bibinfo
  {journal} {Phys. Rev. D}\ }\textbf {\bibinfo {volume} {83}},\ \bibinfo
  {pages} {104022} (\bibinfo {year} {2011})},\ \bibinfo {note} {[Erratum:
  Phys.Rev.D 85, 089906 (2012)]},\ \Eprint {https://arxiv.org/abs/1104.0819}
  {arXiv:1104.0819 [gr-qc]} \BibitemShut {NoStop}%
\bibitem [{\citenamefont {Ezquiaga}\ and\ \citenamefont
  {Zumalac\'arregui}(2020)}]{Ezquiaga:2020dao}%
  \BibitemOpen
  \bibfield  {author} {\bibinfo {author} {\bibfnamefont {J.~M.}\ \bibnamefont
  {Ezquiaga}}\ and\ \bibinfo {author} {\bibfnamefont {M.}~\bibnamefont
  {Zumalac\'arregui}},\ }\bibfield  {title} {\bibinfo {title} {{Gravitational
  wave lensing beyond general relativity: birefringence, echoes and shadows}},\
  }\href {https://doi.org/10.1103/PhysRevD.102.124048} {\bibfield  {journal}
  {\bibinfo  {journal} {Phys. Rev. D}\ }\textbf {\bibinfo {volume} {102}},\
  \bibinfo {pages} {124048} (\bibinfo {year} {2020})},\ \Eprint
  {https://arxiv.org/abs/2009.12187} {arXiv:2009.12187 [gr-qc]} \BibitemShut
  {NoStop}%
\bibitem [{\citenamefont {Goyal}\ \emph {et~al.}(2021)\citenamefont {Goyal},
  \citenamefont {Haris}, \citenamefont {Mehta},\ and\ \citenamefont
  {Ajith}}]{Goyal:2020bkm}%
  \BibitemOpen
  \bibfield  {author} {\bibinfo {author} {\bibfnamefont {S.}~\bibnamefont
  {Goyal}}, \bibinfo {author} {\bibfnamefont {K.}~\bibnamefont {Haris}},
  \bibinfo {author} {\bibfnamefont {A.~K.}\ \bibnamefont {Mehta}},\ and\
  \bibinfo {author} {\bibfnamefont {P.}~\bibnamefont {Ajith}},\ }\bibfield
  {title} {\bibinfo {title} {{Testing the nature of gravitational-wave
  polarizations using strongly lensed signals}},\ }\href
  {https://doi.org/10.1103/PhysRevD.103.024038} {\bibfield  {journal} {\bibinfo
   {journal} {Phys. Rev. D}\ }\textbf {\bibinfo {volume} {103}},\ \bibinfo
  {pages} {024038} (\bibinfo {year} {2021})},\ \Eprint
  {https://arxiv.org/abs/2008.07060} {arXiv:2008.07060 [gr-qc]} \BibitemShut
  {NoStop}%
\bibitem [{\citenamefont {Will}(2018{\natexlab{a}})}]{Will:2018gku}%
  \BibitemOpen
  \bibfield  {author} {\bibinfo {author} {\bibfnamefont {C.~M.}\ \bibnamefont
  {Will}},\ }\bibfield  {title} {\bibinfo {title} {{Solar system versus
  gravitational-wave bounds on the graviton mass}},\ }\href
  {https://doi.org/10.1088/1361-6382/aad13c} {\bibfield  {journal} {\bibinfo
  {journal} {Class. Quant. Grav.}\ }\textbf {\bibinfo {volume} {35}},\ \bibinfo
  {pages} {17LT01} (\bibinfo {year} {2018}{\natexlab{a}})},\ \Eprint
  {https://arxiv.org/abs/1805.10523} {arXiv:1805.10523 [gr-qc]} \BibitemShut
  {NoStop}%
\bibitem [{\citenamefont {Liang}\ \emph {et~al.}(2017)\citenamefont {Liang},
  \citenamefont {Gong}, \citenamefont {Hou},\ and\ \citenamefont
  {Liu}}]{Liang:2017ahj}%
  \BibitemOpen
  \bibfield  {author} {\bibinfo {author} {\bibfnamefont {D.}~\bibnamefont
  {Liang}}, \bibinfo {author} {\bibfnamefont {Y.}~\bibnamefont {Gong}},
  \bibinfo {author} {\bibfnamefont {S.}~\bibnamefont {Hou}},\ and\ \bibinfo
  {author} {\bibfnamefont {Y.}~\bibnamefont {Liu}},\ }\bibfield  {title}
  {\bibinfo {title} {{Polarizations of gravitational waves in $f(R)$
  gravity}},\ }\href {https://doi.org/10.1103/PhysRevD.95.104034} {\bibfield
  {journal} {\bibinfo  {journal} {Phys. Rev. D}\ }\textbf {\bibinfo {volume}
  {95}},\ \bibinfo {pages} {104034} (\bibinfo {year} {2017})},\ \Eprint
  {https://arxiv.org/abs/1701.05998} {arXiv:1701.05998 [gr-qc]} \BibitemShut
  {NoStop}%
\bibitem [{\citenamefont {Nishizawa}(2016)}]{Nishizawa:2016kba}%
  \BibitemOpen
  \bibfield  {author} {\bibinfo {author} {\bibfnamefont {A.}~\bibnamefont
  {Nishizawa}},\ }\bibfield  {title} {\bibinfo {title} {{Constraining the
  propagation speed of gravitational waves with compact binaries at
  cosmological distances}},\ }\href
  {https://doi.org/10.1103/PhysRevD.93.124036} {\bibfield  {journal} {\bibinfo
  {journal} {Phys. Rev. D}\ }\textbf {\bibinfo {volume} {93}},\ \bibinfo
  {pages} {124036} (\bibinfo {year} {2016})},\ \Eprint
  {https://arxiv.org/abs/1601.01072} {arXiv:1601.01072 [gr-qc]} \BibitemShut
  {NoStop}%
\bibitem [{\citenamefont {Alves}\ \emph {et~al.}(2009)\citenamefont {Alves},
  \citenamefont {Miranda},\ and\ \citenamefont {de~Araujo}}]{Alves:2009eg}%
  \BibitemOpen
  \bibfield  {author} {\bibinfo {author} {\bibfnamefont {M.~E.~S.}\
  \bibnamefont {Alves}}, \bibinfo {author} {\bibfnamefont {O.~D.}\ \bibnamefont
  {Miranda}},\ and\ \bibinfo {author} {\bibfnamefont {J.~C.~N.}\ \bibnamefont
  {de~Araujo}},\ }\bibfield  {title} {\bibinfo {title} {{Probing the f(R)
  formalism through gravitational wave polarizations}},\ }\href
  {https://doi.org/10.1016/j.physletb.2009.08.005} {\bibfield  {journal}
  {\bibinfo  {journal} {Phys. Lett. B}\ }\textbf {\bibinfo {volume} {679}},\
  \bibinfo {pages} {401} (\bibinfo {year} {2009})},\ \Eprint
  {https://arxiv.org/abs/0908.0861} {arXiv:0908.0861 [gr-qc]} \BibitemShut
  {NoStop}%
\bibitem [{\citenamefont {Will}(1998)}]{Will:1997bb}%
  \BibitemOpen
  \bibfield  {author} {\bibinfo {author} {\bibfnamefont {C.~M.}\ \bibnamefont
  {Will}},\ }\bibfield  {title} {\bibinfo {title} {{Bounding the mass of the
  graviton using gravitational wave observations of inspiralling compact
  binaries}},\ }\href {https://doi.org/10.1103/PhysRevD.57.2061} {\bibfield
  {journal} {\bibinfo  {journal} {Phys. Rev. D}\ }\textbf {\bibinfo {volume}
  {57}},\ \bibinfo {pages} {2061} (\bibinfo {year} {1998})},\ \Eprint
  {https://arxiv.org/abs/gr-qc/9709011} {arXiv:gr-qc/9709011} \BibitemShut
  {NoStop}%
\bibitem [{\citenamefont {Odintsov}\ \emph {et~al.}(2022)\citenamefont
  {Odintsov}, \citenamefont {Oikonomou},\ and\ \citenamefont
  {Myrzakulov}}]{Odintsov:2022cbm}%
  \BibitemOpen
  \bibfield  {author} {\bibinfo {author} {\bibfnamefont {S.~D.}\ \bibnamefont
  {Odintsov}}, \bibinfo {author} {\bibfnamefont {V.~K.}\ \bibnamefont
  {Oikonomou}},\ and\ \bibinfo {author} {\bibfnamefont {R.}~\bibnamefont
  {Myrzakulov}},\ }\bibfield  {title} {\bibinfo {title} {{Spectrum of
  Primordial Gravitational Waves in Modified Gravities: A Short Overview}},\
  }\href {https://doi.org/10.3390/sym14040729} {\bibfield  {journal} {\bibinfo
  {journal} {Symmetry}\ }\textbf {\bibinfo {volume} {14}},\ \bibinfo {pages}
  {729} (\bibinfo {year} {2022})},\ \Eprint {https://arxiv.org/abs/2204.00876}
  {arXiv:2204.00876 [gr-qc]} \BibitemShut {NoStop}%
\bibitem [{\citenamefont {Mukherjee}\ \emph {et~al.}(2020)\citenamefont
  {Mukherjee}, \citenamefont {Wandelt},\ and\ \citenamefont
  {Silk}}]{Mukherjee:2019wcg}%
  \BibitemOpen
  \bibfield  {author} {\bibinfo {author} {\bibfnamefont {S.}~\bibnamefont
  {Mukherjee}}, \bibinfo {author} {\bibfnamefont {B.~D.}\ \bibnamefont
  {Wandelt}},\ and\ \bibinfo {author} {\bibfnamefont {J.}~\bibnamefont
  {Silk}},\ }\bibfield  {title} {\bibinfo {title} {{Probing the theory of
  gravity with gravitational lensing of gravitational waves and galaxy
  surveys}},\ }\href {https://doi.org/10.1093/mnras/staa827} {\bibfield
  {journal} {\bibinfo  {journal} {Mon. Not. Roy. Astron. Soc.}\ }\textbf
  {\bibinfo {volume} {494}},\ \bibinfo {pages} {1956} (\bibinfo {year}
  {2020})},\ \Eprint {https://arxiv.org/abs/1908.08951} {arXiv:1908.08951
  [astro-ph.CO]} \BibitemShut {NoStop}%
\bibitem [{\citenamefont {Mukherjee}\ \emph {et~al.}(2021)\citenamefont
  {Mukherjee}, \citenamefont {Wandelt},\ and\ \citenamefont
  {Silk}}]{Mukherjee:2020mha}%
  \BibitemOpen
  \bibfield  {author} {\bibinfo {author} {\bibfnamefont {S.}~\bibnamefont
  {Mukherjee}}, \bibinfo {author} {\bibfnamefont {B.~D.}\ \bibnamefont
  {Wandelt}},\ and\ \bibinfo {author} {\bibfnamefont {J.}~\bibnamefont
  {Silk}},\ }\bibfield  {title} {\bibinfo {title} {{Testing the general theory
  of relativity using gravitational wave propagation from dark standard
  sirens}},\ }\href {https://doi.org/10.1093/mnras/stab001} {\bibfield
  {journal} {\bibinfo  {journal} {Mon. Not. Roy. Astron. Soc.}\ }\textbf
  {\bibinfo {volume} {502}},\ \bibinfo {pages} {1136} (\bibinfo {year}
  {2021})},\ \Eprint {https://arxiv.org/abs/2012.15316} {arXiv:2012.15316
  [astro-ph.CO]} \BibitemShut {NoStop}%
\bibitem [{\citenamefont {Liao}\ \emph
  {et~al.}(2017{\natexlab{a}})\citenamefont {Liao}, \citenamefont {Fan},
  \citenamefont {Ding}, \citenamefont {Biesiada},\ and\ \citenamefont
  {Zhu}}]{Liao:2017ioi}%
  \BibitemOpen
  \bibfield  {author} {\bibinfo {author} {\bibfnamefont {K.}~\bibnamefont
  {Liao}}, \bibinfo {author} {\bibfnamefont {X.-L.}\ \bibnamefont {Fan}},
  \bibinfo {author} {\bibfnamefont {X.-H.}\ \bibnamefont {Ding}}, \bibinfo
  {author} {\bibfnamefont {M.}~\bibnamefont {Biesiada}},\ and\ \bibinfo
  {author} {\bibfnamefont {Z.-H.}\ \bibnamefont {Zhu}},\ }\bibfield  {title}
  {\bibinfo {title} {{Precision cosmology from future lensed gravitational wave
  and electromagnetic signals}},\ }\href
  {https://doi.org/10.1038/s41467-017-01152-9} {\bibfield  {journal} {\bibinfo
  {journal} {Nature Commun.}\ }\textbf {\bibinfo {volume} {8}},\ \bibinfo
  {pages} {1148} (\bibinfo {year} {2017}{\natexlab{a}})},\ \bibinfo {note}
  {[Erratum: Nature Commun. 8, 2136 (2017)]},\ \Eprint
  {https://arxiv.org/abs/1703.04151} {arXiv:1703.04151 [astro-ph.CO]}
  \BibitemShut {NoStop}%
\bibitem [{\citenamefont {Meena}(2023)}]{meena2023gravitational}%
  \BibitemOpen
  \bibfield  {author} {\bibinfo {author} {\bibfnamefont {A.~K.}\ \bibnamefont
  {Meena}},\ }\href@noop {} {\bibinfo {title} {Gravitational lensing of
  gravitational waves: Probing intermediate mass black holes in galaxy lenses
  with global minima}} (\bibinfo {year} {2023}),\ \Eprint
  {https://arxiv.org/abs/2305.02880} {arXiv:2305.02880 [astro-ph.CO]}
  \BibitemShut {NoStop}%
\bibitem [{\citenamefont {Takahashi}(2017)}]{Takahashi_2017}%
  \BibitemOpen
  \bibfield  {author} {\bibinfo {author} {\bibfnamefont {R.}~\bibnamefont
  {Takahashi}},\ }\bibfield  {title} {\bibinfo {title} {Arrival time
  differences between gravitational waves and electromagnetic signals due to
  gravitational lensing},\ }\href {https://doi.org/10.3847/1538-4357/835/1/103}
  {\bibfield  {journal} {\bibinfo  {journal} {The Astrophysical Journal}\
  }\textbf {\bibinfo {volume} {835}},\ \bibinfo {pages} {103} (\bibinfo {year}
  {2017})}\BibitemShut {NoStop}%
\bibitem [{\citenamefont {Hou}\ \emph {et~al.}(2020)\citenamefont {Hou},
  \citenamefont {Fan}, \citenamefont {Liao},\ and\ \citenamefont
  {Zhu}}]{Hou_2020}%
  \BibitemOpen
  \bibfield  {author} {\bibinfo {author} {\bibfnamefont {S.}~\bibnamefont
  {Hou}}, \bibinfo {author} {\bibfnamefont {X.-L.}\ \bibnamefont {Fan}},
  \bibinfo {author} {\bibfnamefont {K.}~\bibnamefont {Liao}},\ and\ \bibinfo
  {author} {\bibfnamefont {Z.-H.}\ \bibnamefont {Zhu}},\ }\bibfield  {title}
  {\bibinfo {title} {Gravitational wave interference via gravitational lensing:
  Measurements of luminosity distance, lens mass, and cosmological
  parameters},\ }\href {https://doi.org/10.1103/PhysRevD.101.064011} {\bibfield
   {journal} {\bibinfo  {journal} {Phys. Rev. D}\ }\textbf {\bibinfo {volume}
  {101}},\ \bibinfo {pages} {064011} (\bibinfo {year} {2020})}\BibitemShut
  {NoStop}%
\bibitem [{\citenamefont {Eardley}\ \emph
  {et~al.}(1973{\natexlab{a}})\citenamefont {Eardley}, \citenamefont {Lee},
  \citenamefont {Lightman}, \citenamefont {Wagoner},\ and\ \citenamefont
  {Will}}]{Eardley73}%
  \BibitemOpen
  \bibfield  {author} {\bibinfo {author} {\bibfnamefont {D.~M.}\ \bibnamefont
  {Eardley}}, \bibinfo {author} {\bibfnamefont {D.~L.}\ \bibnamefont {Lee}},
  \bibinfo {author} {\bibfnamefont {A.~P.}\ \bibnamefont {Lightman}}, \bibinfo
  {author} {\bibfnamefont {R.~V.}\ \bibnamefont {Wagoner}},\ and\ \bibinfo
  {author} {\bibfnamefont {C.~M.}\ \bibnamefont {Will}},\ }\bibfield  {title}
  {\bibinfo {title} {Gravitational-wave observations as a tool for testing
  relativistic gravity},\ }\href {https://doi.org/10.1103/PhysRevLett.30.884}
  {\bibfield  {journal} {\bibinfo  {journal} {Phys. Rev. Lett.}\ }\textbf
  {\bibinfo {volume} {30}},\ \bibinfo {pages} {884} (\bibinfo {year}
  {1973}{\natexlab{a}})}\BibitemShut {NoStop}%
\bibitem [{\citenamefont {Newman}\ and\ \citenamefont
  {Penrose}(1962)}]{Newman:1961qr}%
  \BibitemOpen
  \bibfield  {author} {\bibinfo {author} {\bibfnamefont {E.}~\bibnamefont
  {Newman}}\ and\ \bibinfo {author} {\bibfnamefont {R.}~\bibnamefont
  {Penrose}},\ }\bibfield  {title} {\bibinfo {title} {{An Approach to
  gravitational radiation by a method of spin coefficients}},\ }\href
  {https://doi.org/10.1063/1.1724257} {\bibfield  {journal} {\bibinfo
  {journal} {J. Math. Phys.}\ }\textbf {\bibinfo {volume} {3}},\ \bibinfo
  {pages} {566} (\bibinfo {year} {1962})}\BibitemShut {NoStop}%
\bibitem [{\citenamefont {Nishizawa}(2018)}]{Nishizawa:2017nef}%
  \BibitemOpen
  \bibfield  {author} {\bibinfo {author} {\bibfnamefont {A.}~\bibnamefont
  {Nishizawa}},\ }\bibfield  {title} {\bibinfo {title} {{Generalized framework
  for testing gravity with gravitational-wave propagation. I. Formulation}},\
  }\href {https://doi.org/10.1103/PhysRevD.97.104037} {\bibfield  {journal}
  {\bibinfo  {journal} {Phys. Rev. D}\ }\textbf {\bibinfo {volume} {97}},\
  \bibinfo {pages} {104037} (\bibinfo {year} {2018})},\ \Eprint
  {https://arxiv.org/abs/1710.04825} {arXiv:1710.04825 [gr-qc]} \BibitemShut
  {NoStop}%
\bibitem [{\citenamefont {Arun}\ and\ \citenamefont {Pai}(2013)}]{Arun:2013bp}%
  \BibitemOpen
  \bibfield  {author} {\bibinfo {author} {\bibfnamefont {K.~G.}\ \bibnamefont
  {Arun}}\ and\ \bibinfo {author} {\bibfnamefont {A.}~\bibnamefont {Pai}},\
  }\bibfield  {title} {\bibinfo {title} {{Tests of General Relativity and
  Alternative theories of gravity using Gravitational Wave observations}},\
  }\href {https://doi.org/10.1142/S0218271813410125} {\bibfield  {journal}
  {\bibinfo  {journal} {Int. J. Mod. Phys. D}\ }\textbf {\bibinfo {volume}
  {22}},\ \bibinfo {pages} {1341012} (\bibinfo {year} {2013})},\ \Eprint
  {https://arxiv.org/abs/1302.2198} {arXiv:1302.2198 [gr-qc]} \BibitemShut
  {NoStop}%
\bibitem [{\citenamefont {Abbott}\ \emph {et~al.}(2018)\citenamefont {Abbott}
  \emph {et~al.}}]{LIGOScientific:2017ous}%
  \BibitemOpen
  \bibfield  {author} {\bibinfo {author} {\bibfnamefont {B.~P.}\ \bibnamefont
  {Abbott}} \emph {et~al.} (\bibinfo {collaboration} {LIGO Scientific,
  Virgo}),\ }\bibfield  {title} {\bibinfo {title} {{First search for
  nontensorial gravitational waves from known pulsars}},\ }\href
  {https://doi.org/10.1103/PhysRevLett.120.031104} {\bibfield  {journal}
  {\bibinfo  {journal} {Phys. Rev. Lett.}\ }\textbf {\bibinfo {volume} {120}},\
  \bibinfo {pages} {031104} (\bibinfo {year} {2018})},\ \Eprint
  {https://arxiv.org/abs/1709.09203} {arXiv:1709.09203 [gr-qc]} \BibitemShut
  {NoStop}%
\bibitem [{\citenamefont {Fesik}\ \emph {et~al.}(2017)\citenamefont {Fesik},
  \citenamefont {Baryshev}, \citenamefont {Sokolov},\ and\ \citenamefont
  {Paturel}}]{Fesik2017LIGOVirgoEL}%
  \BibitemOpen
  \bibfield  {author} {\bibinfo {author} {\bibfnamefont {L.}~\bibnamefont
  {Fesik}}, \bibinfo {author} {\bibfnamefont {Y.~V.}\ \bibnamefont {Baryshev}},
  \bibinfo {author} {\bibfnamefont {V.~V.}\ \bibnamefont {Sokolov}},\ and\
  \bibinfo {author} {\bibfnamefont {G.}~\bibnamefont {Paturel}},\ }\bibfield
  {title} {\bibinfo {title} {Ligo-virgo events localization as a test of
  gravitational wave polarization state},\ }\href
  {https://api.semanticscholar.org/CorpusID:119455948} {\bibfield  {journal}
  {\bibinfo  {journal} {arXiv: General Relativity and Quantum Cosmology}\ }
  (\bibinfo {year} {2017})}\BibitemShut {NoStop}%
\bibitem [{\citenamefont {Takeda}\ \emph {et~al.}(2021)\citenamefont {Takeda},
  \citenamefont {Morisaki},\ and\ \citenamefont {Nishizawa}}]{Takeda:2020tjj}%
  \BibitemOpen
  \bibfield  {author} {\bibinfo {author} {\bibfnamefont {H.}~\bibnamefont
  {Takeda}}, \bibinfo {author} {\bibfnamefont {S.}~\bibnamefont {Morisaki}},\
  and\ \bibinfo {author} {\bibfnamefont {A.}~\bibnamefont {Nishizawa}},\
  }\bibfield  {title} {\bibinfo {title} {{Pure polarization test of GW170814
  and GW170817 using waveforms consistent with modified theories of gravity}},\
  }\href {https://doi.org/10.1103/PhysRevD.103.064037} {\bibfield  {journal}
  {\bibinfo  {journal} {Phys. Rev. D}\ }\textbf {\bibinfo {volume} {103}},\
  \bibinfo {pages} {064037} (\bibinfo {year} {2021})},\ \Eprint
  {https://arxiv.org/abs/2010.14538} {arXiv:2010.14538 [gr-qc]} \BibitemShut
  {NoStop}%
\bibitem [{\citenamefont {Hyun}\ \emph {et~al.}(2019)\citenamefont {Hyun},
  \citenamefont {Kim},\ and\ \citenamefont {Lee}}]{Hyun:2018pgn}%
  \BibitemOpen
  \bibfield  {author} {\bibinfo {author} {\bibfnamefont {Y.-H.}\ \bibnamefont
  {Hyun}}, \bibinfo {author} {\bibfnamefont {Y.}~\bibnamefont {Kim}},\ and\
  \bibinfo {author} {\bibfnamefont {S.}~\bibnamefont {Lee}},\ }\bibfield
  {title} {\bibinfo {title} {{Exact amplitudes of six polarization modes for
  gravitational waves}},\ }\href {https://doi.org/10.1103/PhysRevD.99.124002}
  {\bibfield  {journal} {\bibinfo  {journal} {Phys. Rev. D}\ }\textbf {\bibinfo
  {volume} {99}},\ \bibinfo {pages} {124002} (\bibinfo {year} {2019})},\
  \Eprint {https://arxiv.org/abs/1810.09316} {arXiv:1810.09316 [gr-qc]}
  \BibitemShut {NoStop}%
\bibitem [{\citenamefont {Shoom}\ \emph {et~al.}(2022)\citenamefont {Shoom},
  \citenamefont {Kumar},\ and\ \citenamefont {Krishnendu}}]{Shoom:2022cmo}%
  \BibitemOpen
  \bibfield  {author} {\bibinfo {author} {\bibfnamefont {A.~A.}\ \bibnamefont
  {Shoom}}, \bibinfo {author} {\bibfnamefont {S.}~\bibnamefont {Kumar}},\ and\
  \bibinfo {author} {\bibfnamefont {N.~V.}\ \bibnamefont {Krishnendu}},\
  }\href@noop {} {\bibinfo {title} {Constraining mass of the graviton with
  gw170817}} (\bibinfo {year} {2022}),\ \Eprint
  {https://arxiv.org/abs/2205.15432} {arXiv:2205.15432 [gr-qc]} \BibitemShut
  {NoStop}%
\bibitem [{\citenamefont
  {Pi\'orkowska-Kurpas}(2022)}]{Piorkowska-Kurpas:2022xmb}%
  \BibitemOpen
  \bibfield  {author} {\bibinfo {author} {\bibfnamefont {A.}~\bibnamefont
  {Pi\'orkowska-Kurpas}},\ }\bibfield  {title} {\bibinfo {title} {{Graviton
  Mass in the Era of Multi-Messenger Astronomy}},\ }\href
  {https://doi.org/10.3390/universe8020083} {\bibfield  {journal} {\bibinfo
  {journal} {Universe}\ }\textbf {\bibinfo {volume} {8}},\ \bibinfo {pages}
  {83} (\bibinfo {year} {2022})}\BibitemShut {NoStop}%
\bibitem [{\citenamefont {Jana}\ and\ \citenamefont
  {Mohanty}(2019)}]{Jana:2018djs}%
  \BibitemOpen
  \bibfield  {author} {\bibinfo {author} {\bibfnamefont {S.}~\bibnamefont
  {Jana}}\ and\ \bibinfo {author} {\bibfnamefont {S.}~\bibnamefont {Mohanty}},\
  }\bibfield  {title} {\bibinfo {title} {{Constraints on $f(R)$ theories of
  gravity from GW170817}},\ }\href {https://doi.org/10.1103/PhysRevD.99.044056}
  {\bibfield  {journal} {\bibinfo  {journal} {Phys. Rev. D}\ }\textbf {\bibinfo
  {volume} {99}},\ \bibinfo {pages} {044056} (\bibinfo {year} {2019})},\
  \Eprint {https://arxiv.org/abs/1807.04060} {arXiv:1807.04060 [gr-qc]}
  \BibitemShut {NoStop}%
\bibitem [{\citenamefont {Svidzinsky}\ and\ \citenamefont
  {Hilborn}(2021)}]{Svidzinsky:2018hnx}%
  \BibitemOpen
  \bibfield  {author} {\bibinfo {author} {\bibfnamefont {A.~A.}\ \bibnamefont
  {Svidzinsky}}\ and\ \bibinfo {author} {\bibfnamefont {R.~C.}\ \bibnamefont
  {Hilborn}},\ }\bibfield  {title} {\bibinfo {title} {{GW170817 event rules out
  general relativity in favor of vector gravity}},\ }\href
  {https://doi.org/10.1140/epjs/s11734-021-00080-6} {\bibfield  {journal}
  {\bibinfo  {journal} {Eur. Phys. J. ST}\ }\textbf {\bibinfo {volume} {230}},\
  \bibinfo {pages} {1149} (\bibinfo {year} {2021})},\ \Eprint
  {https://arxiv.org/abs/1804.03520} {arXiv:1804.03520 [physics.gen-ph]}
  \BibitemShut {NoStop}%
\bibitem [{\citenamefont {Nakamura}\ and\ \citenamefont
  {Deguchi}(1999)}]{Nakamura:1999uwi}%
  \BibitemOpen
  \bibfield  {author} {\bibinfo {author} {\bibfnamefont {T.~T.}\ \bibnamefont
  {Nakamura}}\ and\ \bibinfo {author} {\bibfnamefont {S.}~\bibnamefont
  {Deguchi}},\ }\bibfield  {title} {\bibinfo {title} {{Wave Optics in
  Gravitational Lensing}},\ }\href {https://doi.org/10.1143/ptps.133.137}
  {\bibfield  {journal} {\bibinfo  {journal} {Prog. Theor. Phys. Suppl.}\
  }\textbf {\bibinfo {volume} {133}},\ \bibinfo {pages} {137} (\bibinfo {year}
  {1999})}\BibitemShut {NoStop}%
\bibitem [{\citenamefont {Takahashi}\ and\ \citenamefont
  {Nakamura}(2003)}]{Takahashi:2003ix}%
  \BibitemOpen
  \bibfield  {author} {\bibinfo {author} {\bibfnamefont {R.}~\bibnamefont
  {Takahashi}}\ and\ \bibinfo {author} {\bibfnamefont {T.}~\bibnamefont
  {Nakamura}},\ }\bibfield  {title} {\bibinfo {title} {{Wave effects in
  gravitational lensing of gravitational waves from chirping binaries}},\
  }\href {https://doi.org/10.1086/377430} {\bibfield  {journal} {\bibinfo
  {journal} {Astrophys. J.}\ }\textbf {\bibinfo {volume} {595}},\ \bibinfo
  {pages} {1039} (\bibinfo {year} {2003})},\ \Eprint
  {https://arxiv.org/abs/astro-ph/0305055} {arXiv:astro-ph/0305055}
  \BibitemShut {NoStop}%
\bibitem [{\citenamefont {Liao}\ \emph
  {et~al.}(2017{\natexlab{b}})\citenamefont {Liao}, \citenamefont {Fan},
  \citenamefont {Ding}, \citenamefont {Biesiada},\ and\ \citenamefont
  {Zhu}}]{liao2017precision}%
  \BibitemOpen
  \bibfield  {author} {\bibinfo {author} {\bibfnamefont {K.}~\bibnamefont
  {Liao}}, \bibinfo {author} {\bibfnamefont {X.-L.}\ \bibnamefont {Fan}},
  \bibinfo {author} {\bibfnamefont {X.}~\bibnamefont {Ding}}, \bibinfo {author}
  {\bibfnamefont {M.}~\bibnamefont {Biesiada}},\ and\ \bibinfo {author}
  {\bibfnamefont {Z.-H.}\ \bibnamefont {Zhu}},\ }\bibfield  {title} {\bibinfo
  {title} {Precision cosmology from future lensed gravitational wave and
  electromagnetic signals},\ }\href@noop {} {\bibfield  {journal} {\bibinfo
  {journal} {Nature communications}\ }\textbf {\bibinfo {volume} {8}},\
  \bibinfo {pages} {1} (\bibinfo {year} {2017}{\natexlab{b}})}\BibitemShut
  {NoStop}%
\bibitem [{\citenamefont {Sereno}\ \emph {et~al.}(2011)\citenamefont {Sereno},
  \citenamefont {Jetzer}, \citenamefont {Sesana},\ and\ \citenamefont
  {Volonteri}}]{Sereno:2011ty}%
  \BibitemOpen
  \bibfield  {author} {\bibinfo {author} {\bibfnamefont {M.}~\bibnamefont
  {Sereno}}, \bibinfo {author} {\bibfnamefont {P.}~\bibnamefont {Jetzer}},
  \bibinfo {author} {\bibfnamefont {A.}~\bibnamefont {Sesana}},\ and\ \bibinfo
  {author} {\bibfnamefont {M.}~\bibnamefont {Volonteri}},\ }\bibfield  {title}
  {\bibinfo {title} {{Cosmography with strong lensing of LISA gravitational
  wave sources}},\ }\href {https://doi.org/10.1111/j.1365-2966.2011.18895.x}
  {\bibfield  {journal} {\bibinfo  {journal} {Mon. Not. Roy. Astron. Soc.}\
  }\textbf {\bibinfo {volume} {415}},\ \bibinfo {pages} {2773} (\bibinfo {year}
  {2011})},\ \Eprint {https://arxiv.org/abs/1104.1977} {arXiv:1104.1977
  [astro-ph.CO]} \BibitemShut {NoStop}%
\bibitem [{\citenamefont {Li}\ \emph {et~al.}(2019)\citenamefont {Li},
  \citenamefont {Fan},\ and\ \citenamefont {Gou}}]{Li:2019rns}%
  \BibitemOpen
  \bibfield  {author} {\bibinfo {author} {\bibfnamefont {Y.}~\bibnamefont
  {Li}}, \bibinfo {author} {\bibfnamefont {X.}~\bibnamefont {Fan}},\ and\
  \bibinfo {author} {\bibfnamefont {L.}~\bibnamefont {Gou}},\ }\bibfield
  {title} {\bibinfo {title} {{Constraining Cosmological Parameters in the FLRW
  Metric with Lensed GW+EM Signals}},\ }\href
  {https://doi.org/10.3847/1538-4357/ab037e} {\bibfield  {journal} {\bibinfo
  {journal} {Astrophys. J.}\ }\textbf {\bibinfo {volume} {873}},\ \bibinfo
  {pages} {37} (\bibinfo {year} {2019})},\ \Eprint
  {https://arxiv.org/abs/1901.10638} {arXiv:1901.10638 [astro-ph.CO]}
  \BibitemShut {NoStop}%
\bibitem [{\citenamefont {Diego}(2020)}]{Diego:2019rzc}%
  \BibitemOpen
  \bibfield  {author} {\bibinfo {author} {\bibfnamefont {J.~M.}\ \bibnamefont
  {Diego}},\ }\bibfield  {title} {\bibinfo {title} {{Constraining the abundance
  of primordial black holes with gravitational lensing of gravitational waves
  at LIGO frequencies}},\ }\href {https://doi.org/10.1103/PhysRevD.101.123512}
  {\bibfield  {journal} {\bibinfo  {journal} {Phys. Rev. D}\ }\textbf {\bibinfo
  {volume} {101}},\ \bibinfo {pages} {123512} (\bibinfo {year} {2020})},\
  \Eprint {https://arxiv.org/abs/1911.05736} {arXiv:1911.05736 [astro-ph.CO]}
  \BibitemShut {NoStop}%
\bibitem [{\citenamefont {Oguri}\ and\ \citenamefont
  {Takahashi}(2020)}]{Oguri:2020ldf}%
  \BibitemOpen
  \bibfield  {author} {\bibinfo {author} {\bibfnamefont {M.}~\bibnamefont
  {Oguri}}\ and\ \bibinfo {author} {\bibfnamefont {R.}~\bibnamefont
  {Takahashi}},\ }\bibfield  {title} {\bibinfo {title} {{Probing Dark Low-mass
  Halos and Primordial Black Holes with Frequency-dependent Gravitational
  Lensing Dispersions of Gravitational Waves}},\ }\href
  {https://doi.org/10.3847/1538-4357/abafab} {\bibfield  {journal} {\bibinfo
  {journal} {Astrophys. J.}\ }\textbf {\bibinfo {volume} {901}},\ \bibinfo
  {pages} {58} (\bibinfo {year} {2020})},\ \Eprint
  {https://arxiv.org/abs/2007.01936} {arXiv:2007.01936 [astro-ph.CO]}
  \BibitemShut {NoStop}%
\bibitem [{\citenamefont {Lai}\ \emph {et~al.}(2018)\citenamefont {Lai},
  \citenamefont {Hannuksela}, \citenamefont {Herrera-Mart\'\i{}n},
  \citenamefont {Diego}, \citenamefont {Broadhurst},\ and\ \citenamefont
  {Li}}]{Lai:2018rto}%
  \BibitemOpen
  \bibfield  {author} {\bibinfo {author} {\bibfnamefont {K.-H.}\ \bibnamefont
  {Lai}}, \bibinfo {author} {\bibfnamefont {O.~A.}\ \bibnamefont {Hannuksela}},
  \bibinfo {author} {\bibfnamefont {A.}~\bibnamefont {Herrera-Mart\'\i{}n}},
  \bibinfo {author} {\bibfnamefont {J.~M.}\ \bibnamefont {Diego}}, \bibinfo
  {author} {\bibfnamefont {T.}~\bibnamefont {Broadhurst}},\ and\ \bibinfo
  {author} {\bibfnamefont {T.~G.~F.}\ \bibnamefont {Li}},\ }\bibfield  {title}
  {\bibinfo {title} {{Discovering intermediate-mass black hole lenses through
  gravitational wave lensing}},\ }\href
  {https://doi.org/10.1103/PhysRevD.98.083005} {\bibfield  {journal} {\bibinfo
  {journal} {Phys. Rev. D}\ }\textbf {\bibinfo {volume} {98}},\ \bibinfo
  {pages} {083005} (\bibinfo {year} {2018})},\ \Eprint
  {https://arxiv.org/abs/1801.07840} {arXiv:1801.07840 [gr-qc]} \BibitemShut
  {NoStop}%
\bibitem [{\citenamefont {Auclair}\ \emph {et~al.}(2023)\citenamefont {Auclair}
  \emph {et~al.}}]{LISACosmologyWorkingGroup:2022jok}%
  \BibitemOpen
  \bibfield  {author} {\bibinfo {author} {\bibfnamefont {P.}~\bibnamefont
  {Auclair}} \emph {et~al.} (\bibinfo {collaboration} {LISA Cosmology Working
  Group}),\ }\bibfield  {title} {\bibinfo {title} {{Cosmology with the Laser
  Interferometer Space Antenna}},\ }\href
  {https://doi.org/10.1007/s41114-023-00045-2} {\bibfield  {journal} {\bibinfo
  {journal} {Living Rev. Rel.}\ }\textbf {\bibinfo {volume} {26}},\ \bibinfo
  {pages} {5} (\bibinfo {year} {2023})},\ \Eprint
  {https://arxiv.org/abs/2204.05434} {arXiv:2204.05434 [astro-ph.CO]}
  \BibitemShut {NoStop}%
\bibitem [{\citenamefont {Fan}\ \emph {et~al.}(2017)\citenamefont {Fan},
  \citenamefont {Liao}, \citenamefont {Biesiada}, \citenamefont
  {Piorkowska-Kurpas},\ and\ \citenamefont {Zhu}}]{Fan:2016swi}%
  \BibitemOpen
  \bibfield  {author} {\bibinfo {author} {\bibfnamefont {X.-L.}\ \bibnamefont
  {Fan}}, \bibinfo {author} {\bibfnamefont {K.}~\bibnamefont {Liao}}, \bibinfo
  {author} {\bibfnamefont {M.}~\bibnamefont {Biesiada}}, \bibinfo {author}
  {\bibfnamefont {A.}~\bibnamefont {Piorkowska-Kurpas}},\ and\ \bibinfo
  {author} {\bibfnamefont {Z.-H.}\ \bibnamefont {Zhu}},\ }\bibfield  {title}
  {\bibinfo {title} {{Speed of Gravitational Waves from Strongly Lensed
  Gravitational Waves and Electromagnetic Signals}},\ }\href
  {https://doi.org/10.1103/PhysRevLett.118.091102} {\bibfield  {journal}
  {\bibinfo  {journal} {Phys. Rev. Lett.}\ }\textbf {\bibinfo {volume} {118}},\
  \bibinfo {pages} {091102} (\bibinfo {year} {2017})},\ \Eprint
  {https://arxiv.org/abs/1612.04095} {arXiv:1612.04095 [gr-qc]} \BibitemShut
  {NoStop}%
\bibitem [{\citenamefont {Collett}\ and\ \citenamefont
  {Bacon}(2017)}]{Collett:2016dey}%
  \BibitemOpen
  \bibfield  {author} {\bibinfo {author} {\bibfnamefont {T.~E.}\ \bibnamefont
  {Collett}}\ and\ \bibinfo {author} {\bibfnamefont {D.}~\bibnamefont
  {Bacon}},\ }\bibfield  {title} {\bibinfo {title} {{Testing the speed of
  gravitational waves over cosmological distances with strong gravitational
  lensing}},\ }\href {https://doi.org/10.1103/PhysRevLett.118.091101}
  {\bibfield  {journal} {\bibinfo  {journal} {Phys. Rev. Lett.}\ }\textbf
  {\bibinfo {volume} {118}},\ \bibinfo {pages} {091101} (\bibinfo {year}
  {2017})},\ \Eprint {https://arxiv.org/abs/1602.05882} {arXiv:1602.05882
  [astro-ph.HE]} \BibitemShut {NoStop}%
\bibitem [{\citenamefont {Amaro-Seoane}\ \emph {et~al.}(2017)\citenamefont
  {Amaro-Seoane} \emph {et~al.}}]{amaro2017laser}%
  \BibitemOpen
  \bibfield  {author} {\bibinfo {author} {\bibfnamefont {P.}~\bibnamefont
  {Amaro-Seoane}} \emph {et~al.},\ }\bibfield  {title} {\bibinfo {title} {Laser
  interferometer space antenna},\ }\href@noop {} {\bibfield  {journal}
  {\bibinfo  {journal} {arXiv preprint arXiv:1702.00786}\ } (\bibinfo {year}
  {2017})}\BibitemShut {NoStop}%
\bibitem [{\citenamefont {Maggiore}\ \emph {et~al.}(2020)\citenamefont
  {Maggiore} \emph {et~al.}}]{Maggiore:2019uih}%
  \BibitemOpen
  \bibfield  {author} {\bibinfo {author} {\bibfnamefont {M.}~\bibnamefont
  {Maggiore}} \emph {et~al.},\ }\bibfield  {title} {\bibinfo {title} {{Science
  Case for the Einstein Telescope}},\ }\href
  {https://doi.org/10.1088/1475-7516/2020/03/050} {\bibfield  {journal}
  {\bibinfo  {journal} {JCAP}\ }\textbf {\bibinfo {volume} {03}},\ \bibinfo
  {pages} {050}},\ \Eprint {https://arxiv.org/abs/1912.02622} {arXiv:1912.02622
  [astro-ph.CO]} \BibitemShut {NoStop}%
\bibitem [{\citenamefont {{Evans}}\ \emph {et~al.}(2021)\citenamefont
  {{Evans}}, \citenamefont {{Adhikari}}, \citenamefont {{Afle}}, \citenamefont
  {{Ballmer}}, \citenamefont {{Biscoveanu}}, \citenamefont {{Borhanian}},
  \citenamefont {{Brown}}, \citenamefont {{Chen}}, \citenamefont
  {{Eisenstein}}, \citenamefont {{Gruson}}, \citenamefont {{Gupta}},
  \citenamefont {{Hall}}, \citenamefont {{Huxford}}, \citenamefont {{Kamai}},
  \citenamefont {{Kashyap}}, \citenamefont {{Kissel}}, \citenamefont {{Kuns}},
  \citenamefont {{Landry}}, \citenamefont {{Lenon}}, \citenamefont
  {{Lovelace}}, \citenamefont {{McCuller}}, \citenamefont {{Ng}}, \citenamefont
  {{Nitz}}, \citenamefont {{Read}}, \citenamefont {{Sathyaprakash}},
  \citenamefont {{Shoemaker}}, \citenamefont {{Slagmolen}}, \citenamefont
  {{Smith}}, \citenamefont {{Srivastava}}, \citenamefont {{Sun}}, \citenamefont
  {{Vitale}},\ and\ \citenamefont {{Weiss}}}]{Evans:2021gyd}%
  \BibitemOpen
  \bibfield  {author} {\bibinfo {author} {\bibfnamefont {M.}~\bibnamefont
  {{Evans}}}, \bibinfo {author} {\bibfnamefont {R.~X.}\ \bibnamefont
  {{Adhikari}}}, \bibinfo {author} {\bibfnamefont {C.}~\bibnamefont {{Afle}}},
  \bibinfo {author} {\bibfnamefont {S.~W.}\ \bibnamefont {{Ballmer}}}, \bibinfo
  {author} {\bibfnamefont {S.}~\bibnamefont {{Biscoveanu}}}, \bibinfo {author}
  {\bibfnamefont {S.}~\bibnamefont {{Borhanian}}}, \bibinfo {author}
  {\bibfnamefont {D.~A.}\ \bibnamefont {{Brown}}}, \bibinfo {author}
  {\bibfnamefont {Y.}~\bibnamefont {{Chen}}}, \bibinfo {author} {\bibfnamefont
  {R.}~\bibnamefont {{Eisenstein}}}, \bibinfo {author} {\bibfnamefont
  {A.}~\bibnamefont {{Gruson}}}, \bibinfo {author} {\bibfnamefont
  {A.}~\bibnamefont {{Gupta}}}, \bibinfo {author} {\bibfnamefont {E.~D.}\
  \bibnamefont {{Hall}}}, \bibinfo {author} {\bibfnamefont {R.}~\bibnamefont
  {{Huxford}}}, \bibinfo {author} {\bibfnamefont {B.}~\bibnamefont {{Kamai}}},
  \bibinfo {author} {\bibfnamefont {R.}~\bibnamefont {{Kashyap}}}, \bibinfo
  {author} {\bibfnamefont {J.~S.}\ \bibnamefont {{Kissel}}}, \bibinfo {author}
  {\bibfnamefont {K.}~\bibnamefont {{Kuns}}}, \bibinfo {author} {\bibfnamefont
  {P.}~\bibnamefont {{Landry}}}, \bibinfo {author} {\bibfnamefont
  {A.}~\bibnamefont {{Lenon}}}, \bibinfo {author} {\bibfnamefont
  {G.}~\bibnamefont {{Lovelace}}}, \bibinfo {author} {\bibfnamefont
  {L.}~\bibnamefont {{McCuller}}}, \bibinfo {author} {\bibfnamefont {K.~K.~Y.}\
  \bibnamefont {{Ng}}}, \bibinfo {author} {\bibfnamefont {A.~H.}\ \bibnamefont
  {{Nitz}}}, \bibinfo {author} {\bibfnamefont {J.}~\bibnamefont {{Read}}},
  \bibinfo {author} {\bibfnamefont {B.~S.}\ \bibnamefont {{Sathyaprakash}}},
  \bibinfo {author} {\bibfnamefont {D.~H.}\ \bibnamefont {{Shoemaker}}},
  \bibinfo {author} {\bibfnamefont {B.~J.~J.}\ \bibnamefont {{Slagmolen}}},
  \bibinfo {author} {\bibfnamefont {J.~R.}\ \bibnamefont {{Smith}}}, \bibinfo
  {author} {\bibfnamefont {V.}~\bibnamefont {{Srivastava}}}, \bibinfo {author}
  {\bibfnamefont {L.}~\bibnamefont {{Sun}}}, \bibinfo {author} {\bibfnamefont
  {S.}~\bibnamefont {{Vitale}}},\ and\ \bibinfo {author} {\bibfnamefont
  {R.}~\bibnamefont {{Weiss}}},\ }\bibfield  {title} {\bibinfo {title} {{A
  Horizon Study for Cosmic Explorer: Science, Observatories, and Community}},\
  }\href {https://doi.org/10.48550/arXiv.2109.09882} {\bibfield  {journal}
  {\bibinfo  {journal} {arXiv e-prints}\ ,\ \bibinfo {eid} {arXiv:2109.09882}}
  (\bibinfo {year} {2021})},\ \Eprint {https://arxiv.org/abs/2109.09882}
  {arXiv:2109.09882 [astro-ph.IM]} \BibitemShut {NoStop}%
\bibitem [{\citenamefont {Kawamura}\ \emph {et~al.}(2021)\citenamefont
  {Kawamura} \emph {et~al.}}]{Kawamura:2020pcg}%
  \BibitemOpen
  \bibfield  {author} {\bibinfo {author} {\bibfnamefont {S.}~\bibnamefont
  {Kawamura}} \emph {et~al.},\ }\bibfield  {title} {\bibinfo {title} {{Current
  status of space gravitational wave antenna DECIGO and B-DECIGO}},\ }\href
  {https://doi.org/10.1093/ptep/ptab019} {\bibfield  {journal} {\bibinfo
  {journal} {PTEP}\ }\textbf {\bibinfo {volume} {2021}},\ \bibinfo {pages}
  {05A105} (\bibinfo {year} {2021})},\ \Eprint
  {https://arxiv.org/abs/2006.13545} {arXiv:2006.13545 [gr-qc]} \BibitemShut
  {NoStop}%
\bibitem [{\citenamefont {Biesiada}\ \emph {et~al.}(2014)\citenamefont
  {Biesiada}, \citenamefont {Ding}, \citenamefont {Piorkowska},\ and\
  \citenamefont {Zhu}}]{Biesiada:2014kwa}%
  \BibitemOpen
  \bibfield  {author} {\bibinfo {author} {\bibfnamefont {M.}~\bibnamefont
  {Biesiada}}, \bibinfo {author} {\bibfnamefont {X.}~\bibnamefont {Ding}},
  \bibinfo {author} {\bibfnamefont {A.}~\bibnamefont {Piorkowska}},\ and\
  \bibinfo {author} {\bibfnamefont {Z.-H.}\ \bibnamefont {Zhu}},\ }\bibfield
  {title} {\bibinfo {title} {{Strong gravitational lensing of gravitational
  waves from double compact binaries - perspectives for the Einstein
  Telescope}},\ }\href {https://doi.org/10.1088/1475-7516/2014/10/080}
  {\bibfield  {journal} {\bibinfo  {journal} {JCAP}\ }\textbf {\bibinfo
  {volume} {10}},\ \bibinfo {pages} {080}},\ \Eprint
  {https://arxiv.org/abs/1409.8360} {arXiv:1409.8360 [astro-ph.HE]}
  \BibitemShut {NoStop}%
\bibitem [{\citenamefont {Pi{\'{o}}rkowska-Kurpas}\ \emph
  {et~al.}(2021)\citenamefont {Pi{\'{o}}rkowska-Kurpas}, \citenamefont {Hou},
  \citenamefont {Biesiada}, \citenamefont {Ding}, \citenamefont {Cao},
  \citenamefont {Fan}, \citenamefont {Kawamura},\ and\ \citenamefont
  {Zhu}}]{Piorkowska2021}%
  \BibitemOpen
  \bibfield  {author} {\bibinfo {author} {\bibfnamefont {A.}~\bibnamefont
  {Pi{\'{o}}rkowska-Kurpas}}, \bibinfo {author} {\bibfnamefont
  {S.}~\bibnamefont {Hou}}, \bibinfo {author} {\bibfnamefont {M.}~\bibnamefont
  {Biesiada}}, \bibinfo {author} {\bibfnamefont {X.}~\bibnamefont {Ding}},
  \bibinfo {author} {\bibfnamefont {S.}~\bibnamefont {Cao}}, \bibinfo {author}
  {\bibfnamefont {X.}~\bibnamefont {Fan}}, \bibinfo {author} {\bibfnamefont
  {S.}~\bibnamefont {Kawamura}},\ and\ \bibinfo {author} {\bibfnamefont
  {Z.-H.}\ \bibnamefont {Zhu}},\ }\bibfield  {title} {\bibinfo {title}
  {Inspiraling double compact object detection and lensing rate: Forecast for
  {DECIGO} and b-{DECIGO}},\ }\href {https://doi.org/10.3847/1538-4357/abd482}
  {\bibfield  {journal} {\bibinfo  {journal} {Astrophys. J}\ }\textbf {\bibinfo
  {volume} {908}},\ \bibinfo {pages} {196} (\bibinfo {year}
  {2021})}\BibitemShut {NoStop}%
\bibitem [{\citenamefont {Chung}\ and\ \citenamefont
  {Li}(2021)}]{Chung:2021rcu}%
  \BibitemOpen
  \bibfield  {author} {\bibinfo {author} {\bibfnamefont {A.~K.-W.}\
  \bibnamefont {Chung}}\ and\ \bibinfo {author} {\bibfnamefont {T.~G.~F.}\
  \bibnamefont {Li}},\ }\bibfield  {title} {\bibinfo {title} {{Lensing of
  gravitational waves as a novel probe of graviton mass}},\ }\href
  {https://doi.org/10.1103/PhysRevD.104.124060} {\bibfield  {journal} {\bibinfo
   {journal} {Phys. Rev. D}\ }\textbf {\bibinfo {volume} {104}},\ \bibinfo
  {pages} {124060} (\bibinfo {year} {2021})},\ \Eprint
  {https://arxiv.org/abs/2106.09630} {arXiv:2106.09630 [gr-qc]} \BibitemShut
  {NoStop}%
\bibitem [{\citenamefont {Grespan}\ and\ \citenamefont
  {Biesiada}(2023)}]{Grespan:2023cpa}%
  \BibitemOpen
  \bibfield  {author} {\bibinfo {author} {\bibfnamefont {M.}~\bibnamefont
  {Grespan}}\ and\ \bibinfo {author} {\bibfnamefont {M.}~\bibnamefont
  {Biesiada}},\ }\bibfield  {title} {\bibinfo {title} {{Strong Gravitational
  Lensing of Gravitational Waves: A Review}},\ }\href
  {https://doi.org/10.3390/universe9050200} {\bibfield  {journal} {\bibinfo
  {journal} {Universe}\ }\textbf {\bibinfo {volume} {9}},\ \bibinfo {pages}
  {200} (\bibinfo {year} {2023})}\BibitemShut {NoStop}%
\bibitem [{\citenamefont {Biesiada}\ and\ \citenamefont
  {Harikumar}(2021)}]{Biesiada:2021pzo}%
  \BibitemOpen
  \bibfield  {author} {\bibinfo {author} {\bibfnamefont {M.}~\bibnamefont
  {Biesiada}}\ and\ \bibinfo {author} {\bibfnamefont {S.}~\bibnamefont
  {Harikumar}},\ }\bibfield  {title} {\bibinfo {title} {{Gravitational Lensing
  of Continuous Gravitational Waves}},\ }\href
  {https://doi.org/10.3390/universe7120502} {\bibfield  {journal} {\bibinfo
  {journal} {Universe}\ }\textbf {\bibinfo {volume} {7}},\ \bibinfo {pages}
  {502} (\bibinfo {year} {2021})},\ \Eprint {https://arxiv.org/abs/2111.05963}
  {arXiv:2111.05963 [gr-qc]} \BibitemShut {NoStop}%
\bibitem [{\citenamefont {Eardley}\ \emph
  {et~al.}(1973{\natexlab{b}})\citenamefont {Eardley}, \citenamefont {Lee},
  \citenamefont {Lightman}, \citenamefont {Wagoner},\ and\ \citenamefont
  {Will}}]{Eardley:1973br}%
  \BibitemOpen
  \bibfield  {author} {\bibinfo {author} {\bibfnamefont {D.~M.}\ \bibnamefont
  {Eardley}}, \bibinfo {author} {\bibfnamefont {D.~L.}\ \bibnamefont {Lee}},
  \bibinfo {author} {\bibfnamefont {A.~P.}\ \bibnamefont {Lightman}}, \bibinfo
  {author} {\bibfnamefont {R.~V.}\ \bibnamefont {Wagoner}},\ and\ \bibinfo
  {author} {\bibfnamefont {C.~M.}\ \bibnamefont {Will}},\ }\bibfield  {title}
  {\bibinfo {title} {{Gravitational-wave observations as a tool for testing
  relativistic gravity}},\ }\href {https://doi.org/10.1103/PhysRevLett.30.884}
  {\bibfield  {journal} {\bibinfo  {journal} {Phys. Rev. Lett.}\ }\textbf
  {\bibinfo {volume} {30}},\ \bibinfo {pages} {884} (\bibinfo {year}
  {1973}{\natexlab{b}})}\BibitemShut {NoStop}%
\bibitem [{\citenamefont {collaboration}\ \emph {et~al.}(2023)\citenamefont
  {collaboration} \emph {et~al.}}]{LIGOScientific:2023bwz}%
  \BibitemOpen
  \bibfield  {author} {\bibinfo {author} {\bibfnamefont {T.~L.}\ \bibnamefont
  {collaboration}} \emph {et~al.},\ }\href@noop {} {\bibinfo {title} {Search
  for gravitational-lensing signatures in the full third observing run of the
  ligo-virgo network}} (\bibinfo {year} {2023}),\ \Eprint
  {https://arxiv.org/abs/2304.08393} {arXiv:2304.08393 [gr-qc]} \BibitemShut
  {NoStop}%
\bibitem [{\citenamefont {Abbott}\ \emph {et~al.}(2021)\citenamefont {Abbott}
  \emph {et~al.}}]{LIGOScientific:2021izm}%
  \BibitemOpen
  \bibfield  {author} {\bibinfo {author} {\bibfnamefont {R.}~\bibnamefont
  {Abbott}} \emph {et~al.} (\bibinfo {collaboration} {LIGO Scientific,
  VIRGO}),\ }\bibfield  {title} {\bibinfo {title} {{Search for Lensing
  Signatures in the Gravitational-Wave Observations from the First Half of
  LIGO\textendash{}Virgo\textquoteright{}s Third Observing Run}},\ }\href
  {https://doi.org/10.3847/1538-4357/ac23db} {\bibfield  {journal} {\bibinfo
  {journal} {Astrophys. J.}\ }\textbf {\bibinfo {volume} {923}},\ \bibinfo
  {pages} {14} (\bibinfo {year} {2021})},\ \Eprint
  {https://arxiv.org/abs/2105.06384} {arXiv:2105.06384 [gr-qc]} \BibitemShut
  {NoStop}%
\bibitem [{\citenamefont {Lovelock}(1971)}]{Lovelock:1971yv}%
  \BibitemOpen
  \bibfield  {author} {\bibinfo {author} {\bibfnamefont {D.}~\bibnamefont
  {Lovelock}},\ }\bibfield  {title} {\bibinfo {title} {{The Einstein tensor and
  its generalizations}},\ }\href {https://doi.org/10.1063/1.1665613} {\bibfield
   {journal} {\bibinfo  {journal} {J. Math. Phys.}\ }\textbf {\bibinfo {volume}
  {12}},\ \bibinfo {pages} {498} (\bibinfo {year} {1971})}\BibitemShut
  {NoStop}%
\bibitem [{\citenamefont {Weinberg}(2008)}]{Weinberg:2008zzc}%
  \BibitemOpen
  \bibfield  {author} {\bibinfo {author} {\bibfnamefont {S.}~\bibnamefont
  {Weinberg}},\ }\href@noop {} {\emph {\bibinfo {title} {{Cosmology}}}}\
  (\bibinfo {year} {2008})\BibitemShut {NoStop}%
\bibitem [{\citenamefont {Brax}\ \emph {et~al.}(2008)\citenamefont {Brax},
  \citenamefont {van~de Bruck}, \citenamefont {Davis},\ and\ \citenamefont
  {Shaw}}]{Brax:2008hh}%
  \BibitemOpen
  \bibfield  {author} {\bibinfo {author} {\bibfnamefont {P.}~\bibnamefont
  {Brax}}, \bibinfo {author} {\bibfnamefont {C.}~\bibnamefont {van~de Bruck}},
  \bibinfo {author} {\bibfnamefont {A.-C.}\ \bibnamefont {Davis}},\ and\
  \bibinfo {author} {\bibfnamefont {D.~J.}\ \bibnamefont {Shaw}},\ }\bibfield
  {title} {\bibinfo {title} {{f(R) Gravity and Chameleon Theories}},\ }\href
  {https://doi.org/10.1103/PhysRevD.78.104021} {\bibfield  {journal} {\bibinfo
  {journal} {Phys. Rev. D}\ }\textbf {\bibinfo {volume} {78}},\ \bibinfo
  {pages} {104021} (\bibinfo {year} {2008})},\ \Eprint
  {https://arxiv.org/abs/0806.3415} {arXiv:0806.3415 [astro-ph]} \BibitemShut
  {NoStop}%
\bibitem [{\citenamefont {Muller}\ \emph {et~al.}(1988)\citenamefont {Muller},
  \citenamefont {Schmidt},\ and\ \citenamefont {Starobinsky}}]{Muller:1987hp}%
  \BibitemOpen
  \bibfield  {author} {\bibinfo {author} {\bibfnamefont {V.}~\bibnamefont
  {Muller}}, \bibinfo {author} {\bibfnamefont {H.~J.}\ \bibnamefont
  {Schmidt}},\ and\ \bibinfo {author} {\bibfnamefont {A.~A.}\ \bibnamefont
  {Starobinsky}},\ }\bibfield  {title} {\bibinfo {title} {{The Stability of the
  De Sitter Space-time in Fourth Order Gravity}},\ }\href
  {https://doi.org/10.1016/0370-2693(88)90007-X} {\bibfield  {journal}
  {\bibinfo  {journal} {Phys. Lett. B}\ }\textbf {\bibinfo {volume} {202}},\
  \bibinfo {pages} {198} (\bibinfo {year} {1988})}\BibitemShut {NoStop}%
\bibitem [{\citenamefont {Kehagias}\ \emph {et~al.}(2015)\citenamefont
  {Kehagias}, \citenamefont {Kounnas}, \citenamefont {L\"ust},\ and\
  \citenamefont {Riotto}}]{Kehagias:2015ata}%
  \BibitemOpen
  \bibfield  {author} {\bibinfo {author} {\bibfnamefont {A.}~\bibnamefont
  {Kehagias}}, \bibinfo {author} {\bibfnamefont {C.}~\bibnamefont {Kounnas}},
  \bibinfo {author} {\bibfnamefont {D.}~\bibnamefont {L\"ust}},\ and\ \bibinfo
  {author} {\bibfnamefont {A.}~\bibnamefont {Riotto}},\ }\bibfield  {title}
  {\bibinfo {title} {{Black hole solutions in $R^{2}$ gravity}},\ }\href
  {https://doi.org/10.1007/JHEP05(2015)143} {\bibfield  {journal} {\bibinfo
  {journal} {JHEP}\ }\textbf {\bibinfo {volume} {05}},\ \bibinfo {pages}
  {143}},\ \Eprint {https://arxiv.org/abs/1502.04192} {arXiv:1502.04192
  [hep-th]} \BibitemShut {NoStop}%
\bibitem [{\citenamefont {Dalang}\ \emph {et~al.}(2021)\citenamefont {Dalang},
  \citenamefont {Fleury},\ and\ \citenamefont {Lombriser}}]{Dalang_2021}%
  \BibitemOpen
  \bibfield  {author} {\bibinfo {author} {\bibfnamefont {C.}~\bibnamefont
  {Dalang}}, \bibinfo {author} {\bibfnamefont {P.}~\bibnamefont {Fleury}},\
  and\ \bibinfo {author} {\bibfnamefont {L.}~\bibnamefont {Lombriser}},\
  }\bibfield  {title} {\bibinfo {title} {Scalar and tensor gravitational
  waves},\ }\href {https://doi.org/10.1103/PhysRevD.103.064075} {\bibfield
  {journal} {\bibinfo  {journal} {Phys. Rev. D}\ }\textbf {\bibinfo {volume}
  {103}},\ \bibinfo {pages} {064075} (\bibinfo {year} {2021})}\BibitemShut
  {NoStop}%
\bibitem [{\citenamefont {Misner}\ \emph {et~al.}(1973)\citenamefont {Misner},
  \citenamefont {Thorne},\ and\ \citenamefont {Wheeler}}]{Misner:1973prb}%
  \BibitemOpen
  \bibfield  {author} {\bibinfo {author} {\bibfnamefont {C.~W.}\ \bibnamefont
  {Misner}}, \bibinfo {author} {\bibfnamefont {K.~S.}\ \bibnamefont {Thorne}},\
  and\ \bibinfo {author} {\bibfnamefont {J.~A.}\ \bibnamefont {Wheeler}},\
  }\href@noop {} {\emph {\bibinfo {title} {Gravitation}}}\ (\bibinfo
  {publisher} {W. H. Freeman},\ \bibinfo {address} {San Francisco},\ \bibinfo
  {year} {1973})\BibitemShut {NoStop}%
\bibitem [{\citenamefont {Will}(2018{\natexlab{b}})}]{Will:2018bme}%
  \BibitemOpen
  \bibfield  {author} {\bibinfo {author} {\bibfnamefont {C.~M.}\ \bibnamefont
  {Will}},\ }\href@noop {} {\emph {\bibinfo {title} {Theory and Experiment in
  Gravitational Physics}}}\ (\bibinfo  {publisher} {Cambridge University
  Press},\ \bibinfo {year} {2018})\BibitemShut {NoStop}%
\bibitem [{\citenamefont {Mirshekari}\ \emph {et~al.}(2012)\citenamefont
  {Mirshekari}, \citenamefont {Yunes},\ and\ \citenamefont
  {Will}}]{Mirshekari:2011yq}%
  \BibitemOpen
  \bibfield  {author} {\bibinfo {author} {\bibfnamefont {S.}~\bibnamefont
  {Mirshekari}}, \bibinfo {author} {\bibfnamefont {N.}~\bibnamefont {Yunes}},\
  and\ \bibinfo {author} {\bibfnamefont {C.~M.}\ \bibnamefont {Will}},\
  }\bibfield  {title} {\bibinfo {title} {{Constraining Generic Lorentz
  Violation and the Speed of the Graviton with Gravitational Waves}},\ }\href
  {https://doi.org/10.1103/PhysRevD.85.024041} {\bibfield  {journal} {\bibinfo
  {journal} {Phys. Rev. D}\ }\textbf {\bibinfo {volume} {85}},\ \bibinfo
  {pages} {024041} (\bibinfo {year} {2012})},\ \Eprint
  {https://arxiv.org/abs/1110.2720} {arXiv:1110.2720 [gr-qc]} \BibitemShut
  {NoStop}%
\bibitem [{\citenamefont {Nishizawa}\ and\ \citenamefont
  {Nakamura}(2014)}]{Nishizawa:2014zna}%
  \BibitemOpen
  \bibfield  {author} {\bibinfo {author} {\bibfnamefont {A.}~\bibnamefont
  {Nishizawa}}\ and\ \bibinfo {author} {\bibfnamefont {T.}~\bibnamefont
  {Nakamura}},\ }\bibfield  {title} {\bibinfo {title} {{Measuring Speed of
  Gravitational Waves by Observations of Photons and Neutrinos from Compact
  Binary Mergers and Supernovae}},\ }\href
  {https://doi.org/10.1103/PhysRevD.90.044048} {\bibfield  {journal} {\bibinfo
  {journal} {Phys. Rev. D}\ }\textbf {\bibinfo {volume} {90}},\ \bibinfo
  {pages} {044048} (\bibinfo {year} {2014})},\ \Eprint
  {https://arxiv.org/abs/1406.5544} {arXiv:1406.5544 [gr-qc]} \BibitemShut
  {NoStop}%
\bibitem [{\citenamefont {Yunes}\ \emph {et~al.}(2016)\citenamefont {Yunes},
  \citenamefont {Yagi},\ and\ \citenamefont {Pretorius}}]{Yunes:2016jcc}%
  \BibitemOpen
  \bibfield  {author} {\bibinfo {author} {\bibfnamefont {N.}~\bibnamefont
  {Yunes}}, \bibinfo {author} {\bibfnamefont {K.}~\bibnamefont {Yagi}},\ and\
  \bibinfo {author} {\bibfnamefont {F.}~\bibnamefont {Pretorius}},\ }\bibfield
  {title} {\bibinfo {title} {{Theoretical Physics Implications of the Binary
  Black-Hole Mergers GW150914 and GW151226}},\ }\href
  {https://doi.org/10.1103/PhysRevD.94.084002} {\bibfield  {journal} {\bibinfo
  {journal} {Phys. Rev. D}\ }\textbf {\bibinfo {volume} {94}},\ \bibinfo
  {pages} {084002} (\bibinfo {year} {2016})},\ \Eprint
  {https://arxiv.org/abs/1603.08955} {arXiv:1603.08955 [gr-qc]} \BibitemShut
  {NoStop}%
\bibitem [{\citenamefont {Morisaki}\ and\ \citenamefont
  {Suyama}(2019)}]{Morisaki:2018htj}%
  \BibitemOpen
  \bibfield  {author} {\bibinfo {author} {\bibfnamefont {S.}~\bibnamefont
  {Morisaki}}\ and\ \bibinfo {author} {\bibfnamefont {T.}~\bibnamefont
  {Suyama}},\ }\bibfield  {title} {\bibinfo {title} {{Detectability of
  ultralight scalar field dark matter with gravitational-wave detectors}},\
  }\href {https://doi.org/10.1103/PhysRevD.100.123512} {\bibfield  {journal}
  {\bibinfo  {journal} {Phys. Rev. D}\ }\textbf {\bibinfo {volume} {100}},\
  \bibinfo {pages} {123512} (\bibinfo {year} {2019})},\ \Eprint
  {https://arxiv.org/abs/1811.05003} {arXiv:1811.05003 [hep-ph]} \BibitemShut
  {NoStop}%
\bibitem [{\citenamefont {Brito}\ \emph {et~al.}(2017)\citenamefont {Brito},
  \citenamefont {Ghosh}, \citenamefont {Barausse}, \citenamefont {Berti},
  \citenamefont {Cardoso}, \citenamefont {Dvorkin}, \citenamefont {Klein},\
  and\ \citenamefont {Pani}}]{Brito:2017zvb}%
  \BibitemOpen
  \bibfield  {author} {\bibinfo {author} {\bibfnamefont {R.}~\bibnamefont
  {Brito}}, \bibinfo {author} {\bibfnamefont {S.}~\bibnamefont {Ghosh}},
  \bibinfo {author} {\bibfnamefont {E.}~\bibnamefont {Barausse}}, \bibinfo
  {author} {\bibfnamefont {E.}~\bibnamefont {Berti}}, \bibinfo {author}
  {\bibfnamefont {V.}~\bibnamefont {Cardoso}}, \bibinfo {author} {\bibfnamefont
  {I.}~\bibnamefont {Dvorkin}}, \bibinfo {author} {\bibfnamefont
  {A.}~\bibnamefont {Klein}},\ and\ \bibinfo {author} {\bibfnamefont
  {P.}~\bibnamefont {Pani}},\ }\bibfield  {title} {\bibinfo {title}
  {{Gravitational wave searches for ultralight bosons with LIGO and LISA}},\
  }\href {https://doi.org/10.1103/PhysRevD.96.064050} {\bibfield  {journal}
  {\bibinfo  {journal} {Phys. Rev. D}\ }\textbf {\bibinfo {volume} {96}},\
  \bibinfo {pages} {064050} (\bibinfo {year} {2017})},\ \Eprint
  {https://arxiv.org/abs/1706.06311} {arXiv:1706.06311 [gr-qc]} \BibitemShut
  {NoStop}%
\bibitem [{\citenamefont {Harry}\ and\ \citenamefont
  {Noller}(2022)}]{Harry:2022zey}%
  \BibitemOpen
  \bibfield  {author} {\bibinfo {author} {\bibfnamefont {I.}~\bibnamefont
  {Harry}}\ and\ \bibinfo {author} {\bibfnamefont {J.}~\bibnamefont {Noller}},\
  }\bibfield  {title} {\bibinfo {title} {{Probing the speed of gravity with
  LVK, LISA, and joint observations}},\ }\href
  {https://doi.org/10.1007/s10714-022-03016-0} {\bibfield  {journal} {\bibinfo
  {journal} {Gen. Rel. Grav.}\ }\textbf {\bibinfo {volume} {54}},\ \bibinfo
  {pages} {133} (\bibinfo {year} {2022})},\ \Eprint
  {https://arxiv.org/abs/2207.10096} {arXiv:2207.10096 [gr-qc]} \BibitemShut
  {NoStop}%
\bibitem [{\citenamefont {Yagi}\ and\ \citenamefont
  {Tanaka}(2010)}]{Yagi:2009zm}%
  \BibitemOpen
  \bibfield  {author} {\bibinfo {author} {\bibfnamefont {K.}~\bibnamefont
  {Yagi}}\ and\ \bibinfo {author} {\bibfnamefont {T.}~\bibnamefont {Tanaka}},\
  }\bibfield  {title} {\bibinfo {title} {{Constraining alternative theories of
  gravity by gravitational waves from precessing eccentric compact binaries
  with LISA}},\ }\href {https://doi.org/10.1103/PhysRevD.81.109902} {\bibfield
  {journal} {\bibinfo  {journal} {Phys. Rev. D}\ }\textbf {\bibinfo {volume}
  {81}},\ \bibinfo {pages} {064008} (\bibinfo {year} {2010})},\ \bibinfo {note}
  {[Erratum: Phys.Rev.D 81, 109902 (2010)]},\ \Eprint
  {https://arxiv.org/abs/0906.4269} {arXiv:0906.4269 [gr-qc]} \BibitemShut
  {NoStop}%
\bibitem [{\citenamefont {Baker}\ \emph {et~al.}(2022)\citenamefont {Baker}
  \emph {et~al.}}]{LISACosmologyWorkingGroup:2022wjo}%
  \BibitemOpen
  \bibfield  {author} {\bibinfo {author} {\bibfnamefont {T.}~\bibnamefont
  {Baker}} \emph {et~al.} (\bibinfo {collaboration} {LISA Cosmology Working
  Group}),\ }\bibfield  {title} {\bibinfo {title} {{Measuring the propagation
  speed of gravitational waves with LISA}},\ }\href
  {https://doi.org/10.1088/1475-7516/2022/08/031} {\bibfield  {journal}
  {\bibinfo  {journal} {JCAP}\ }\textbf {\bibinfo {volume} {08}}\bibfield
  {number} {\bibinfo  {number} { (08)},\ \bibinfo {pages} {031}},\ }\Eprint
  {https://arxiv.org/abs/2203.00566} {arXiv:2203.00566 [gr-qc]} \BibitemShut
  {NoStop}%
\bibitem [{\citenamefont {Sato}\ \emph {et~al.}(2017)\citenamefont {Sato} \emph
  {et~al.}}]{Sato:2017dkf}%
  \BibitemOpen
  \bibfield  {author} {\bibinfo {author} {\bibfnamefont {S.}~\bibnamefont
  {Sato}} \emph {et~al.},\ }\bibfield  {title} {\bibinfo {title} {{The status
  of DECIGO}},\ }\href {https://doi.org/10.1088/1742-6596/840/1/012010}
  {\bibfield  {journal} {\bibinfo  {journal} {J. Phys. Conf. Ser.}\ }\textbf
  {\bibinfo {volume} {840}},\ \bibinfo {pages} {012010} (\bibinfo {year}
  {2017})}\BibitemShut {NoStop}%
\bibitem [{\citenamefont {Eardley}\ \emph
  {et~al.}(1973{\natexlab{c}})\citenamefont {Eardley}, \citenamefont {Lee},\
  and\ \citenamefont {Lightman}}]{Eardley:1973zuo}%
  \BibitemOpen
  \bibfield  {author} {\bibinfo {author} {\bibfnamefont {D.~M.}\ \bibnamefont
  {Eardley}}, \bibinfo {author} {\bibfnamefont {D.~L.}\ \bibnamefont {Lee}},\
  and\ \bibinfo {author} {\bibfnamefont {A.~P.}\ \bibnamefont {Lightman}},\
  }\bibfield  {title} {\bibinfo {title} {{Gravitational-wave observations as a
  tool for testing relativistic gravity}},\ }\href
  {https://doi.org/10.1103/PhysRevD.8.3308} {\bibfield  {journal} {\bibinfo
  {journal} {Phys. Rev. D}\ }\textbf {\bibinfo {volume} {8}},\ \bibinfo {pages}
  {3308} (\bibinfo {year} {1973}{\natexlab{c}})}\BibitemShut {NoStop}%
\bibitem [{\citenamefont {Rizwana~Kausar}\ \emph {et~al.}(2016)\citenamefont
  {Rizwana~Kausar}, \citenamefont {Philippoz},\ and\ \citenamefont
  {Jetzer}}]{RizwanaKausar:2016zgi}%
  \BibitemOpen
  \bibfield  {author} {\bibinfo {author} {\bibfnamefont {H.}~\bibnamefont
  {Rizwana~Kausar}}, \bibinfo {author} {\bibfnamefont {L.}~\bibnamefont
  {Philippoz}},\ and\ \bibinfo {author} {\bibfnamefont {P.}~\bibnamefont
  {Jetzer}},\ }\bibfield  {title} {\bibinfo {title} {{Gravitational Wave
  Polarization Modes in $f(R)$ Theories}},\ }\href
  {https://doi.org/10.1103/PhysRevD.93.124071} {\bibfield  {journal} {\bibinfo
  {journal} {Phys. Rev. D}\ }\textbf {\bibinfo {volume} {93}},\ \bibinfo
  {pages} {124071} (\bibinfo {year} {2016})},\ \Eprint
  {https://arxiv.org/abs/1606.07000} {arXiv:1606.07000 [gr-qc]} \BibitemShut
  {NoStop}%
\bibitem [{\citenamefont {Isi}\ \emph {et~al.}(2017)\citenamefont {Isi},
  \citenamefont {Pitkin},\ and\ \citenamefont {Weinstein}}]{Isi:2017equ}%
  \BibitemOpen
  \bibfield  {author} {\bibinfo {author} {\bibfnamefont {M.}~\bibnamefont
  {Isi}}, \bibinfo {author} {\bibfnamefont {M.}~\bibnamefont {Pitkin}},\ and\
  \bibinfo {author} {\bibfnamefont {A.~J.}\ \bibnamefont {Weinstein}},\
  }\bibfield  {title} {\bibinfo {title} {{Probing Dynamical Gravity with the
  Polarization of Continuous Gravitational Waves}},\ }\href
  {https://doi.org/10.1103/PhysRevD.96.042001} {\bibfield  {journal} {\bibinfo
  {journal} {Phys. Rev. D}\ }\textbf {\bibinfo {volume} {96}},\ \bibinfo
  {pages} {042001} (\bibinfo {year} {2017})},\ \Eprint
  {https://arxiv.org/abs/1703.07530} {arXiv:1703.07530 [gr-qc]} \BibitemShut
  {NoStop}%
\bibitem [{\citenamefont {Brans}\ and\ \citenamefont
  {Dicke}(1961)}]{Brans:1961sx}%
  \BibitemOpen
  \bibfield  {author} {\bibinfo {author} {\bibfnamefont {C.}~\bibnamefont
  {Brans}}\ and\ \bibinfo {author} {\bibfnamefont {R.~H.}\ \bibnamefont
  {Dicke}},\ }\bibfield  {title} {\bibinfo {title} {{Mach's principle and a
  relativistic theory of gravitation}},\ }\href
  {https://doi.org/10.1103/PhysRev.124.925} {\bibfield  {journal} {\bibinfo
  {journal} {Phys. Rev.}\ }\textbf {\bibinfo {volume} {124}},\ \bibinfo {pages}
  {925} (\bibinfo {year} {1961})}\BibitemShut {NoStop}%
\bibitem [{\citenamefont {Vilhena}\ \emph {et~al.}(2021)\citenamefont
  {Vilhena}, \citenamefont {Medeiros},\ and\ \citenamefont
  {Cuzinatto}}]{Vilhena:2021bsx}%
  \BibitemOpen
  \bibfield  {author} {\bibinfo {author} {\bibfnamefont {S.~G.}\ \bibnamefont
  {Vilhena}}, \bibinfo {author} {\bibfnamefont {L.~G.}\ \bibnamefont
  {Medeiros}},\ and\ \bibinfo {author} {\bibfnamefont {R.~R.}\ \bibnamefont
  {Cuzinatto}},\ }\bibfield  {title} {\bibinfo {title} {{Gravitational waves in
  higher-order R2 gravity}},\ }\href
  {https://doi.org/10.1103/PhysRevD.104.084061} {\bibfield  {journal} {\bibinfo
   {journal} {Phys. Rev. D}\ }\textbf {\bibinfo {volume} {104}},\ \bibinfo
  {pages} {084061} (\bibinfo {year} {2021})},\ \Eprint
  {https://arxiv.org/abs/2108.06874} {arXiv:2108.06874 [gr-qc]} \BibitemShut
  {NoStop}%
\bibitem [{\citenamefont {Dalang}\ \emph {et~al.}(2022)\citenamefont {Dalang},
  \citenamefont {Cusin},\ and\ \citenamefont {Lagos}}]{Dalang:2021qhu}%
  \BibitemOpen
  \bibfield  {author} {\bibinfo {author} {\bibfnamefont {C.}~\bibnamefont
  {Dalang}}, \bibinfo {author} {\bibfnamefont {G.}~\bibnamefont {Cusin}},\ and\
  \bibinfo {author} {\bibfnamefont {M.}~\bibnamefont {Lagos}},\ }\bibfield
  {title} {\bibinfo {title} {{Polarization distortions of lensed gravitational
  waves}},\ }\href {https://doi.org/10.1103/PhysRevD.105.024005} {\bibfield
  {journal} {\bibinfo  {journal} {Phys. Rev. D}\ }\textbf {\bibinfo {volume}
  {105}},\ \bibinfo {pages} {024005} (\bibinfo {year} {2022})},\ \Eprint
  {https://arxiv.org/abs/2104.10119} {arXiv:2104.10119 [gr-qc]} \BibitemShut
  {NoStop}%
\bibitem [{\citenamefont {Miller}\ and\ \citenamefont
  {Mendes}(2023)}]{Miller:2023kkd}%
  \BibitemOpen
  \bibfield  {author} {\bibinfo {author} {\bibfnamefont {A.~L.}\ \bibnamefont
  {Miller}}\ and\ \bibinfo {author} {\bibfnamefont {L.}~\bibnamefont
  {Mendes}},\ }\bibfield  {title} {\bibinfo {title} {{First search for
  ultralight dark matter with a space-based gravitational-wave antenna: LISA
  Pathfinder}},\ }\href {https://doi.org/10.1103/PhysRevD.107.063015}
  {\bibfield  {journal} {\bibinfo  {journal} {Phys. Rev. D}\ }\textbf {\bibinfo
  {volume} {107}},\ \bibinfo {pages} {063015} (\bibinfo {year} {2023})},\
  \Eprint {https://arxiv.org/abs/2301.08736} {arXiv:2301.08736 [gr-qc]}
  \BibitemShut {NoStop}%
\bibitem [{\citenamefont {{Brito}}\ \emph {et~al.}(2022)\citenamefont
  {{Brito}}, \citenamefont {{Chakrabarti}}, \citenamefont {{Clesse}},
  \citenamefont {{Dvorkin}}, \citenamefont {{Garcia-Bellido}}, \citenamefont
  {{Meyers}}, \citenamefont {{Ng}}, \citenamefont {{Miller}}, \citenamefont
  {{Shandera}},\ and\ \citenamefont {{Sun}}}]{Brito:2022lmd}%
  \BibitemOpen
  \bibfield  {author} {\bibinfo {author} {\bibfnamefont {R.}~\bibnamefont
  {{Brito}}}, \bibinfo {author} {\bibfnamefont {S.}~\bibnamefont
  {{Chakrabarti}}}, \bibinfo {author} {\bibfnamefont {S.}~\bibnamefont
  {{Clesse}}}, \bibinfo {author} {\bibfnamefont {C.}~\bibnamefont {{Dvorkin}}},
  \bibinfo {author} {\bibfnamefont {J.}~\bibnamefont {{Garcia-Bellido}}},
  \bibinfo {author} {\bibfnamefont {J.}~\bibnamefont {{Meyers}}}, \bibinfo
  {author} {\bibfnamefont {K.~K.~Y.}\ \bibnamefont {{Ng}}}, \bibinfo {author}
  {\bibfnamefont {A.~L.}\ \bibnamefont {{Miller}}}, \bibinfo {author}
  {\bibfnamefont {S.}~\bibnamefont {{Shandera}}},\ and\ \bibinfo {author}
  {\bibfnamefont {L.}~\bibnamefont {{Sun}}},\ }\bibfield  {title} {\bibinfo
  {title} {{Snowmass2021 Cosmic Frontier White Paper: Probing dark matter with
  small-scale astrophysical observations}},\ }\href
  {https://doi.org/10.48550/arXiv.2203.15954} {\bibfield  {journal} {\bibinfo
  {journal} {arXiv e-prints}\ ,\ \bibinfo {eid} {arXiv:2203.15954}} (\bibinfo
  {year} {2022})},\ \Eprint {https://arxiv.org/abs/2203.15954}
  {arXiv:2203.15954 [hep-ph]} \BibitemShut {NoStop}%
\bibitem [{\citenamefont {Yu}\ \emph {et~al.}(2023)\citenamefont {Yu},
  \citenamefont {Yao}, \citenamefont {Tang},\ and\ \citenamefont
  {Wu}}]{Yu:2023iog}%
  \BibitemOpen
  \bibfield  {author} {\bibinfo {author} {\bibfnamefont {J.-C.}\ \bibnamefont
  {Yu}}, \bibinfo {author} {\bibfnamefont {Y.-H.}\ \bibnamefont {Yao}},
  \bibinfo {author} {\bibfnamefont {Y.}~\bibnamefont {Tang}},\ and\ \bibinfo
  {author} {\bibfnamefont {Y.-L.}\ \bibnamefont {Wu}},\ }\bibfield  {title}
  {\bibinfo {title} {{Sensitivity of space-based gravitational-wave
  interferometers to ultralight bosonic fields and dark matter}},\ }\href
  {https://doi.org/10.1103/PhysRevD.108.083007} {\bibfield  {journal} {\bibinfo
   {journal} {Phys. Rev. D}\ }\textbf {\bibinfo {volume} {108}},\ \bibinfo
  {pages} {083007} (\bibinfo {year} {2023})},\ \Eprint
  {https://arxiv.org/abs/2307.09197} {arXiv:2307.09197 [gr-qc]} \BibitemShut
  {NoStop}%
\bibitem [{\citenamefont {Garoffolo}\ \emph {et~al.}(2020)\citenamefont
  {Garoffolo}, \citenamefont {Tasinato}, \citenamefont {Carbone}, \citenamefont
  {Bertacca},\ and\ \citenamefont {Matarrese}}]{Garoffolo:2019mna}%
  \BibitemOpen
  \bibfield  {author} {\bibinfo {author} {\bibfnamefont {A.}~\bibnamefont
  {Garoffolo}}, \bibinfo {author} {\bibfnamefont {G.}~\bibnamefont {Tasinato}},
  \bibinfo {author} {\bibfnamefont {C.}~\bibnamefont {Carbone}}, \bibinfo
  {author} {\bibfnamefont {D.}~\bibnamefont {Bertacca}},\ and\ \bibinfo
  {author} {\bibfnamefont {S.}~\bibnamefont {Matarrese}},\ }\bibfield  {title}
  {\bibinfo {title} {{Gravitational waves and geometrical optics in
  scalar-tensor theories}},\ }\href
  {https://doi.org/10.1088/1475-7516/2020/11/040} {\bibfield  {journal}
  {\bibinfo  {journal} {JCAP}\ }\textbf {\bibinfo {volume} {11}},\ \bibinfo
  {pages} {040}},\ \Eprint {https://arxiv.org/abs/1912.08093} {arXiv:1912.08093
  [gr-qc]} \BibitemShut {NoStop}%
\bibitem [{\citenamefont {Hui}(2021)}]{Hui:2021tkt}%
  \BibitemOpen
  \bibfield  {author} {\bibinfo {author} {\bibfnamefont {L.}~\bibnamefont
  {Hui}},\ }\bibfield  {title} {\bibinfo {title} {{Wave Dark Matter}},\ }\href
  {https://doi.org/10.1146/annurev-astro-120920-010024} {\bibfield  {journal}
  {\bibinfo  {journal} {Ann. Rev. Astron. Astrophys.}\ }\textbf {\bibinfo
  {volume} {59}},\ \bibinfo {pages} {247} (\bibinfo {year} {2021})},\ \Eprint
  {https://arxiv.org/abs/2101.11735} {arXiv:2101.11735 [astro-ph.CO]}
  \BibitemShut {NoStop}%
\bibitem [{\citenamefont {{Schneider}}\ \emph {et~al.}(1992)\citenamefont
  {{Schneider}}, \citenamefont {{Ehlers}},\ and\ \citenamefont
  {{Falco}}}]{Gravitational_lenses1992}%
  \BibitemOpen
  \bibfield  {author} {\bibinfo {author} {\bibfnamefont {P.}~\bibnamefont
  {{Schneider}}}, \bibinfo {author} {\bibfnamefont {J.}~\bibnamefont
  {{Ehlers}}},\ and\ \bibinfo {author} {\bibfnamefont {E.~E.}\ \bibnamefont
  {{Falco}}},\ }\href {https://doi.org/10.1007/978-3-662-03758-4} {\emph
  {\bibinfo {title} {{Gravitational Lenses}}}}\ (\bibinfo {year}
  {1992})\BibitemShut {NoStop}%
\bibitem [{\citenamefont {Garg}\ and\ \citenamefont
  {Dodin}(2022{\natexlab{a}})}]{Garg:2022wdm}%
  \BibitemOpen
  \bibfield  {author} {\bibinfo {author} {\bibfnamefont {D.}~\bibnamefont
  {Garg}}\ and\ \bibinfo {author} {\bibfnamefont {I.~Y.}\ \bibnamefont
  {Dodin}},\ }\bibfield  {title} {\bibinfo {title} {{Gravitational wave modes
  in matter}},\ }\href {https://doi.org/10.1088/1475-7516/2022/08/017}
  {\bibfield  {journal} {\bibinfo  {journal} {JCAP}\ }\textbf {\bibinfo
  {volume} {08}}\bibfield  {number} {\bibinfo  {number} { (08)},\ \bibinfo
  {pages} {017}},\ }\Eprint {https://arxiv.org/abs/2204.09095}
  {arXiv:2204.09095 [gr-qc]} \BibitemShut {NoStop}%
\bibitem [{\citenamefont {Garg}\ and\ \citenamefont
  {Dodin}(2022{\natexlab{b}})}]{Garg:2021jss}%
  \BibitemOpen
  \bibfield  {author} {\bibinfo {author} {\bibfnamefont {D.}~\bibnamefont
  {Garg}}\ and\ \bibinfo {author} {\bibfnamefont {I.~Y.}\ \bibnamefont
  {Dodin}},\ }\bibfield  {title} {\bibinfo {title} {{Gauge invariants of
  linearized gravity with a general background metric}},\ }\href
  {https://doi.org/10.1088/1361-6382/aca067} {\bibfield  {journal} {\bibinfo
  {journal} {Class. Quant. Grav.}\ }\textbf {\bibinfo {volume} {39}},\ \bibinfo
  {pages} {245003} (\bibinfo {year} {2022}{\natexlab{b}})},\ \Eprint
  {https://arxiv.org/abs/2105.04680} {arXiv:2105.04680 [gr-qc]} \BibitemShut
  {NoStop}%
\end{thebibliography}%

\end{document}